\documentclass[a4paper,fleqn,usenatbib]{mnras}

\usepackage{mathptmx}

\usepackage[T1]{fontenc}
\usepackage{ae,aecompl}

\usepackage{pdflscape}

\usepackage{graphicx}	
\usepackage{amsmath}	
\usepackage{amssymb}	

\title[The UTMOST pulsar timing programme I]{The UTMOST pulsar timing programme I: overview and first results}

\author[F.~Jankowski et al.]{F.~Jankowski,$^{1,2,3}$\thanks{E-mail: fjankowsk@gmail.com} M.~Bailes,$^{2}$ W.~van~Straten,$^{4,2,3}$ E.~F.~Keane,$^{5,2,3}$
C.~Flynn,$^{2}$\newauthor
E.~D.~Barr,$^{6,2}$
T.~Bateman,$^{7}$
S.~Bhandari,$^{2,3,8}$
M.~Caleb,$^{1,2,3}$
D.~Campbell-Wilson,$^{3,7}$\newauthor
W.~Farah,$^{2}$
A.~J.~Green,$^{3,7}$
R.~W.~Hunstead,$^{7}$
A.~Jameson,$^{2,3}$
S.~Os{\l}owski,$^{2}$\newauthor
A.~Parthasarathy,$^{2,3}$
P.~A.~Rosado,$^{2}$ and V.~Venkatraman~Krishnan$^{2,3,6}$
\\
$^{1}$Jodrell Bank Centre for Astrophysics, School of Physics and Astronomy, The University of Manchester, Manchester M13 9PL, UK\\
$^{2}$Centre for Astrophysics and Supercomputing, Swinburne University of Technology, PO Box 218, Hawthorn, VIC 3122, Australia\\
$^{3}$ARC Centre of Excellence for All-Sky Astrophysics (CAASTRO)\\
$^{4}$Institute for Radio Astronomy and Space Research, Auckland University of Technology, Private Bag 92006, Auckland 1142, New Zealand\\
$^{5}$SKA Organisation, Jodrell Bank Observatory, Cheshire, SK11 9DL, UK\\
$^{6}$Max-Planck-Institut f\"ur Radioastronomie, Auf dem H\"ugel 69, D-53121 Bonn, Germany\\
$^{7}$Sydney Institute for Astronomy (SIfA), School of Physics, The University of Sydney, NSW 2006, Australia\\
$^{8}$CSIRO Astronomy and Space Science, PO Box 76, Epping, NSW 1710, Australia}

\date{Accepted XXX. Received YYY; in original form ZZZ}

\pubyear{2018}

\begin{document}
\label{firstpage}
\pagerange{\pageref{firstpage}--\pageref{lastpage}}
\maketitle

\begin{abstract}
We present an overview and the first results from a large-scale pulsar timing programme that is part of the UTMOST project at the refurbished Molonglo Observatory Synthesis Radio Telescope (MOST) near Canberra, Australia. We currently observe more than 400 mainly bright southern radio pulsars with up to daily cadences. For 205 (8 in binaries, 4 millisecond pulsars) we publish updated timing models, together with their flux densities, flux density variability, and pulse widths at 843~MHz, derived from observations spanning between 1.4 and 3 yr. In comparison with the ATNF pulsar catalogue, we improve the precision of the rotational and astrometric parameters for 123 pulsars, for 47 by at least an order of magnitude. The time spans between our measurements and those in the literature are up to 48 yr, which allows us to investigate their long-term spin-down history and to estimate proper motions for 60 pulsars, of which 24 are newly determined and most are major improvements. The results are consistent with interferometric measurements from the literature. A model with two Gaussian components centred at 139 and $463~\text{km} \: \text{s}^{-1}$ fits the transverse velocity distribution best. The pulse duty cycle distributions at 50 and 10 per cent maximum are best described by log-normal distributions with medians of 2.3 and 4.4 per cent, respectively. We discuss two pulsars that exhibit spin-down rate changes and drifting subpulses. Finally, we describe the autonomous observing system and the dynamic scheduler that has increased the observing efficiency by a factor of 2--3 in comparison with static scheduling.
\end{abstract}




\section{Introduction}
\label{sec:Introduction}

The scientific motivations for pulsar timing studies are manyfold and range from precision astrometry, to understanding the physics of the pulsar emission, the properties of ultra-dense matter and the structure of neutron stars, to tests of the laws of General Relativity and the search for low-frequency gravitational waves (e.g.\ \citealt{1983Rankin, 2014Wex, 2015Watts, 2016Verbiest, 2017Wang}). The UTMOST project is a major upgrade of the Molonglo Observatory Synthesis Radio Telescope (MOST) near Canberra, Australia. The project has transformed the telescope into a powerful instrument for large-scale pulsar observations and the search for single pulse events such as fast radio bursts (FRBs; \citealt{2017Bailes}). Eight FRBs have been discovered so far (\citealt{2017Caleb}; \citealt{2018Farah}; Farah et al., in prep.). Here we report on the pulsar timing component of the project, the UTMOST pulsar timing programme. We focus on questions that are best suited to a high-cadence programme of intermediate sensitivity and observing duration, which are the search for and characterisation of pulsar glitches and the monitoring of a sample of intermittent pulsars. More generally, the aim of the UTMOST timing programme is to monitor a large sample of pulsars with relatively high precision and cadence in order to extract knowledge about the fundamental physical processes that determine their rotation and emission. As such, this programme aims to continue, complement and in many aspects improve the historic and contemporaneous efforts at other radio telescopes (e.g.\ \citealt{1988Cordes, 1993DAlessandro, 2004Hobbs, 2013Manchester}). A major aim of the project is to act as a technology test-bed for upcoming telescopes, such as MeerKAT (e.g.\ MeerTIME; \citealt{2016Bailes}) or the Square Kilometre Array (SKA).

Pulsar glitches are sudden spin-up events that interrupt the otherwise steady spin-down of mainly young and middle-aged pulsars. They are thought to be caused by processes inside the neutron star and are usually explained using the superfluid vortex model \citep{1975Anderson, 1984Alpar, 1985Pines}, in which the outer crust and superfluid interior rotate differentially. Once the differential rotation exceeds a threshold, the pinned superfluid vortices unpin and some of the angular momentum that is stored in the superfluid is released to the crust, and a spin-up is observed. Another explanation is the cracking of the star's crust and a resulting change in moment of inertia \citep{1969Ruderman}. Observations of glitches provide a unique\footnote{Apart from various oscillatory modes of the star.} opportunity to relate the neutron star's rotation to its bulk properties and structure \citep{1969Baym, 1999Link}. A recent theoretical review is given by \citet{2015Haskell}. 529 glitches in 188 pulsars are currently reported in the Jodrell Bank glitch table \citep{2011Espinoza}, indicating that only about 7 per cent of the discovered pulsar population are known to glitch. This fraction might be significantly underestimated as a result from lack of monitoring. In addition, most of these glitches are from a small set of pulsars that exhibit exceptionally high glitch rates and the majority of glitches are poorly sampled by timing observations. The detection of glitches in high-cadence observations is therefore crucial to improve our understanding of matter at the highest densities.

The second topic that we focus on is emission intermittency, by which we mean the cessation of pulsar emission for one or multiple rotations (nulling), the change between two or more stable emission modes (mode-changing) and the absence of emission for extended periods (long-term intermittency) \citep{1992Biggs, 2006Kramer, 2007Wang, 2010Lyne, 2017Lyne}. These phenomena are thought to originate in the pulsar's magnetosphere and provide insights into the plasma physics of the pulsar emission. \citet{2016Melrose} give a recent theoretical overview of possible pulsar emission mechanisms. In the case of mode-changing, simultaneous X-ray and radio observations have recently shown that bright and quiet modes switch simultaneously in the radio and X-ray regimes. Whether this indicates a rapid change of the magnetosphere as a whole, or whether this can be explained successfully by a particular emission mechanism is a matter of current debate \citep{2013Hermsen, 2016Mereghetti, 2017Archibald, 2018Hermsen}. Further radio timing observations together with single-pulse recording might hopefully advance this topic.

The main aspects of our programme are:
\begin{enumerate}
	\item Timing of hundreds of pulsars with high cadences (up to daily) with typical observing times between five to ten minutes
	
	\item Timing a large number of pulsars that have not been observed elsewhere since their initial timing observations many years/decades ago
	
	\item A dedicated search programme for pulsar glitches
	
	\item A dedicated monitoring programme of intermittent pulsars
	
	\item A modern, dynamic and fully autonomous telescope scheduling system.
\end{enumerate}

We have described the main focus of the timing programme above. However, in this paper, we mainly consider the scientific verification of the instrument and of the timing infrastructure, and pulsar properties that can be derived reliably from UTMOST measurements together with historical data. These are the timing stability of pulsars over decades and their proper motions. Proper motion measurements are interesting, as they allow one to estimate the transverse velocities of astronomical objects, under the requirement that somewhat accurate distances to these objects are available. In the case of pulsars, which were early on identified to be high-velocity objects \citep{1970Gunn}, it was proposed that they receive a birth kick of maybe a few hundred $\text{km} \: \text{s}^{-1}$, possibly because of slight asymmetries in supernova explosions, that often disrupt the binary system \citep{1989Bailes, 1991Bhattacharya, 1996Tauris}. Analysing their transverse velocities might therefore allow to constrain models for pulsar birth kicks, supernova properties and possibly binary evolution. Proper motions of pulsars have been studied using interferometric methods (e.g.\ \citealt{1982Lyne, 1997Fomalont, 2003Brisken, 2009Chatterjee, 2013Deller}) and pulsar timing techniques (e.g.\ \citealt{2005Hobbs, 2005Zou}). In this paper, we combine pulsar timing position data obtained at UTMOST with multiple historical position measurements from the literature, spanning up to 48 yr, to derive reliable inferred proper motions.

Together with their calibrated flux densities, we analyse the integrated stable profiles of pulsars. They are of importance because they represent the intersection of the pulsar's beam with the line-of-sight and therefore reflect the particular geometry of the pulsar and the configuration of the beam. Together with the rotating vector model \citep{1969Radhakrishnan}, various beam configurations have been proposed (e.g.\ \citealt{1988Lyne}) and empirical classifications have been developed (\citealt{1983Rankin} and later publications in that series). Multi-frequency pulse profile measurements also allow to test the radius-to-frequency mapping picture \citep{1978Cordes}, in which the emission altitude scales inversely with frequency, i.e.\ low-frequency radio emission is supposed to be created higher in pulsar's magnetosphere than high-frequency radiation. Also of interested is the presence and separation of profile components and their scaling with frequency (e.g.\ \citealt{1996Xilouris}).

The paper is structured as follows: in Section~\ref{sec:Observations}, we describe the design of the timing programme and relevant technical details. In Section~\ref{sec:Analysis}, we describe the data analysis pipeline, the pulsar timing method and the flux density calibration methodology. In Section~\ref{sec:Results}, we present the science verification of the system and a selection of our first results. Finally, we give our conclusions and ideas for future work in Section~\ref{sec:Conclusions}. Appendix~\ref{sec:DynamicTelescopeScheduling} contains a detailed description of our algorithm for optimal dynamic telescope scheduling, including a performance evaluation. The full tables with best-fitting pulsar ephemerides and pulse widths and flux densities are presented in Appendices~\ref{sec:BestFittingPulsarEphemerides} and \ref{sec:FluxDensitiesAndPulseWidths}, respectively.

\section{Observations}
\label{sec:Observations}

\subsection{The refurbished telescope system (UTMOST)}
\label{sec:TheRefurbishedTelescopeSystem}

The technical details of the refurbished telescope system and its new digital backend are presented by \citet{2017Bailes}. Here we summarise the project focussing on the properties relevant for the pulsar timing project. The MOST is a successor of the Mills-Cross telescope \citep{1963Mills}, re-engineered to operate at a centre frequency of 843~MHz \citep{1981Mills}. It consists of two arms aligned orthogonally in north-south and east-west\footnote{The whole east-west arm has a slope of $11 \arcmin 51.5 \arcsec$ to the west. Additionally, the east arm alone has a small offset of $4.9 \arcsec$ north of true east.} direction, of which currently only the east-west arm is operational. The telescope is located about $35 \: \text{km}$ south-east of Canberra, Australia. The interferometric array comprises 352 modules arranged in groups of four, which are termed bays. Each module contains 22 ring antennas that receive right-hand circularly polarized radio waves. The voltages from each antenna are combined in phase in a resonant cavity that then leads to a low-noise amplifier. Each module has a geometric area of $4.42 \times 11.60 \: \text{m}^2$ and the antenna efficiency is estimated to be $0.55$. 44 bays belong to the east arm, which is separated by a gap of $15 \: \text{m}$ from the west arm. Each arm is a cylindrical paraboloid with the feed line and the ring antennas in its focus. Combined they are $1.556 \: \text{km}$ long, have a total collecting area of about $18,000 \: \text{m}^2$ and contain 7744 ring antennas. The primary beam has an approximate size of $4.25 \degr \times 2.8 \degr$ and is steered in the following way: in the north-south (NS) direction the arms can be mechanically tilted and in the east-west (EW) direction the beam is steered away from the meridian by differentially rotating the ring antennas with respect to each other. Its zenith limits are $\pm 54 \degr$ in NS, at which the telescope hits mechanical end switches. We adopt a software limit of $\pm 60 \degr$ in meridian distance (MD). In practice, we conduct the vast majority of pulsar observations between $\pm 45 \degr$ from the meridian. The telescope's slew rates are about $5 \degr \: \text{min}^{-1}$ in NS and $2.5 \degr \: \text{min}^{-1}$ in MD.

UTMOST operates at a centre frequency of $843 \: \text{MHz}$ and samples a bandwidth of $31.25 \: \text{MHz}$. The beamformer synthesises narrow fan beams that tile the primary beam in the EW direction. The number of fan beams and their spacing can be configured. The initial configuration was 352 fan beams across $4.0 \degr$, but recently we switched to 512 fan beams. In addition, the beamformer can currently synthesise up to four tied-array beams that track pulsars or other objects as they move inside the primary beam. This is the main mode for pulsar observations. The data from each tied-array beam is then folded by \textsc{dspsr} \citep{2011VanStraten} using the most recent pulsar ephemeris from the project's local ephemeris repository, before which the signal is coherently dedispersed. \textsc{dspsr} uses spectral kurtosis \citep{2010Nita} to identify and replace radio frequency interference (RFI) with Gaussian noise in the voltage stream.

An important feature that separates this timing project from others is that we retain filterbank data for all tied-array beams, in addition to the folded archives, irrespective of the pulsar observed. The single-polarization filterbank data are saved at a lower temporal resolution, currently about $655.36~\mu \text{s}$. This has the advantage that we can analyse data at a single-pulse level with a reasonable number of phase bins ($\geq 64$) for most normal (i.e.\ non-millisecond) pulsars with periods in excess of about $50 \: \text{ms}$. The data set, therefore, allows the study of single-pulse properties, which is necessary to understand pulse nulling, intermittency, sub-pulse drifting, giant pulse emission, or other single-pulse phenomena. The fact that the filterbank data are saved unconditionally allows us to investigate changes in single-pulse behaviour in retrospect, for example when a change in pulsar rotation is discovered during subsequent analysis.

\subsection{Station clock and reference position}
\label{sec:StationClock}

We use a Brandywine NFS--200 Plus station clock as a time reference. It houses a Rubidium atomic clock that is tied to the UTC time standard received via the Global Positioning System (GPS). It provides a 1 pulse-per-second signal that is distributed to the correlator and the receiver boards that digitise the radio signal in the field. The quoted timing accuracy is $100 \: \text{ns}$ absolute UTC and the clock has an Allen variance of less than $5 \times 10^{-12}$ in a day. We use a concrete pillar near the centre of the telescope (in the gap between the east and west arms) as the reference position. Its location as determined by the 2012 geodetic survey of the observatory \citep{2013Garthwaite} is:
\begin{equation}
	\text{lat} = -35 \degr 22 \arcmin 14.5518 \arcsec,\\
	\text{lon} = 149 \degr 25 \arcmin 28.8906 \arcsec
	\label{eq:UTMOSTLocation}
\end{equation}
in ITRF08\footnote{International Terrestrial Reference Frame 2008} coordinates assuming a GRS80\footnote{Geodetic Reference System 1980} ellipsoid. We measured its elevation as $741 \: \text{m}$ using a consumer GPS device. The uncertainties of the horizontal coordinates are unknown, as they were not stated in the geodetic survey, but are believed to be of the order of 10~cm, a typical value achievable with modern professional satellite surveying equipment. The elevation is much less certain, because it is inherently more difficult to determine it using GPS than it is to determine the horizontal coordinates and because a consumer-grade GPS device was used to measure it. We expect the vertical uncertainty to be about 3~m, which is 1.5 times the value of the 50~\% circular error probable reported by the device.

\subsection{The automatic observing system}
\label{TheAutomaticObservingSystem}

\begin{figure*}
	\centering
	\includegraphics[width=0.8\textwidth]{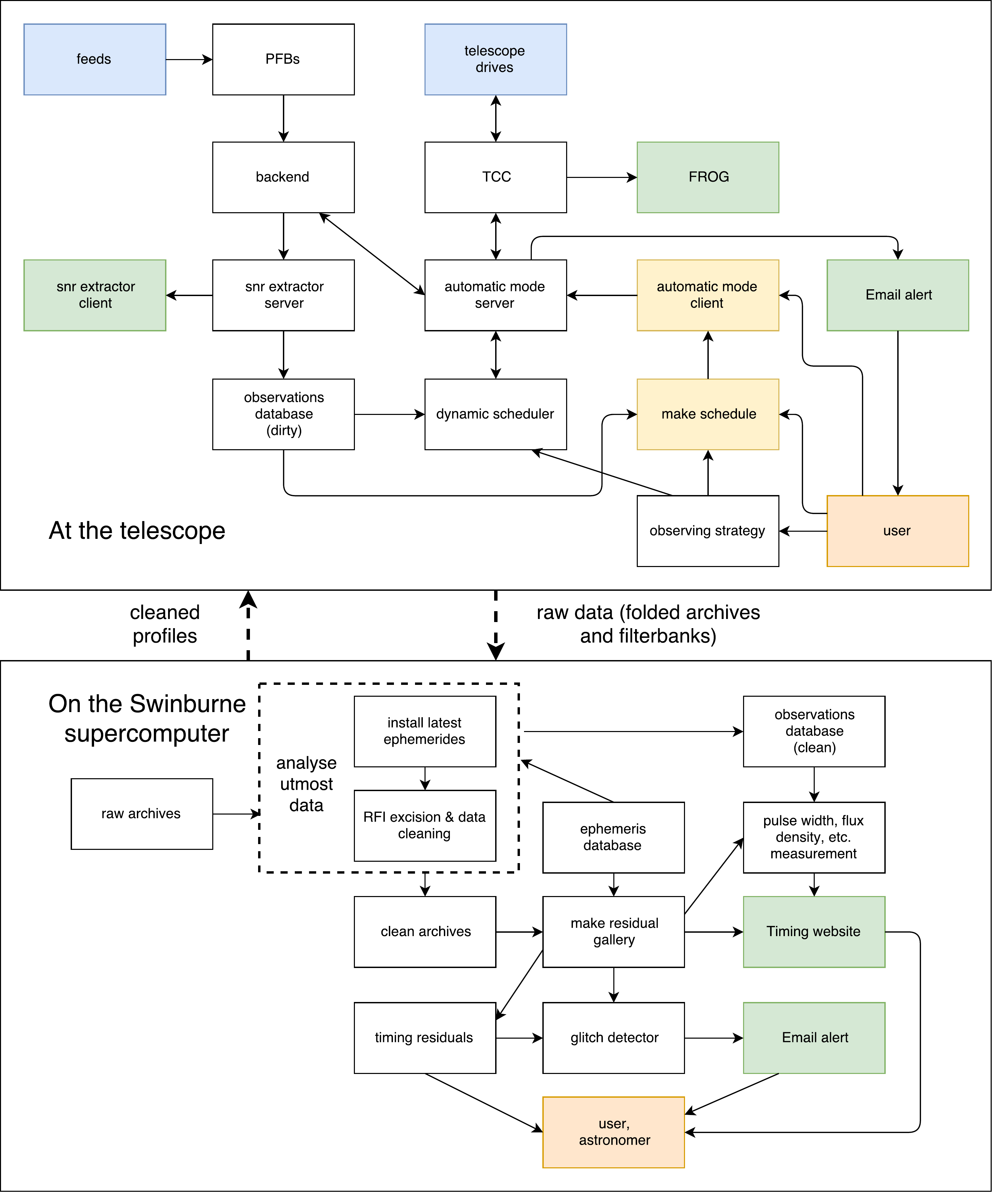}
	\caption{Schematic overview of the hardware and software components that constitute the UTMOST pulsar timing system at the telescope and on Swinburne University's gSTAR supercomputer. Telescope hardware is marked in blue, visualisation and status reporting tools in green and manual control software in yellow. The system uses a modern, modular design approach. The dynamic observing system represents a large fraction of the software in use at the telescope.}
	\label{fig:MolongloSoftware}
\end{figure*}

The components of the automatic observing system at the telescope are depicted in the upper part of Fig.~\ref{fig:MolongloSoftware}. Central to its operation is the \textsc{automatic mode server}, that directly controls the backend and the telescope control system (\textsc{tcc}), which is responsible for the low-level operation of the telescope drives and rotation of the ring antennas. It allows static, schedule file-based and fully autonomous operation using the dynamic scheduler. The \textsc{tcc} provides the current pointing position; statistics about the observations, such as signal-to-noise ratio (S/N) or pulse time-of-arrival (ToA) precision, are extracted live by \textsc{snr extractor server}. A safety system detects and stops unexpected behaviour of the telescope. The current state of the telescope and the observation are reported by monitoring programs called \textsc{frog} and \textsc{snr extractor client}. Our software is available online in the following repositories: \textsc{tcc}\footnote{\url{https://github.com/ewanbarr/}} and all other high-level software\footnote{\url{https://bitbucket.org/jankowsk/}}. We describe the details of the dynamic scheduler and its performance separately in Appendix~\ref{sec:DynamicTelescopeScheduling}.

\subsection{Target selection}
\label{sec:TargetSelection}

The pulsars reported on in this paper are mainly relatively bright southern pulsars. Most of them are isolated, canonical or slow, i.e.\ not millisecond pulsars (MSPs). The target selection was primarily influenced by the instantaneous sensitivity of the system and the given scientific priorities as outlined in the introduction. As a consequence, many of them are relatively young pulsars. Suitably bright candidate pulsars were selected based on the absolute calibrated spectral measurements presented by \citet{2018Jankowski}. As the telescope's sensitivity increased, more and more candidates could be added to the regular timing programme. The observing times range from a few minutes on bright sources to multiple hours on fainter pulsars, e.g.\ to record a full orbit of a binary system, often at much reduced cadences. Typical observing times are about five to ten minutes. In the new transit operation mode, maximum observing times are restricted to the transit time of a source through the primary beam, with typical values around ten to fifteen minutes. The observing times per pulsar are chosen so that reasonably stable integrated profiles are recorded, averaging multiple hundreds to many thousands of single pulses.

\subsection{The current dataset}
\label{sec:TheCurrentDataset}

The earliest tied-array beam observations occurred in 2014 October. Prior to this, we folded the signals from a small subset of modules individually and combined them incoherently. Those data date back to 2013 November but are of lower quality. The timing programme started in earnest around 2015 October. Up to 2017 June, the telescope was operated in tracking mode, after which we switched to a transit-mode operation, as we found that the significant increase in sensitivity far outweighed the ability to slew in the EW direction. The majority of observations reported here were obtained in tracking mode, with about 10 per cent from transit mode operation. For the analysis presented in this paper, we first removed non-detections and corrupted observations, then selected pulsars with at least 1.4~yr of timing data to ensure the reliability of our results. The fraction of data lost due to RFI excision is highly variable as a function of time of day, with observations during the night generally being the cleanest, as expected. On average, about 10 per cent of data get excised post-folding. This fraction has improved significantly during the progress of the project, as well as the sensitivity and stability of the telescope system, as the project matured.

\section{Analysis}
\label{sec:Analysis}

\subsection{Pulsar data analysis and timing pipeline}
\label{sec:PulsarTimingPipeline}

We rely on standard techniques and software for the data analysis and timing pipeline. In particular, we use  \textsc{tempo2}, \textsc{dspsr}, \textsc{psrchive} and its \textsc{python} bindings \citep{2006Hobbs, 2011VanStraten, 2004Hotan}. A schematic overview of the pipeline is presented in the lower part of Fig.~\ref{fig:MolongloSoftware}. We use the ``Fourier domain with Markov chain Monte Carlo" (FDM) algorithm for ToA and uncertainty estimation and an International Pulsar Timing Array (IPTA) compatible file format \citep{2016Verbiest}. The site arrival times are transformed to the Solar System barycentre using the Jet Propulsion Laboratory (JPL) DE430 Solar System ephemeris \citep{2014Folkner}, which is tied to the International Celestial Reference Frame (ICRF), and we use the TT(TAI) terrestrial time scale. We employ smoothed versions of high-S/N pulse profiles as standard templates, where the smoothing is usually performed using wavelets. Use of analytic standard templates did not improve the timing precision significantly and was deemed unnecessary.

We began timing from the ephemerides published in the most recent version of the Australia Telescope National Facility (ATNF) pulsar catalogue at the time \citep{2005Manchester}, manually established phase connection when required and refined the timing models. We report all pulsar parameters in Barycentric Coordinate Time (BCT, SI units) and correct for the tropospheric and Shapiro delay due to the planets and the Sun, in addition to all usual delay corrections in \textsc{tempo2}. We report all uncertainties at the $1 \sigma$ level, if not noted otherwise. We account for white measurement noise in the following way: we find that the FDM algorithm determines more realistic ToA uncertainties than the default ``Fourier phase gradient" (PGS) algorithm \citep{1992Taylor}, especially in the case of very low-S/N observations. As a result, the reduced $\chi^2$ values are close to unity for most of the normal pulsars, except for those that show interesting rotational behaviour. In the case of bright and millisecond pulsars (MSPs), we use the \textsc{efac/equad} plugin \citep{2015Wang} for \textsc{tempo2} to correct for underestimated ToA uncertainties and arrive at reduced $\chi^2$ values near unity. This is done by multiplying the ToA uncertainties by a constant factor (EFAC), and/or by adding a systematic contribution (EQUAD) in quadrature to the ToA uncertainties. This is a standard procedure and is justified because of the presence of pulse jitter (for EQUAD) and imperfections in the algorithm that determines the ToA uncertainties (for EFAC; \citealt{2016Verbiest}). Characterisation of the red timing noise parameters of the pulsars will be possible once longer timing baselines of 5--10~yr are available. Apart from visual inspection, we formally test whether the residuals are white using the Shapiro--Wilk test for normality \citep{1965Shapiro, 2014Ivezic}. Unless otherwise stated all further analysis is based on those best-fitting ephemerides.

\subsection{Flux density calibration}
\label{sec:FluxDensityCalibration}

\begin{table*}
\caption{Parameters of the flux density reference pulsars used in this paper. We list their DMs, their pulse-averaged reference flux densities at 843~MHz, interpolated from absolute flux density calibrated observations at the Parkes telescope \citep{2018Jankowski}, their pulse widths at 50 and 10 per cent maximum estimated from UTMOST data, the long-term flux density variability $\epsilon$ at 610~MHz \citep{2000Stinebring}, where available, and the expected modulation indices at the MOST for 10 min integrations $m_\text{tot,10}$ computed using the \textsc{ne2001} model.}
\label{tab:FluxDensityReferences}
\begin{tabular}{lccccccl}
\hline
PSRJ			& DM 							& $S_{843}$			& $\text{W}_{50, 843}$	& $\text{W}_{10, 843}$		& $\epsilon$	& $m_\text{tot,10}$	& Comment\\
			& ($\text{pc} \: \text{cm}^{-3}$)	& (mJy)				& (ms)				& (ms)					& (per cent)		&			&\\
\hline
J1056$-$6258	& 320.3							& $48 \pm 5$			& $17.4 \pm 0.7$		& $37 \pm 10$			& --			& 0.08			&\\
J1243$-$6423	& 297.3							& $103 \pm 33$		& $5.71 \pm 0.01$	& $8.93 \pm 0.05$		& --			& 0.11			& nulling (see text)\\
J1327$-$6222	& 318.8							& $93 \pm 20$		& $15.2 \pm 0.2$		& $40 \pm 2$				& --			& 0.07			&\\
J1359$-$6038	& 293.7							& $40 \pm 10$			& $3.2 \pm 0.1$		& $9 \pm 1$				& --			& 0.10			&\\
J1600$-$5044	& 260.6							& $61 \pm 5$			& $12.2 \pm 0.8$		& $40 \pm 6$				& --			& 0.05			&\\
J1644$-$4559	& 478.8							& $920 \pm 70$		& $25.19 \pm 0.03$	& $67.4 \pm 0.2$			& $10$		& 0.05			&\\
J1833$-$0338	& 234.5						& $13 \pm 3$			& $5.7 \pm 0.1$		& $18 \pm 4$				& $25$		& 0.09			&\\
J1848$-$0123	& 159.5						& $27 \pm 6$			& $16.1 \pm 0.8$		& $41 \pm 2$				& --			& 0.07			&\\
J1901+0331	& 402.1						& $22 \pm 5$			& $10.5 \pm 0.2$		& $37.8 \pm 0.8$			& $20$		& 0.06			&\\
J1903+0135	& 245.2						& $19 \pm 3$			& $7.5 \pm 0.1$		& $17.7 \pm 0.5$			& --			& 0.08			&\\
\hline
\end{tabular}
\end{table*}

We use an ensemble of high-dispersion measure (DM) pulsars that have stable flux densities as performance references. We selected them so as to maximise the coverage of flux calibrator sources in right ascension and declination. The parameters of the reference pulsars are listed in Table~\ref{tab:FluxDensityReferences}. Their flux densities at 843~MHz are interpolated from absolute calibrated measurements at the Parkes telescope \citep{2018Jankowski}. We determined the pulse widths at 10 and 50 per cent maximum ($\text{W}_{10}$ and $\text{W}_{50}$) from high-S/N UTMOST observations. We also list the long-term variability in flux density $\epsilon$ measured at a frequency of 610~MHz \citep{2000Stinebring}, where available, and the total expected modulation index due to the combined effect of strong diffractive, refractive, or weak scintillation, for 10 min integrations, computed using standard formulae and the \textsc{ne2001} Galactic free electron-density model \citep{2002Cordes}. PSR J1243$-$6423 is a known nulling pulsar, but its nulling fraction is only about 2 per cent \citep{1992Biggs, 2007Wang}, which does not affect our calibration procedure. Its relatively high flux density at 843~MHz makes it a valuable part of the calibration ensemble. We observe the reference pulsars with the highest cadence, to guarantee calibrated flux densities in each observing run.

The flux density calibration works as follows: as the pulsars are tracked both physically and by the synthesised tied-array beam within the primary beam, we assume that they reside in the centre of the primary beam, or very close to it, over the span of an observation. They are therefore always observed near the maximum sensitivity of the primary beam at that meridian distance (MD) and north-south (NS) angle. Consequently, we can neglect to model the exact sensitivity curve of the primary beam. However, the gain of the system depends on the pointing position of the telescope, i.e.\ it is a function of meridian angle $m$ and north-south angle $n$. We refer to this dependence as the telescope gain curve $\eta = \eta(m, n)$. The gain curve also depends on the fraction of the 352 modules contributing to the tied-array beam and their relative weights in it, which varies as a result of changes to the hardware by the site crew, during the ongoing telescope upgrade. This means that the effective collecting area and the absolute gain of the telescope change over time, which we denote as $\xi = \xi(i)$, where $i$ indicates the preceding re-phasing observation of the array on a quasar. $\xi$ has no effect on the flux density calibration of a pulsar, as it cancels out, but needs to be accounted for in the measurement of the telescope gain curve by normalising the data appropriately. The total gain $G$ can be written as:
\begin{equation}
	G = \eta(m, n) \: \xi(i).
	\label{eq:VariableGain}
\end{equation}
We assume that the system temperature is constant at 300~K and that all variability is contained in the gain. It is equivalent to assuming that the system temperature varies and the gain stays constant. While the gain of a telescope and its system temperature are different quantities and could be measured independently, for our calibration method they are interchangeable as long as the other parameter is held fixed. We present a parametrized form of the telescope gain curve in Section~\ref{sec:MeridianAndNSGainCurve}.

We use the radiometer equation to relate a pulsar's pulse averaged flux density $S_\nu$ with the measured S/N in a folded observation of observing time $t$:
\begin{equation}
	S_\nu = \text{S/N} \: \beta \: \frac{ T_\text{sys} + T_{sky} }{ G \sqrt{B N_\text{p} t} } \sqrt{ \frac{\delta}{1 - \delta} },
	\label{eq:RadiometerEquation}
\end{equation}
where $G$ is the gain, $B = 31.25~\text{MHz}$ is the bandwidth, $N_\text{p} = 1$ is the number of polarizations, $T_\text{sys}$ and $T_\text{sky}$ are the system and sky temperature at the centre frequency $\nu = 843~\text{MHz}$, $\delta = W/P$ is the pulsar's duty cycle, $W$ is its pulse width and $P$ is its period \citep{1985Dewey, 2012Lorimer}. $\beta$ is a degradation factor because of imperfections in the digitisation of the signal, which we assume to be $1.2$. We determine the sky temperature from the 408~MHz all-sky atlas of \citet{1982Haslam} and scale it to 843~MHz using a power law with exponent $-2.6$ \citep{1987Lawson}. Our flux density calibration technique involves the following steps:
\begin{enumerate}
	\item Find all observations of flux density reference pulsars within 6 hours of an observation
	
	\item Look up the sky temperature at 843~MHz for those pulsars
	
	\item Transform the S/Ns of the reference pulsars to S/Ns at zenith using the telescope gain curve
	
	\item Compute the median gain from the S/Ns at zenith of all reference pulsars
	
	\item Compute the S/N at zenith for the pulsar observation using the telescope gain curve
	
	\item Derive $S_{843}$ from the S/N at zenith and the median gain using the radiometer equation.
\end{enumerate}

\section{Results}
\label{sec:Results}

\subsection{Science verification of the system}
\label{sec:ScienceVerification}

\begin{figure}
	\centering
	\includegraphics[width=\columnwidth]{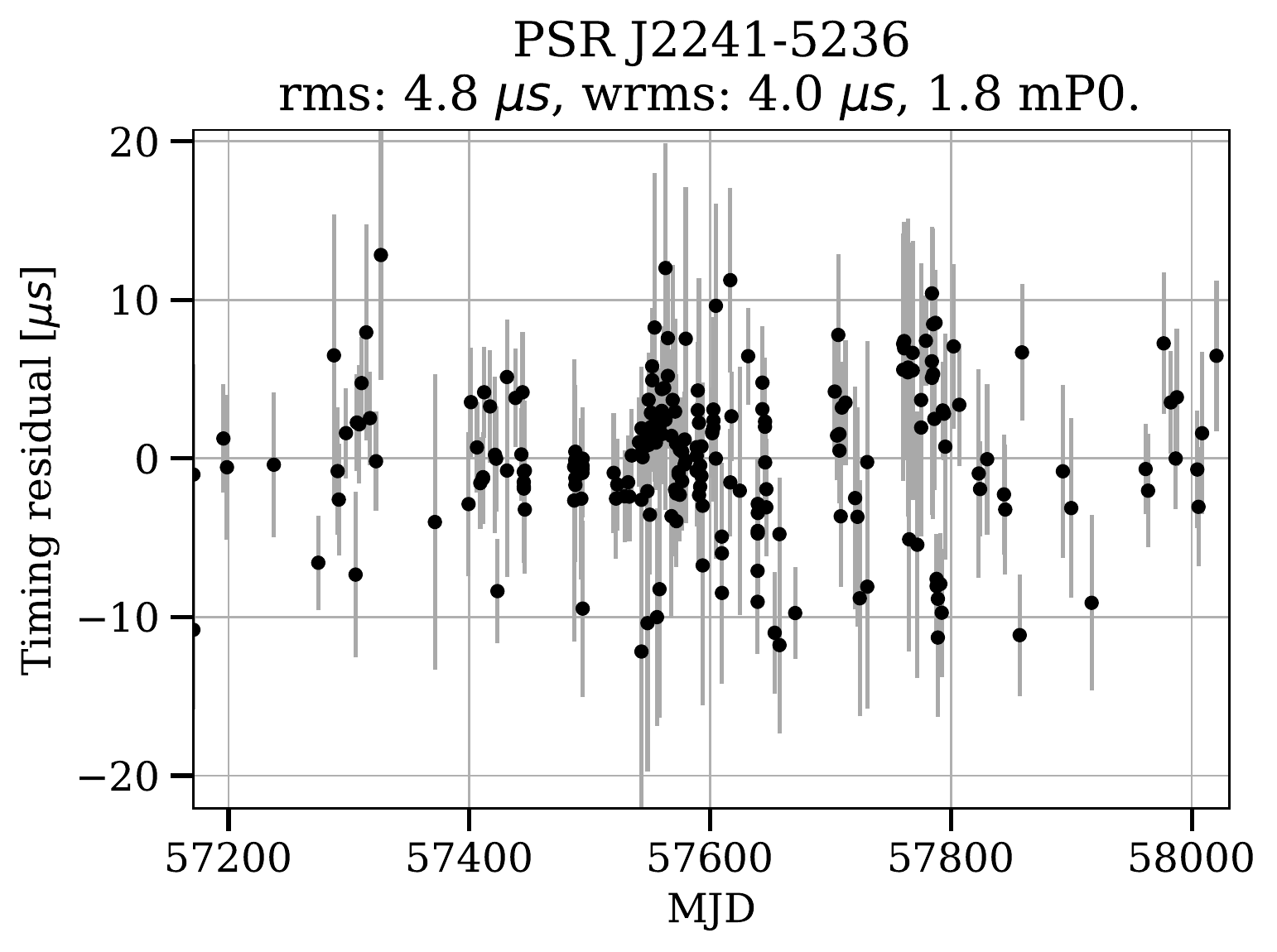}
	\caption{UTMOST timing residuals of the MSP J2241$-$5236 spanning more than two yr, using our best-fitting ephemeris. The weighted rms residual is slightly less than $2 \mu \text{s}$ for the raw ToAs and $4 \mu \text{s}$,  when the ToA uncertainties are corrected for underestimation due to pulse jitter. The stability of the system is verified to a precision of about $5 \mu \text{s}$.}
	\label{fig:MSPTiming}
\end{figure}

To verify the telescope system and to ensure the stability of the station clock, the backend and the signal chain, we regularly monitor various MSPs using the tied-array beam fold mode of the backend. This is an integral part of the pulsar timing programme. The aim is to characterise any systematic effects that might be present in the timing data. In particular, we use the MSPs J0437$-$4715 and J2241$-$5236 as primary and the MSP J1909$-$3744 as secondary timing references, because the latter is often faint at 843~MHz. These pulsars are observed as part of the Parkes Pulsar Timing Array (PPTA) project \citep{2013Manchester}. We obtained the most recent ephemerides for those pulsars derived independently from observations at the Parkes telescope by \citet{2016Reardon} and us (in the case of PSR J2241$-$5236) and applied them to the UTMOST data without fitting for any model parameters. They describe our data well with weighted rms errors of about $6$, $5$ and $5 \mu \text{s}$ for PSRs J0437$-$4715, J2241$-$5236 and J1909$-$3744, respectively. We, therefore, conclude that the pulsar timing system is free from significant systematic effects to a precision of about $5 \mu \text{s}$.

The rms errors are dominated by how well we can determine clock jumps in the data, as they occur. Another issue that affects the timing performance is that the reference module and therefore the phase centre of the array can change between re-phasing observations on a quasar. This can occur when the usual reference module undergoes servicing and means that the location of the phase centre deviates from the timing reference position. The induced change in arrival time is determined by the light travel time along the array, with a maximum of about $5.2 \: \mu \text{s}$. These offsets are accounted for in the clock correction chain. We expect that pulsars can be timed at the microsecond level, or slightly below if the above issues can be handled.

To demonstrate the stability of the system we show the timing residuals of the MSP J2241$-$5236 over the span of more than two yr in Fig.~\ref{fig:MSPTiming}, using our best ephemeris derived from UTMOST data. The weighted rms residual is slightly less than $2 \mu \text{s}$ for the raw ToAs and $4 \mu \text{s}$, when the ToA uncertainties are corrected for underestimation due to pulse jitter. The former corresponds to about 1 milliperiods and the latter to about twice this. A comparison with other timing programmes is complicated due to the fact that each has different science goals, and often vastly different observing cadences and observing times per pulsar, which result in significant differences in S/N of the data sets. A further difficulty is that the observing spans reported on in the literature (e.g.\ from observing programmes at Parkes, or Jodrell Bank) are usually much longer than our current data set. With these caveats in mind, we find that the rms residuals reported by \citet{2004Hobbs} from Jodrell Bank are often comparable with those presented here. In addition, the UTMOST residuals for most of the MSPs are within an order of magnitude of those presented by \citet{2016Reardon} for the PPTA project at the Parkes telescope, which is a project focussed on the highest precision-timing of a small number of MSPs with often hour-long observations and careful calibration techniques.

\subsubsection{Impact of a single polarization instrument}
\label{sec:SinglePolarization}

\begin{figure}
	\centering
	\includegraphics[width=\columnwidth]{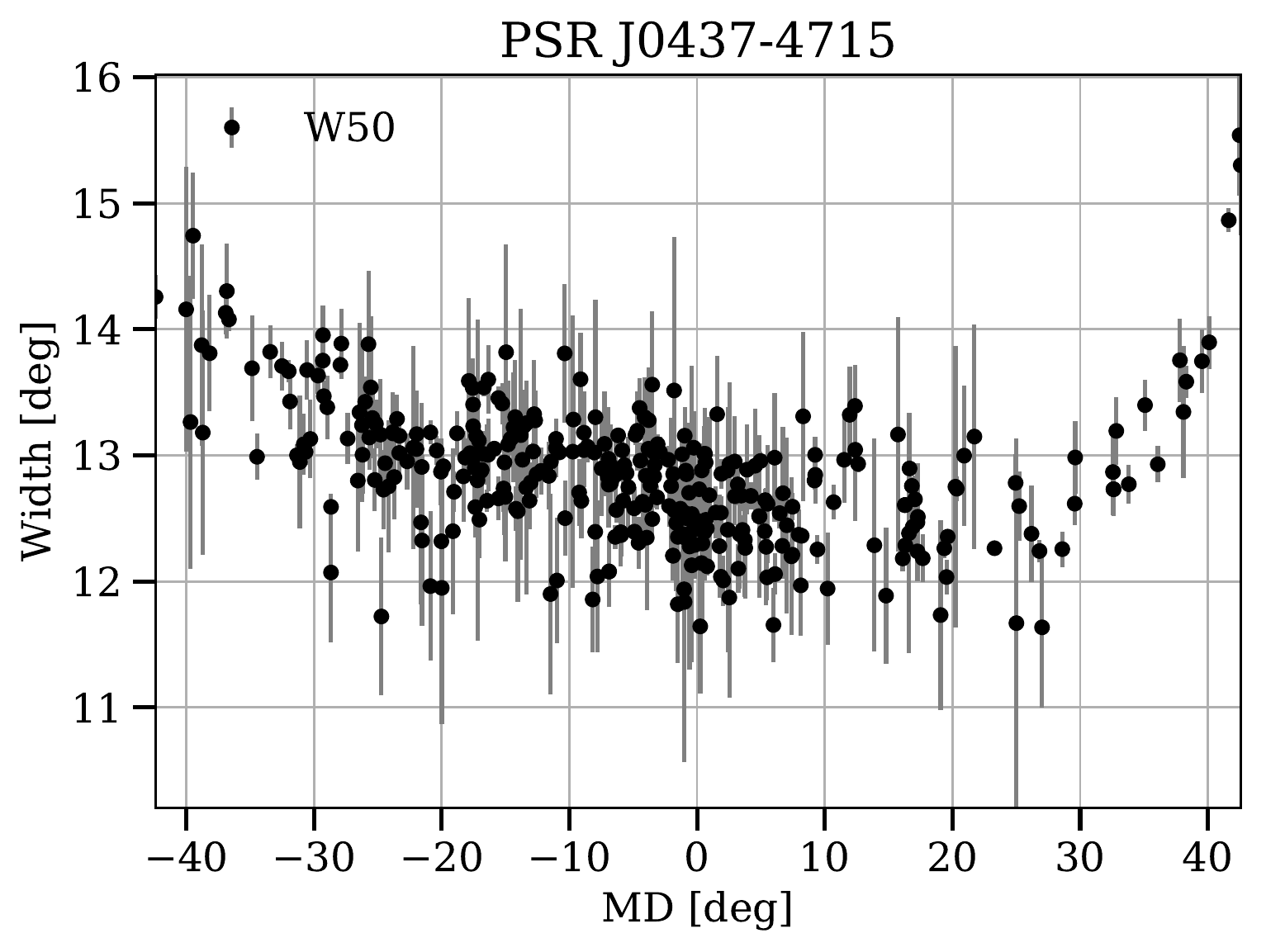}
	\caption{Dependence of PSR J0437$-$4715's 50 per cent pulse width with meridian distance. The pulse width measurements are flat within $\pm 30 \degr$, where we conduct the vast majority of observations, but increase significantly at angles beyond that.}
	\label{fig:PolarizationEffect}
\end{figure}

Only a single polarization of radiation, right-hand circular (RCP), is received by the ring antennas, meaning studies that require full polarization information are infeasible with the current feeds. We investigated whether this had any impact on the achievable timing precision. We conducted long tracking observations covering wide ranges in MD of many pulsars, in particular, the bright MSP J0437$-$4715 several times during 2016--2017. We find that PSR J0437$-$4715's timing residuals are flat within our measurement precision of $5 \mu \text{s}$ between about $\pm 45 \degr$ in MD and about $\pm 80 \degr$ in parallactic angle. At angles beyond that, the residuals show a systematic shift that increases with hour angle. We further find that its measured 50 per cent pulse width increases significantly with MD beyond about $\pm 30 \degr$ (see Fig.~\ref{fig:PolarizationEffect}). In addition, we find a complex dependence of the Vela pulsar's 25 and 10 per cent pulse widths with MD. However, the maximum change is only about 10 per cent peak-to-peak of the mean pulse width.

The reason for the MD dependence is a change of the polarimetric response of the feeds with MD, which results in a change in the projection of a pulsar's polarized profile onto the feeds. As PSR J0437$-$4715 has a complex polarimetric profile, it presents a worst case scenario for a single-polarization instrument. To avoid this effect, we limit pulsar timing observations to $\pm 45 \degr$ for slow pulsars and to $\pm 30 \degr$ for MSPs, where the highest timing precision is required. It is also possible to correct for the systematic profile shift with MD, as described in Section~\ref{sec:PolarimetricResponseOfTheFeeds}.

\subsubsection{Polarimetric response of the feeds}
\label{sec:PolarimetricResponseOfTheFeeds}

We use the pulsar J0437$-$4715 as a polarization reference to characterise the polarimetric response of the telescope feeds. A more rigorous approach is used at the Parkes telescope \citep{2006VanStraten}. In particular, we use a full-Stokes, polarization calibrated pulse profile obtained at the Parkes telescope at a frequency of 728~MHz \citep{2015Dai} and model the projection of it onto a circular feed with arbitrary ellipticity and orientation\footnote{Our simulation code is available online in the \textsc{psrsim} program, which is part of \textsc{psrchive}.}. We simulated all combinations of ellipticity and orientation on a grid and compared the simulated profiles with a high-S/N profile obtained from UTMOST observations near the meridian. The best match between the reference and the simulated profile, that minimises the sum of squared differences, has an ellipticity $+36 \degr$ and orientation $-47 \degr$. This means that the feeds are largely sensitive to right-hand circularly polarized (RCP) radiation from the sky, i.e.\ left-hand circular polarization from the mesh, but also respond to linear polarization. A perfect RCP response would have ellipticity $+45 \degr$ (\citealt{1960Chandrasekhar}, see also equation~15 of \citealt{2000Britton}). Therefore, the response of the antenna to a linearly polarized source (like a pulsar) is expected to vary with parallactic angle.

\subsection{Calibrated flux densities}
\label{sec:FluxDensities}

\subsubsection{Telescope gain curve}
\label{sec:MeridianAndNSGainCurve}

We measured the relative sensitivity of the phased array as a function of MD and NS using nearly 70 hours of long pulsar tracks, covering a wide range of meridian and NS angles, of the flux density reference pulsars. The data are well fitted by a function of the form $\zeta (m) = a/\cos( m - m_0 ) + b$ in MD, where $m$ is the MD angle and $m_0$, $a$ and $b$ are free parameters. In NS the data can be fit by a parabolic function. The two-dimensional telescope gain function is a product of both:
\begin{equation}
	\eta (m, n) = \left[ \frac{a_\text{m}}{ \cos( m - m_0 ) } + b_\text{m} \right] \left[ a_\text{n} (n - n_0)^2 + b_\text{n} \right].
	\label{eq:TelescopeGainCurve}
\end{equation}
Physically the first factor is a projection effect, i.e.\ the apparent reduction in geometric collecting area as a source moves away from the meridian. The second factor includes the spill-over from the ground that increases with NS angle. We determine the best-fitting model from a maximum-likelihood fit to the data simultaneously in both dimensions. After normalising the model so that it is unity at the point ($m_0$, $n_0$), we find the following best-fitting parameters: $a_\text{m} = -0.78$, $m_0 = 2.55 \: \degr$, $b_\text{m} = 1.54$, $a_\text{n} = -6.5 \times 10^{-5}$, $n_0 = 20.96 \degr$ and $b_\text{n} = 1.31$. We use this parametrisation in the flux density calibration.

\subsubsection{Median flux densities and validation}
\label{sec:CrossValidationOfFluxDensityCalibration}

\begin{figure}
	\centering
	\includegraphics[width=\columnwidth]{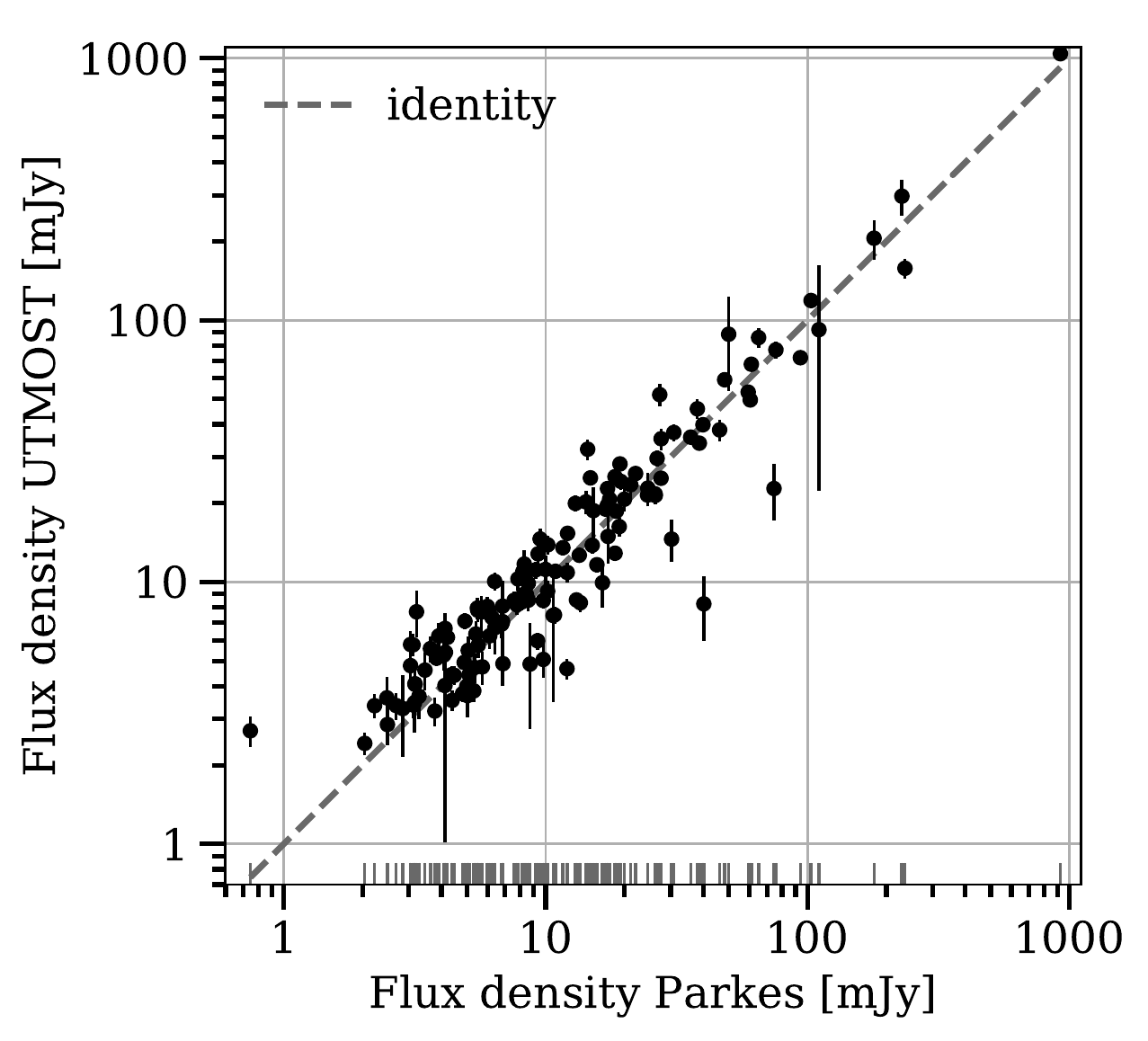}
	\caption{Comparison of the median flux densities of all pulsars in this work to absolute flux density calibrated measurements from the Parkes telescope, interpolated to 843~MHz. We omit the uncertainties of the Parkes data for clarity. The good agreement validates our flux density calibration method.}
	\label{fig:FluxdensityComparison}
\end{figure}

We present the calibrated median flux densities of all analysed pulsars in Appendix~\ref{sec:FluxDensitiesAndPulseWidths}. In addition to the statistical error, we add a 5 per cent systematic uncertainty to reflect the error introduced by the calibration and gain estimation. We validated our flux density calibration technique using two methods:
\begin{enumerate}
	\item where multiple flux density reference pulsars were observed in a single observing run, we iteratively calibrated using all but one of them and compared its derived with its nominal flux density (listed in Table~\ref{tab:FluxDensityReferences})
	
	\item we compared the derived flux densities of all non-reference pulsars with absolute flux density calibrated measurements from the Parkes telescope \citep{2018Jankowski}, interpolated to 843~MHz.
\end{enumerate}
In the first case, we find that the measured flux densities deviate from the nominal values with a median of 14 per cent, indicating that the calibration is self-consistent. In addition, the agreement between the median flux densities obtained in this work and those interpolated from the Parkes data is good, with median and rms differences of 19 and 27 per cent (Fig.~\ref{fig:FluxdensityComparison}).

It has to be kept in mind that our calibration is purely based on the radiometer equation and our knowledge of the system parameters derived from observing calibrator pulsars. Absolute flux density calibration would improve upon the current technique, but is not currently possible at the MOST. Overall, this validates our calibration method, the UTMOST measurements and strengthens the Parkes results presented by \citet{2018Jankowski}.

\subsubsection{Flux density variability}
\label{sec:FluxDensityVariablity}

For each pulsar, we also investigated the flux density time series, computed robust modulation indices (see \citet{2018Jankowski} for its definition) and compared them to the total modulation indices expected due to the combined effect of diffractive and refractive or weak scintillation. We computed the latter from the scintillation times and bandwidths determined using the \textsc{ne2001} model for our observing setup and usual integration times per pulsar. Where the transverse velocity is known we adopt the value from the pulsar catalogue, otherwise we use a default velocity of $100 \: \text{km} \: \text{s}^{-1}$. Flux density modulation beyond what is expected due to scintillation can either be attributed to instrumental effects, underestimation of the influence of scintillation on the data, or intrinsic variability in the pulsar emission. Examples of intrinsic variability include nulling, mode-changing and long-term intermittency. We considered the pulsars that have measured modulation indices that deviate by more than 50 per cent from the expectation due to scintillation separately, from which we selected only those pulsars with well-determined flux density time series with at least 10 measurements. We also adopted a very conservative upper limit of 0.75 for the modulation index due to instrumental effects, which could have been introduced by RFI excision, or by the flux density calibration technique. That is, we consider only those pulsars that have a measured modulation index in excess of the assumed instrumental value.

We find that 27 pulsars have measured modulation indices that exceed the expected values due to scintillation by at least 50 per cent. Most of them are known nulling or long-term intermittent pulsars, such as PSRs J0151$-$0635, J0452$-$1759, J1059$-$5742, J1107$-$5907, J1136+1551, J1502$-$5653, J1717$-$4054, or J2330$-$2005 \citep{1992Biggs, 2007Wang, 2012BurkeSpolaor}. Others are known sub-pulse drifters, such as J0152$-$1637 \citep{2006Weltevrede}, or those which exhibited events in which their flux density changed significantly, such as the glitching high magnetic-field pulsar J1119$-$6127, which suffered a magnetar-like outburst in 2016 July (e.g.\ \citealt{2017Majid}). Other examples include the long-period (5.01~s) pulsar J2033+0042, which seems to be nulling.

\subsection{Updated pulsar models}
\label{sec:UpdatedPulsarEphemerides}

We have updated the timing models of 205 pulsars, for many of which no updated models were published since the initial timing solutions derived shortly after discovery. We present the best-fitting ephemerides for both isolated pulsars and those in binary systems in Appendix~\ref{sec:BestFittingPulsarEphemerides}. The period, position and DM determination epoch is set to MJD 57600 for all pulsars.

\subsubsection{Improvements and changes of spin parameters}
\label{sec:ChangedOfSpinParameters}

\begin{table}
\caption{Pulsar spin and positional data from this work and from the literature that we use in this paper. We list the number of pulsars reported on and the Solar System ephemeris used to transform site arrival times to the Solar System barycentre.}
\label{tab:LiteratureData}
\begin{tabular}{lcl}
\hline
Reference									& \#Pulsars	& Solar ephem\\
\hline
This work									& 205		& JPL DE430\\
\citet{1972Manchester}			& 19			& MIT/CfA\\
\citet{1981Newton}					& 126		& MIT/CfA\\
\citet{1983Downs}					& 24			& JPL DE96\\
\citet{1983Manchester}			& 13			& MIT/CfA\\
\citet{1993Siegman}				& 59			& JPL DE200\\
\citet{2004Hobbs}					& 374		& JPL DE200\\
\citet{2005Zou}						& 74			& JPL DE405\\
ATNF pulsar catalogue (various)	& 2536		& various\\
\multicolumn{3}{l}{(without interferometric positions)}\\
\hline
\end{tabular}
\end{table}

We compare our measurements of the spin parameters with those from version 1.54 of the ATNF pulsar catalogue. The vast majority of our measurements of spin frequency, spin-down rate and second derivative, where available, have smaller or similar relative uncertainties compared to those in the catalogue at their original epochs. Indeed, we reduce the uncertainties of spin frequency and spin-down rate for $61$ and 44 per cent of the pulsars, respectively; for $19$ and 14 per cent by at least an order of magnitude. The period epochs of the ephemerides of these pulsars in the catalogue are on average 25 yr old in comparison with our measurements, with a maximum of nearly 39 yr. This allows us to investigate the spin behaviour of the pulsars over long time spans. To do that we combine our measurements with rotational data from the pulsar catalogue and from a selection of timing programmes from the literature (see Table~\ref{tab:LiteratureData}). While this selection is far from exhaustive, it represents a good cross-section of historic timing efforts and maximises the temporal coverage for a set of mainly southern pulsars. Before comparison, we converted all spin parameters given in Barycentric Dynamic Time (BDT) to BCT/SI units using appropriate powers of the $K$ factor given by \citet{1999Irwin, 2006Hobbs}.

We fit a linear function to the spin frequency data in a maximum likelihood sense to determine the inferred mean spin-down rates and we compare these with the locally determined values at each measurement epoch. We find that the local values differ significantly from the mean spin-down rates for most sources. Positive and negative deviations are distributed as expected if the local measurements are dominated by stochastic red noise processes in the pulsar. In addition, of the pulsars with well-determined spin histories with at least 4 measurement epochs, a few stand out with significant steps in their rotational history or consistently positive or negative deviations from the mean $\dot{\nu}$. Examples are the pulsars J0742$-$2822 (glitched and $\dot{\nu}$ changes), J0922+0638 (glitched and $\dot{\nu}$ changes, $\Delta \dot{\nu} < 0$ always), J1453$-$6413 (changed $\dot{\nu}$), J1731$-$4744 (glitched) and J1825$-$0935 (glitched). Others show evidence of glitches or other rapid changes in rotation, such as PSR J0738$-$4042, which may have interacted with an asteroid or in-falling disk material in late 2005 \citep{2014Brook}, or the pulsars J1600$-$5044 and J1709$-$1640, which exhibit changes in spin frequency of about 1 to $10 \: \text{nHz}$ between the measurements available, after subtraction of the mean spin-down.

Nonetheless, the absolute differences from the mean spin-down rates are small, with a nearly Gaussian distribution with a median of 0.13 per cent. This indicates that the pulsar rotation is generally very stable, as expected, with only small deviations from the mean $\dot{\nu}$ over nearly 50 yr. Possible reasons for the deviations are glitches and their recoveries, $\dot{\nu}$ changes, red timing noise, or changes in spin-down torque due to the pulsar emission switching off, as seen for example in the long-term intermittent pulsar B1931+24 and interpreted as the presence and cessation of the plasma flow of a pulsar wind \citep{2006Kramer}.

PSR J1003$-$4747 provides a good example in which spin parameters and therefore the derived parameters, such as characteristic age and surface magnetic field, have changed significantly between the measurement of \citet{1981Newton} and ours. The pulsar has moved about one order of magnitude towards lower characteristic age and surface magnetic field in the $P-\dot{P}$ diagram. Our $\dot{\nu}$ measurement is close to its inferred long-term spin-down rate and we, therefore, suspect the earlier $\dot{\nu}$ estimate to be in error.

\subsubsection{Second spin frequency derivatives}
\label{sec:MeasuredF2s}

Most of the pulsars have $\ddot{\nu}$ values consistent with zero. However, a small fraction of about 20 per cent shows significant signatures in their residuals that can be modelled by including a $\ddot{\nu}$ term. It is known that those locally derived values are largely a manifestation of stochastic red noise processes that occur inside the neutron star or its magnetosphere. They have been historically explained as a random walk process in pulse phase, spin frequency or spin-down rate (e.g.\ \citealt{1980Cordes}). However, more recent work shows that the timing noise of young pulsars is dominated by the recovery from glitches and that the noise in older pulsars is often quasi-periodic \citep{2010Hobbs}. Another viable explanation is the presence of unresolved micro-glitches, pulse shape variations, or free precession. Consequently, most of the estimated braking indices have unphysically high absolute values.

\subsubsection{Timing positions}
\label{sec:TimingPositions}

We update the astrometric positions for all pulsars analysed in this work. For about 52 per cent we reduce the angular sizes of the $1 \sigma$ uncertainty ellipses in comparison with the ATNF pulsar catalogue, for 25 per cent by at least an order of magnitude. Most of the pulsars for which the UTMOST uncertainty ellipses are larger than from the catalogue are those with positions from interferometric, or high-precision timing measurements. We measure the positions of pulsars that are within $15 \degr$ of the ecliptic plane using ecliptic coordinates to break the covariance that is present if equatorial coordinates are used.

\subsubsection{Inferred proper motions}
\label{sec:InferredProperMotions}

\begin{figure}
	\centering
	\includegraphics[width=\columnwidth]{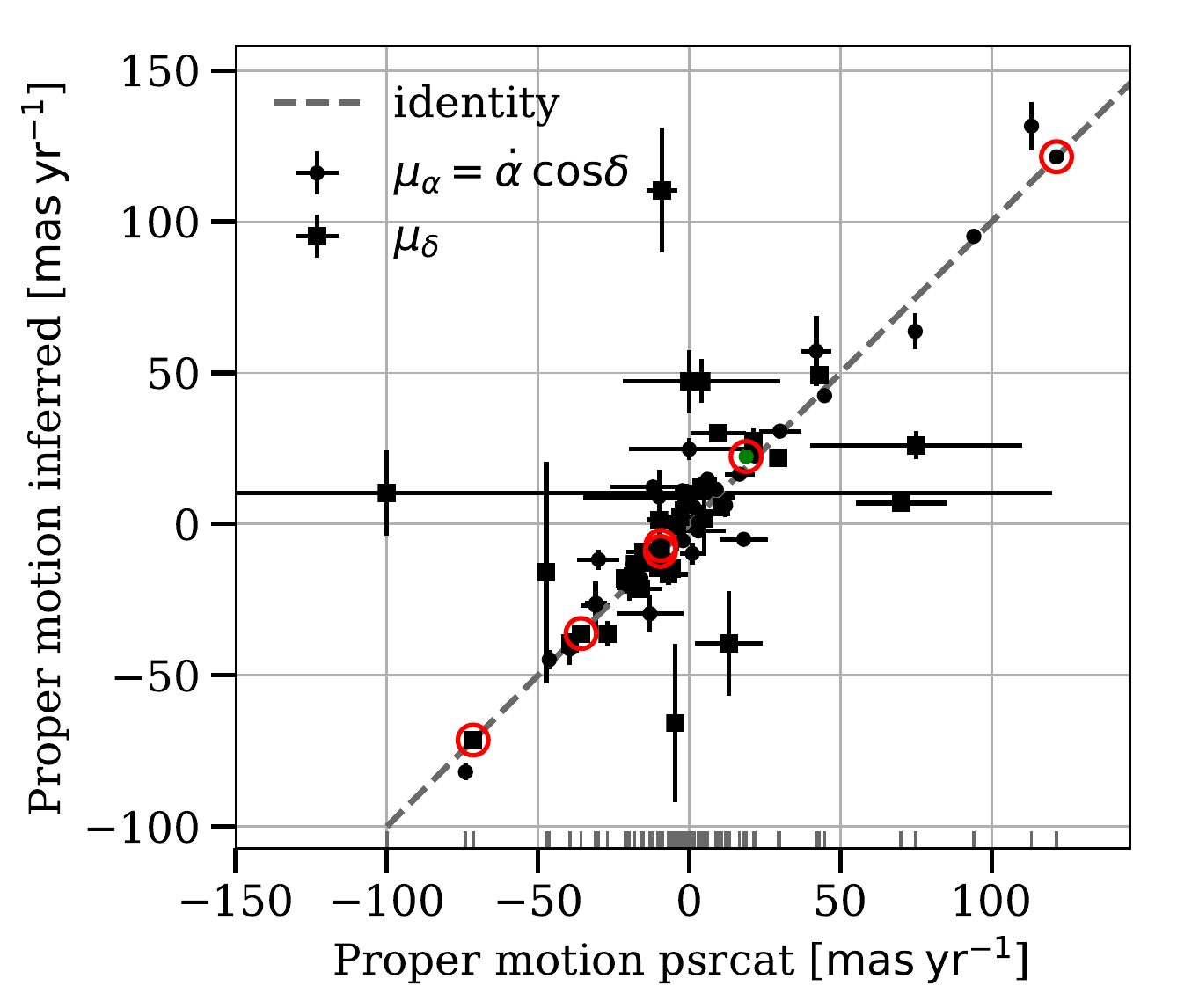}
	\caption{Comparison between the inferred proper motions from this work and those from the pulsar catalogue. We circle the data points for which we also measure significant proper motion signatures in the timing residuals and highlight the data that deviate by at least 5 times the combined uncertainty. The generally good agreement indicates that our method delivers realistic results, especially in comparison with interferometric measurements from the literature. The measurement of PSR~J0922+0638 has been excluded (see text) and the $\mu_\delta$ point of PSR~J1136+1551 was left aside for clarity.}
	\label{fig:ProperMotionComparison}
\end{figure}

\begin{figure}
	\centering
	\includegraphics[width=\columnwidth]{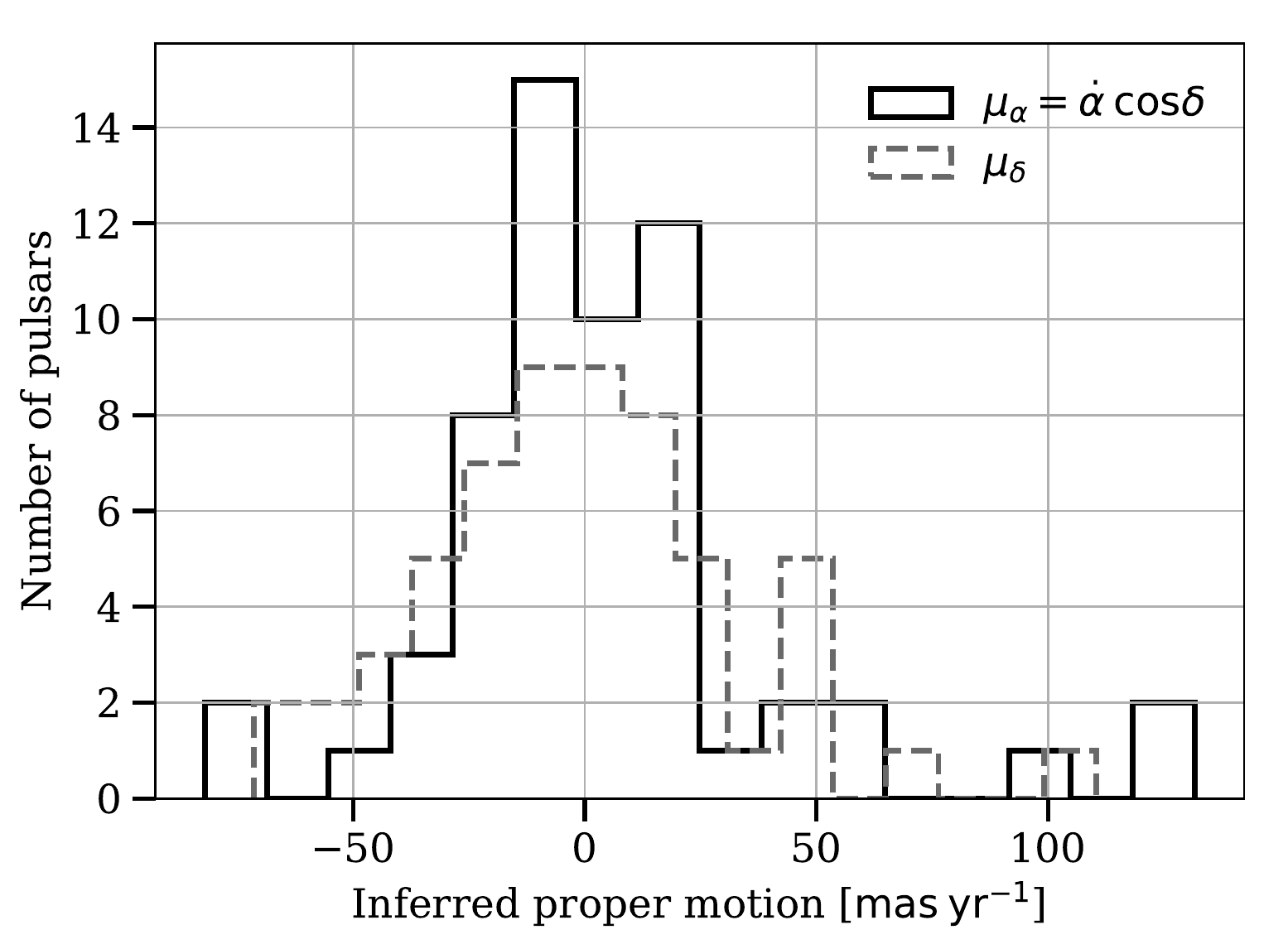}
	\caption{Histogram of all inferred proper motions from this work. The measurement of PSR J0922+0638 has been excluded and the $\mu_\delta$ point of PSR~J1136+1551 was left aside as in Fig.~\ref{fig:ProperMotionComparison}.}
	\label{fig:InferredProperMotions}
\end{figure}

\begin{table*}
\caption{Comparison between the inferred proper motions from this work and from the pulsar catalogue, where available. We list the maximum time spans between our and literature positions, their number, the proper motion references, the pulsar distances and the inferred transverse velocities. Uncertainties are reported at the $1 \sigma$ level in units of the least significant digit. The literature proper motions are from: (1) \citet{2005Hobbs}, (2) \citet{2003Brisken}, (3) \citet{1993Siegman}, (4) \citet{2016Reardon}, (5) \citet{2009Chatterjee}, (6) \citet{1993Harrison}, (7) \citet{2009Deller}, (8) \citet{2002Brisken}, (9) \citet{2010Mignani}, (10) \citet{1990Bailes}, (11) \citet{1997Fomalont}, (12) \citet{2005Zou}, (13) \citet{1999Fomalont}, (14) \citet{2004Chatterjee} and (15) \citet{2013Deller}.}
\label{tab:ProperMotions}
\begin{tabular}{lcccccccccl}
\hline
        & \multicolumn{2}{c}{This work} & \multicolumn{2}{c}{Literature}\\
PSRJ    & $\mu_\alpha \cos \delta$      & $\mu_\delta$  & $\mu_\alpha \cos \delta^\text{c}$     & $\mu_\delta^\text{c}$ & $\Delta t$    & \#Pos & Ref   & $d$   & $V_\text{t}$\\
        & (mas $\text{yr}^{-1}$)        & (mas $\text{yr}^{-1}$)        & (mas $\text{yr}^{-1}$)  & (mas $\text{yr}^{-1}$)      & (yr)  &   &   & (kpc) & (km $\text{s}^{-1})$\\
\hline
J0134$-$2937$^\dagger$ & $17(1)$ & $-9(2)$ & $17(5)$$^\text{e}$ & $-15(6)$$^\text{e}$ & $18.5$ & $2$ & 1 & $2(8)$ & $160(730)$\\
J0152$-$1637 & $1(4)$ & $-36(4)$ & $3(1)$ & $-27(2)$ & $38.5$ & $4$ & 2 & $0.7(1)$ & $120(20)$\\
J0206$-$4028 & $9(4)$ & $26(5)$ & $-10(25)$ & $75(35)$ & $38.5$ & $3$ & 3 & $0.9(2)$ & $110(30)$\\
J0255$-$5304 & $25(4)$ & $7(3)$ & $0(20)$ & $70(15)$ & $38.5$ & $4$ & 3 & $1.1(2)$ & $140(30)$\\
J0437$-$4715$^\text{t}$ & $121.48(2)$ & $-71.43(1)$ & $121.439(2)$ & $-71.475(2)$ & $8.5$ & $2$ & 4 & $0.1568(2)$ & $104.7(2)$\\
J0452$-$1759 & $11.5(9)$ & $6(2)$ & $9(2)$ & $11(2)$ & $29.3$ & $4$ & 5 & $0.4(2)$ & $20(10)$\\
J0525+1115 & $31(2)$ & $-1(8)$ & $30(7)$ & $-4(5)$ & $29.3$ & $3$ & 6 & $3(2)$ & $450(270)$\\
J0536$-$7543 & $3(40)$ & $65(8)$ & -- & -- & $38.5$ & $3$ & -- & $0.8(3)$ & $240(90)$\\
J0601$-$0527 & $-5(2)$ & $-21(3)$ & $18(8)$ & $-16(7)$ & $22.5$ & $2$ & 6 & $4(2)$ & $410(170)$\\
J0630$-$2834 & $-45(3)$ & $28(4)$ & $-46.3(10)$ & $21.3(5)$ & $46.6$ & $5$ & 7 & $0.32(5)$ & $80(10)$\\
J0758$-$1528 & $-10(4)$ & $12(3)$ & $1(4)$ & $4(6)$ & $21.1$ & $2$ & 2 & $3.0(3)$ & $220(50)$\\
J0809$-$4753 & $-19(5)$ & $15(4)$ & -- & -- & $38.4$ & $3$ & -- & $6(4)$ & $730(440)$\\
J0820$-$1350 & $23(2)$ & $-39(3)$ & $21.64(9)$ & $-39.44(5)$ & $29.2$ & $4$ & 5 & $1.9(1)$ & $410(30)$\\
J0837$-$4135 & $11(2)$ & $-13(1)$ & $-2(2)$ & $-18(3)$ & $38.4$ & $4$ & 2 & $2(1)$ & $120(100)$\\
J0922+0638$^\dagger$ & $84(6)$ & $214(15)$ & $18.8(9)$ & $86.4(7)$ & $38.5$ & $4$ & 2 & $1.1(2)$ & $1200(230)$\\
J0942$-$5657 & $-12(7)$ & $12(3)$ & -- & -- & $38.4$ & $3$ & -- & $3(1)$ & $280(140)$\\
J0953+0755 & $-5(1)$ & $22(3)$ & $-2.09(8)$ & $29.46(7)$ & $47.9$ & $5$ & 8 & $0.261(5)$ & $30(4)$\\
J1057$-$5226 & $57(12)$ & $2(7)$ & $42(5)$ & $-3(5)$ & $38.5$ & $3$ & 9 & $0.7(4)$ & $200(120)$\\
J1121$-$5444$^\dagger$ & $-17(6)$ & $18(6)$ & -- & -- & $38.4$ & $2$ & -- & $5(4)$ & $600(490)$\\
J1136+1551 & $-82(3)$ & $358(5)$ & $-74.0(4)$ & $368.1(3)$ & $47.8$ & $5$ & 8 & $0.35(2)$ & $610(40)$\\
J1136$-$5525 & $10(17)$ & $43(9)$ & -- & -- & $38.5$ & $3$ & -- & $2.1(3)$ & $450(120)$\\
J1141$-$6545 & $-33(4)$ & $-6(1)$ & -- & -- & $8.1$ & $2$ & -- & $3(2)$ & $480(330)$\\
J1202$-$5820 & $-14(4)$ & $1(5)$ & -- & -- & $38.5$ & $2$ & -- & $3.0(9)$ & $190(90)$\\
J1224$-$6407 & $20(4)$ & $2(2)$ & -- & -- & $38.5$ & $3$ & -- & $4(2)$ & $390(210)$\\
J1253$-$5820 & $24(2)$ & $13(2)$ & -- & -- & $23.5$ & $2$ & -- & $2.2(4)$ & $280(50)$\\
J1320$-$5359 & $8(4)$ & $49(4)$ & -- & -- & $38.4$ & $3$ & -- & $2.3(7)$ & $550(160)$\\
J1430$-$6623 & $-27(8)$ & $-19(4)$ & $-31(5)$ & $-21(3)$ & $38.5$ & $3$ & 10 & $1.3(2)$ & $210(60)$\\
J1453$-$6413 & $-18(4)$ & $-18(3)$ & $-16(1)$ & $-21.3(8)$ & $39.5$ & $4$ & 10 & $3(1)$ & $340(160)$\\
J1456$-$6843 & $-41(5)$ & $-13(2)$ & $-39.5(4)$ & $-12.3(3)$ & $41.2$ & $3$ & 10 & $0.43(6)$ & $90(20)$\\
J1534$-$5334 & $-27(10)$ & $-12(10)$ & -- & -- & $38.4$ & $3$ & -- & $1.1(1)$ & $160(60)$\\
J1539$-$5626 & $49(13)$ & $-36(16)$ & -- & -- & $25.3$ & $2$ & -- & $3.5(3)$ & $1000(250)$\\
J1557$-$4258 & $-3(1)$ & $10(2)$ & -- & -- & $26.2$ & $2$ & -- & $8(1)$ & $380(90)$\\
J1600$-$5044 & $-5(2)$ & $-13(2)$ & -- & -- & $41.2$ & $4$ & -- & $7(2)$ & $470(140)$\\
J1604$-$4909 & $-12(3)$ & $9(5)$ & $-30(7)$ & $-1(3)$ & $38.4$ & $3$ & 10 & $3.6(6)$ & $250(80)$\\
J1705$-$1906 & $-77(2)$ & $-52(17)$ & -- & -- & $24.3$ & $3$ & -- & $0.9(1)$ & $390(70)$\\
J1709$-$1640 & $-2(1)$ & $47(10)$ & $3(9)$ & $0(14)$ & $46.5$ & $6$ & 11 & $0.8(7)$ & $190(150)$\\
J1744$-$1134$^\text{t}$ & $22.34(8)$ & $-7.1(5)$ & $18.790(6)$ & $-9.40(3)$ & $17.8$ & $3$ & 4 & 0.395 & $43.9(3)$\\
J1745$-$3040 & $15(1)$ & $47(7)$ & $6(3)$ & $4(26)$ & $29.1$ & $4$ & 12 & $0(1)$ & $50(260)$\\
J1751$-$4657 & $-13(5)$ & $-59(7)$ & -- & -- & $38.4$ & $3$ & -- & $0.7(1)$ & $210(40)$\\
J1817$-$3618 & $19(5)$ & $-16(17)$ & -- & -- & $38.4$ & $2$ & -- & $3.8(6)$ & $440(220)$\\
J1820$-$0427 & $-8(1)$ & $30(3)$ & $-10(3)$$^\text{e}$ & $10(9)$$^\text{e}$ & $46.5$ & $6$ & 1 & $0.3(6)$ & $40(90)$\\
J1823$-$3106 & $18(4)$ & $27(18)$ & -- & -- & $20.6$ & $2$ & -- & $1.6(1)$ & $250(120)$\\
J1824$-$1945 & $12.2(10)$ & $10(14)$ & $-12(14)$ & $-100(220)$ & $38.4$ & $5$ & 12 & $4(2)$ & $280(200)$\\
J1825$-$0935 & $-30(6)$ & $110(21)$ & $-13(11)$ & $-9(5)$ & $29.1$ & $5$ & 11 & $0.3(7)$ & $160(380)$\\
J1833$-$0338 & $-16(3)$ & $-3(9)$ & -- & -- & $21.6$ & $2$ & -- & $5.14(3)$ & $400(80)$\\
J1833$-$0827 & $-26(4)$ & $-39(17)$ & $-31(4)$$^\text{e}$ & $13(11)$$^\text{e}$ & $19.5$ & $3$ & 1 & $4.5(5)$ & $1000(330)$\\
J1841+0912 & $-12(5)$ & $38(7)$ & -- & -- & $25.6$ & $2$ & -- & $1.9(2)$ & $370(80)$\\
J1848$-$0123 & $5(2)$ & $-26(6)$ & -- & -- & $29.1$ & $4$ & -- & $4.4(4)$ & $550(130)$\\
J1900$-$2600 & $-21(5)$ & $-16(37)$ & $-19.9(3)$ & $-47.3(9)$ & $23.8$ & $3$ & 13 & $0.7(4)$ & $90(90)$\\
J1901+0331 & $-7(2)$ & $-45(5)$ & -- & -- & $20.7$ & $2$ & -- & $7(2)$ & $1500(460)$\\
J1909+0007 & $5(3)$ & $-32(7)$ & -- & -- & $24.3$ & $2$ & -- & $3.6(3)$ & $550(120)$\\
J1909$-$3744$^\text{t}$ & $-9.0(2)$ & $-36.2(7)$ & $-9.517(5)$ & $-35.80(2)$ & $8.5$ & $2$ & 4 & $1.14(1)$ & $200(4)$\\
J1915+1009 & $6(2)$ & $-17(4)$ & $2(4)$$^\text{e}$ & $-7(6)$$^\text{e}$ & $23.4$ & $3$ & 1 & $7(2)$ & $580(200)$\\
J1932+1059 & $95.2(7)$ & $49(1)$ & $94.1(1)$ & $43.0(2)$ & $47.1$ & $6$ & 14 & $0.31(9)$ & $160(50)$\\
J1941$-$2602 & $6(4)$ & $1(17)$ & $12(2)$ & $-10(4)$ & $29.2$ & $3$ & 2 & $3.6(9)$ & $110(90)$\\
J2048$-$1616 & $132(8)$ & $-66(26)$ & $113.16(2)$ & $-4.6(3)$ & $47.6$ & $7$ & 5 & $0.95(3)$ & $660(70)$\\
\end{tabular}
\end{table*}

\begin{table*}
\contcaption{}
\begin{tabular}{lcccccccccl}
\hline
        & \multicolumn{2}{c}{This work} & \multicolumn{2}{c}{Literature}\\
PSRJ    & $\mu_\alpha \cos \delta$      & $\mu_\delta$  & $\mu_\alpha \cos \delta^\text{c}$     & $\mu_\delta^\text{c}$ & $\Delta t$    & \#Pos & Ref   & $d$   & $V_\text{t}$\\
        & (mas $\text{yr}^{-1}$)        & (mas $\text{yr}^{-1}$)        & (mas $\text{yr}^{-1}$)  & (mas $\text{yr}^{-1}$)      & (yr)  &   &   & (kpc) & (km $\text{s}^{-1})$\\
\hline
J2145$-$0750 & $-9.4(3)$ & $-12.8(8)$ & $-9.59(8)$ & $-8.9(3)$ & $18.7$ & $3$ & 4 & 0.53 & $40(2)$\\
J2222$-$0137 & $42(2)$ & $-15(4)$ & $44.73(2)$ & $-5.68(6)$ & $5.8$ & $2$ & 15 & $0.27(4)$ & $60(10)$\\
J2241$-$5236$^\text{t}$ & $17.1(1)$ & $-3.32(5)$ & -- & -- & $7.0$ & $2$ & -- & $0.7(1)$ & $60(10)$\\
J2330$-$2005 & $64(6)$ & $2(12)$ & $75(2)$ & $5(3)$ & $38.4$ & $4$ & 2 & $0.5(1)$ & $140(50)$\\
\hline
\multicolumn{10}{l}{$^\text{t}$ Significant proper motion signature in UTMOST timing residuals.}\\
\multicolumn{10}{l}{$^\text{e}$ Literature proper motion was converted from ecliptic to equatorial coordinates.}\\
\multicolumn{10}{l}{$^\dagger$ See text.}
\end{tabular}
\end{table*}

The large time spans (up to 48 yr) between our timing observations and those from the literature provide a strong lever arm to determine proper motions. We have compiled pulsar positional data at various measurement epochs from the literature (see Table~\ref{tab:LiteratureData}). We consider positions derived from pulsar timing observations only, as the analysis would be heavily dominated by interferometric measurements with their small uncertainties otherwise. The timing positions depend on, among other factors, the Solar System ephemeris used to transfer the site arrival times to the Solar System barycentre. In our database they range from the early Massachusetts Institute of Technology Lincoln Laboratory, later Harvard Center for Astrophysics (MIT/CfA) ephemerides and the JPL DE96 development ephemeris over the well-used JPL DE200 to the more recent JPL DE430 ephemeris employed in this work \citep{1967Ash, 1982Standish, 1990Standish, 2014Folkner}. Before proper motions can be estimated the positions must be converted to the same reference frame. In principle, this can be done for the earliest ephemerides using the rotation matrices presented by \citet{1996Bartel}. The newer ephemerides (from JPL DE400 onwards) are already referenced to the International Celestial Reference Frame (ICRF). The differences in position are less than 250~mas between the earliest \citep{1984Fomalont, 1996Bartel} and less than 2~mas among the newest Solar System ephemerides \citep{2017Wang}. While the former is certainly a significant systematic error by current standards, the positional uncertainties of the historical data for the mainly normal pulsars in our data set are often larger than that, allowing us to ignore the rotation. We transformed all positions to the ICRF before comparison, where we set the observation time to the period epoch for the positions given in the Fundamental~Katalog~4 reference frame (B1950 coordinates).

From this data set we derive inferred proper motions in the following way: if there are at least three position measurement epochs for a pulsar we determine the rate of change in right ascension (RA) $\dot{\alpha}$ and declination (Dec) $\dot{\delta}$ from a maximum likelihood fit of a linear function to the data separately in both dimensions. The fit is robust against outliers, for which we employ the Huber loss function \citep{1964Huber, 2014Ivezic} with the implementation details given by \citet{2018Jankowski}. In case there are only two measurements (ours and one from the literature), we simply compute the slopes of the lines that join the two points (again, a separate slope is computed for RA and Dec). The proper motions are then $\mu_\alpha = \dot{\alpha} \cos(\delta)$ and $\mu_\delta = \dot{\delta}$. To ensure the reliability of the results we derive proper motions only if the UTMOST timing residuals are white, as judged visually and formally using the Shapiro--Wilk test \citep{1965Shapiro, 2014Ivezic}. This is because timing noise or other phenomena such as glitches, or spin-down rate changes can severely affect the accuracy of the position measurement. \citet{2005Zou} have evaluated this using simulations and more recently, \citet{2015Kerr} studied the effect of timing noise on the positions of \textit{Fermi} pulsars. We additionally require that the positions differ by at least 3 times the combined positional uncertainty in a pairwise sense in at least one dimension. The uncertainties are the formal $1 \sigma$ ones from the fit or calculated using first order error propagation from the position measurements.

The two-point case critically depends on the assumption that our position measurements and those from the literature are free of significant systematic effects. Including multiple measurements from the literature and using a fitting algorithm that is robust against outliers allows us to derive more accurate proper motion values. Generally, we expect that the higher the number of data points and the larger the time spans, the more accurately we can derive individual proper motions. To verify that this method delivers plausible results, we compare our inferred proper motions with the ones from the catalogue, which for these pulsars are mainly from interferometric observations. The agreement is good with median and rms differences of about $16$ and 75 per cent in RA and $59$ and 72 per cent in Dec (see Fig.~\ref{fig:ProperMotionComparison}). Both the absolute and relative differences decrease with increasing characteristic age; however, there is large scatter about this trend. That is expected because timing positions are harder to determine reliably for young pulsars with significant timing noise. We do not see a clear trend with respect to the maximum time span between measurements and their number.

However, the proper motions of the pulsars J0922+0638 and J1744$-$1134 are discrepant by at least 5 times the combined uncertainty in RA, for the former also in Dec. PSR J0922+0638 (B0919+06) is known to exhibit periodic changes in the spin-down rate of around 70 per cent with a periodicity of about 1.6 yr \citep{2010Lyne}. These seem to be magnetospheric in nature but have also been interpreted as slow glitches. In addition, this pulsar has exhibited multiple glitches of normal signature \citep{2010Shabanova}. As such, it seems possible that some of the historic timing position measurements are systematically affected. As yet, we do not resolve the spin-down rate changes in UTMOST data and the residuals appear white. Taken at face value, the fit indicates a proper motion in Dec of about 214~$\text{mas} \: \text{yr}^{-1}$, significantly different from the interferometric measurement of $86.4~\text{mas} \: \text{yr}^{-1}$ by \citet{2003Brisken}. Interferometric techniques are clearly needed in such a complex case (e.g.\ \citealt{1982Lyne}). Regarding the PPTA MSP J1744$-$1134, the proper motion uncertainties are very small and while significant, the absolute difference between our measurements and the published values is about 3~$\text{mas} \: \text{yr}^{-1}$.

Overall, we estimate the total proper motions for 60 pulsars that are significant in at least one coordinate and show them in a histogram in Fig.~\ref{fig:InferredProperMotions}. Out of these, 24 are newly determined\footnote{Not listed in version 1.54 of the pulsar catalogue.} and we improve the precision for others, sometimes significantly. Nearly all measurements are well contained within $\pm 150 \: \text{mas} \: \text{yr}^{-1}$. The proper motion distribution shows a slight positive skew in both dimensions, but the number of high proper motions is low. For the four pulsars for which proper motions can be estimated directly from the UTMOST timing data, they agree with the inferred values within a median difference of 6~$\text{mas} \: \text{yr}^{-1}$ (about 3 times the combined uncertainty). This indicates that longer timing baselines than those presented here are needed to estimate proper motions from timing residuals.

We convert the proper motions to transverse velocities using the relation $V_\text{t} = 4.74 \: \text{km} \: \text{s}^{-1} \mu_{t} \: d$, where the total proper motion $\mu_{t} = \sqrt{\mu_\alpha^2 + \mu_\delta^2} $ is given in $\text{mas} \: \text{yr}^{-1}$ and $d$ is the distance in kpc. We use measured pulsar distances from the pulsar catalogue where available, otherwise, we compute the median distance derived from the DM assuming the \textsc{tc93}, \textsc{ne2001}, and \textsc{ymw16} Galactic free electron-density models \citep{1993Taylor, 2002Cordes, 2017Yao}. Most of the measured distances come from the summary publication of \citet{2012Verbiest} and are corrected for the Lutz--Kelker bias \citep{1973Lutz}. For simplicity, we take the uncertainties as the maximum of the asymmetric confidence intervals. For the DM-derived distances, we use the median among the three models, which most of the time is the \textsc{tc93} value, as there are significant discrepancies for a small number of pulsars, e.g.\ PSRs J0134$-$2937 and J1121$-$5444. For the transverse velocities, we derive the uncertainties using first order error propagation, with the standard error of the median in the DM-derived case. Note that the transverse velocities are with respect to the Solar System barycentre and include Galactic rotation. The majority of pulsars have transverse velocities of less than about $1000 \: \text{km} \: \text{s}^{-1}$, with two outliers near 1200 and 1500~$\text{km} \: \text{s}^{-1}$. The outliers are PSR J0922+0638, as discussed above, and PSR J1901+0331. The velocity uncertainty of the latter is large (nearly $500 \: \text{km} \: \text{s}^{-1}$), as it only has an HI distance estimate \citep{2012Verbiest}. It is therefore consistent at the $1 \sigma$ level. Nonetheless, when excluding both as likely errors, the mean and median transverse velocities are $273(25)  \: \text{km} \: \text{s}^{-1}$ and $215(31) \: \text{km} \: \text{s}^{-1}$ with a standard deviation of $187 \: \text{km} \: \text{s}^{-1}$, which agrees well with the mean 2D speed of normal pulsars of $246(22)  \: \text{km} \: \text{s}^{-1}$ derived by \citet{2005Hobbs}.

\subsubsection{Modality of the pulsar velocity distribution}
\label{sec:PulsarVelocityDistribution}

\begin{figure}
	\centering
	\includegraphics[width=\columnwidth]{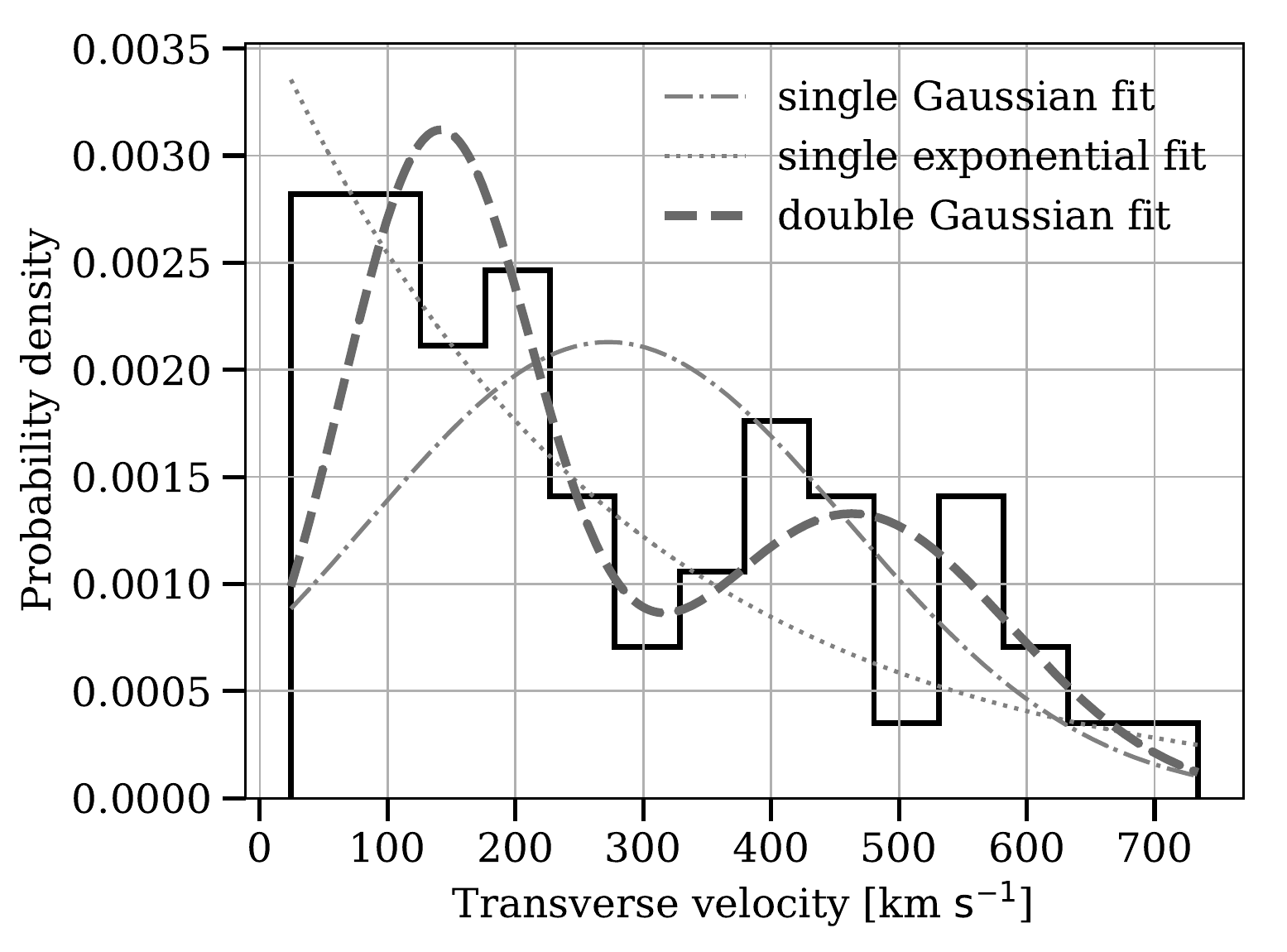}
	\includegraphics[width=\columnwidth]{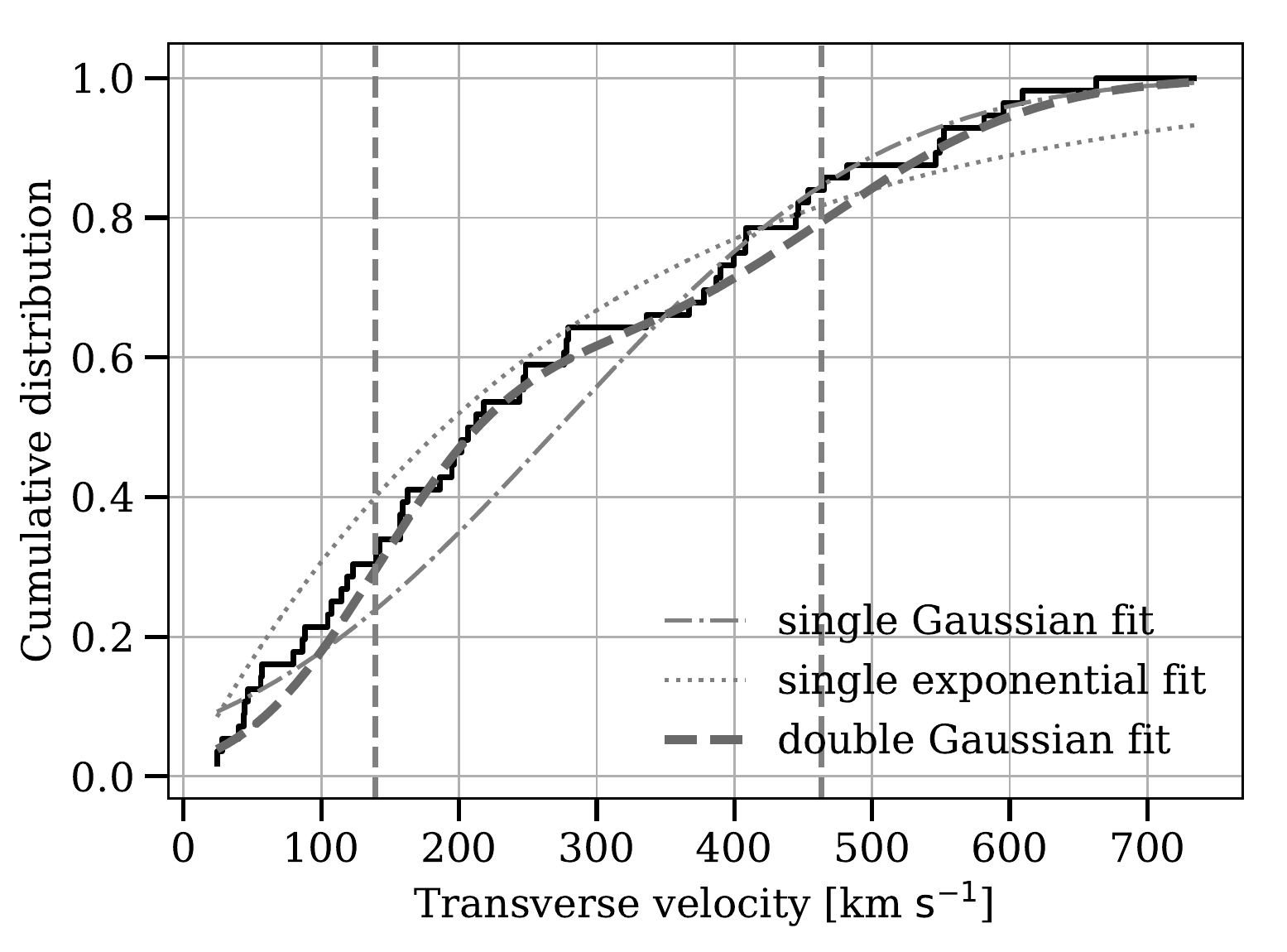}
	\caption{Differential and cumulative distribution of the transverse pulsar velocities from this work with respect to the Solar System barycentre. We show three unbinned maximum likelihood fits to the data: (1) a single Gaussian velocity component, (2) a single exponential velocity component and (3) a two-component Gaussian velocity model. The two-component model (bimodal velocity distribution) is preferred by the data. The vertical lines indicate the mean values of the two velocity components.}
	\label{fig:PulsarVelocityDistribution}
\end{figure}

The pulsar velocity distribution inferred from proper motion measurements has been analysed by various authors, mostly with the aim to understand the pulsar birth velocities, which represent the kick imparted on a neutron star in a supernova explosion. A major question concerns whether the distribution is unimodal, with a single velocity component, or bimodal with a low and high-velocity component. The former has been argued for by \citet{2005Hobbs} for example, who find that the 3D velocities are well described by a single Maxwellian, or \citet{2006Faucher}, who favour a single exponential distribution for the 1D velocity components. Bimodal distributions or the deviation from a single velocity component were suggested for example by \citet{1982Lyne}, \citet{2002Arzoumanian}, \citet{2003Brisken} and more recently \citet{2017Verbunt}, who find two velocity components centred at $90 - 120~\text{km} \: \text{s}^{-1}$ and $300- 540~\text{km} \: \text{s}^{-1}$.

We test whether the transverse velocity distribution is bimodal by fitting three different models to the unbinned velocity data in an iterative maximum likelihood sense: (1) a single Gaussian component, (2) a single exponential component and (3) a bimodal distribution consisting of two Gaussian velocity components (Gaussian mixture model). We select the best-fitting model objectively based on the Akaike information criterion (AIC), corrected for finite sample sizes, and determine the strength of the preference of the best-fitting model over the other models tested using the Akaike weights (e.g.\ \citealt{1974Akaike, 2010Burnham, 2018Jankowski}). That is, the best-fitting model is the one with the lowest AIC. The AIC accounts for the different number of free parameters between the models. For this analysis, we select only the velocities below 1000~$\text{km} \: \text{s}^{-1}$ in order to remove potentially erroneous high-velocity measurements. We find that the measured transverse velocities are best described by a bimodal distribution with two Gaussian velocity components centred at 139 and 463~$\text{km} \: \text{s}^{-1}$ with Gaussian standard deviations of 76 and 124~$\text{km} \: \text{s}^{-1}$, respectively. The weights of the components are 0.59 and 0.41, indicating an abundance of about 1.43/1 for the low-velocity component. The bimodal distribution is preferred over the other models with a probability of 84 per cent, with the single exponential model being second. The data clearly disfavour a single Gaussian component model (probability less than 0.1 per cent). We show the differential and cumulative velocity distribution together with the three fitted models in Fig.~\ref{fig:PulsarVelocityDistribution}. The vertical lines mark the mean values of the two velocity components.

Our analysis confirms the bimodal nature of the transverse velocity distribution and the mean velocities of the two components agree well with previous work (e.g.\ \citealt{2002Arzoumanian, 2003Brisken, 2017Verbunt}). This is remarkable and reassuring, as our velocities are derived using proper motion measurements from timing positions at different epochs, while the latter two authors base their conclusions on interferometric proper motion measurements. However, there is generally good agreement between our proper motion measurements and those from interferometric observations. The analysis is clearly biased, as our data set includes mainly bright pulsars that were discovered early, for which multiple timing position measurements exist. The data set is also relatively small, with only 56 velocity measurements after selection, including both normal and recycled pulsars. Consequently, our conclusions might not be representative for the pulsar population as a whole.

\begin{figure}
	\centering
	\includegraphics[width=\columnwidth]{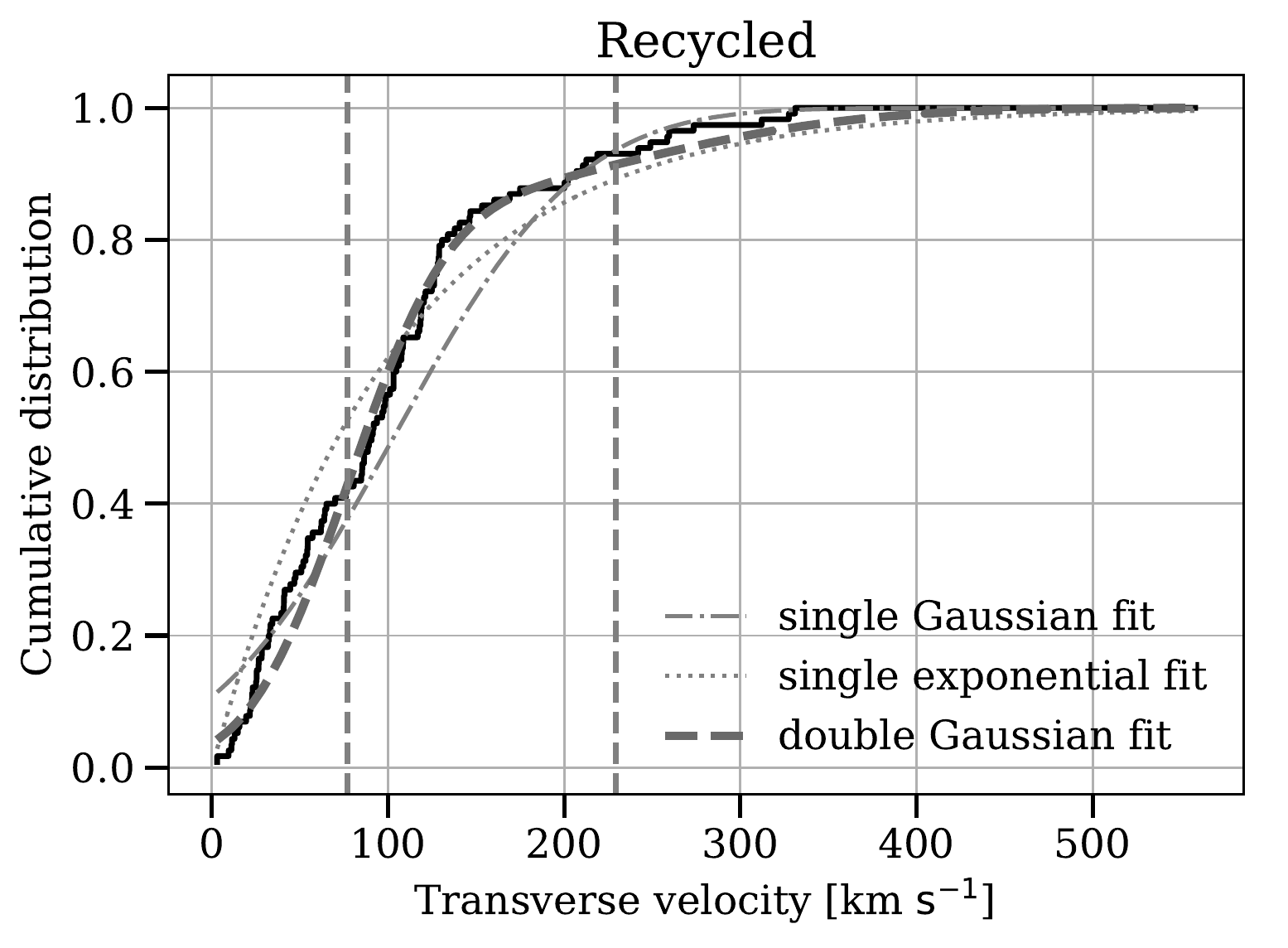}
	\includegraphics[width=\columnwidth]{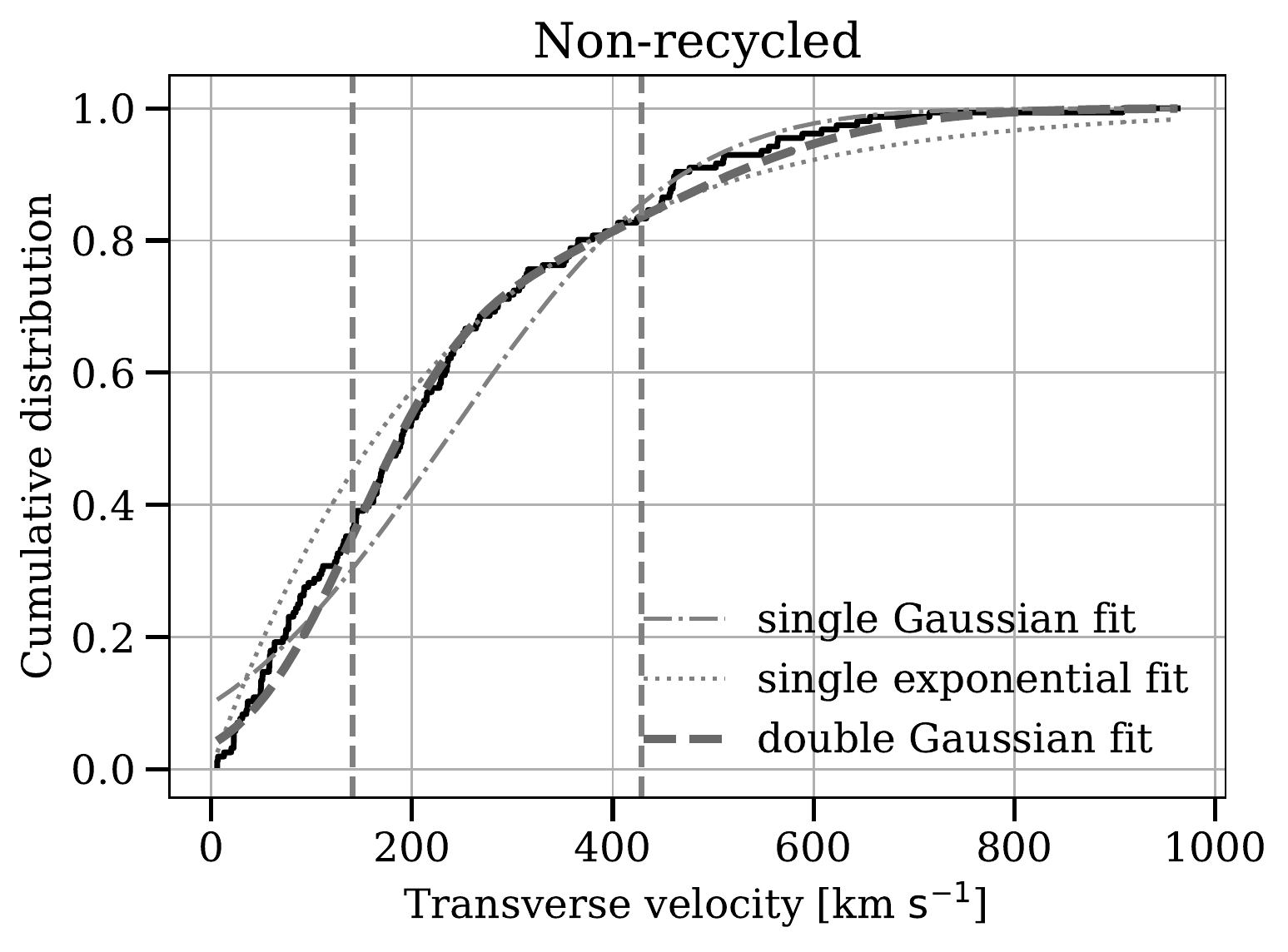}
	\caption{Cumulative distributions of the transverse velocities of 115 recycled (top) and 156 non-recycled pulsars (bottom) using the proper motion data from the pulsar catalogue. We show the three model fits as in Fig.~\ref{fig:PulsarVelocityDistribution}. The transverse velocities of recycled pulsars are much lower than those of non-recycled ones.}
	\label{fig:PsrcatVelocityDistribution}
\end{figure}

In a second step, we repeat the analysis independently for all pulsars that have proper motions listed in version 1.54 of the pulsar catalogue, regardless of the measurement method, in the same way as above. The resulting distribution contains 278 transverse velocities, which is more than a factor of five increase in comparison with our measurements. In addition to the full data set, we analyse the velocity data separately for millisecond pulsars (MSPs) ($P \leq$ 30~ms) and slow pulsars, isolated and pulsars in binary systems and recycled and non-recycled pulsars, as defined using the empirical relation $\frac{\dot{P}}{10^{-17}} \leq 3.23 \left( \frac{P}{100 \text{ms}} \right)^{-2.34}$ for recycled pulsars found by \citet{2012Lee}. A single exponential model clearly fits the data best in each case with a probability near 100 per cent in comparison with the other models. The single and double Gaussian models are clearly disfavoured, with the former most significantly. The velocity distributions of recycled and non-recycled pulsars, determined using the above criteria, show the clearest difference (Fig.~\ref{fig:PsrcatVelocityDistribution}). The velocities of recycled pulsars are much lower than their non-recycled counterparts, with 80 per cent of the former having velocities below $135 \: \text{km} \: \text{s}^{-1}$, while the value for the latter is $365 \: \text{km} \: \text{s}^{-1}$. The mean velocities, determined from the rate parameters of the best-fitting exponential distributions, are $103(10)$ and $235(19) \: \text{km} \: \text{s}^{-1}$ respectively, where the uncertainties are the formal $1 \sigma$ ones from the fit. These are in good agreement with the measurements of 87(13) and $246(22) \: \text{km} \: \text{s}^{-1}$ by \citet{2005Hobbs}.

One might suspect that the difference in best-fitting model between the two analyses (our transverse velocities and those from the pulsar catalogue) is due to different precision of the transverse velocities between the data sets. That is, it could be that the two velocity components are simply smeared out by large uncertainties in proper motion, or distance measurements in the larger data set. However, this does not seem to be the case, as the mean and median absolute and relative velocity uncertainties are comparable. Similarly, the fraction of pulsars that have DM-derived distances is roughly the same of about 60 per cent.

\subsubsection{Potential pulsar birth sites}
\label{sec:PulsarBirthSites}

Using our position and inferred proper motion measurements we integrate the equations of motion of all pulsars with proper motion information back to potential birth sites using \textsc{galpy} \citep{2015Bovy}, assuming a realistic Galactic gravitational potential (\textsc{galpy}'s MWPotential2014). The integration times are determined from their characteristic ages assuming varying breaking indices between 1.5 and 3 that are constant throughout their evolution. We adopt the usual assumption that the birth periods are much smaller than the current periods. We then search for 2D spatial correlations between the varying endpoints of their trajectories with Galactic supernova remnants (SNRs), or sources at X- and $\gamma$-ray energies \citep{2008Wakely, 2014Green, 2015Acero, 2016Rosen, 2017Green}, which could trace yet undetected supernova shocks interacting with dense environments. This approach has various caveats: the characteristic ages generally provide only crude age estimates at best, because reliable braking indices are currently only known for a small number of pulsars and the individual birth periods are essentially unknown. Another complication is that the pulsar radial velocities are unknown and that the distances are uncertain (see Section~\ref{sec:InferredProperMotions}). A further problem is that the average fading times of SNRs in the radio are of the order of 100~kyr, which is often below the characteristic ages of the pulsars studied. A full analysis, as for example presented by \citet{2013Noutsos}, is beyond the scope of this paper. Nonetheless, our analysis can identify potential birth regions and acts as a consistency check for the measured proper motions.

The initial conditions are set to the present day measured values of position, proper motion, distance (Table~\ref{tab:ProperMotions}) and zero radial velocity. We vary the proper motions within their uncertainties and the radial velocities between the three values of $-200$, $0$ and $200 \: \text{km} \: \text{s}^{-1}$ to explore the spread in trajectories. We find that most pulsars are leaving the Galactic plane in either positive or negative latitude direction, as expected. Others are moving mainly in the Galactic disk (PSRs J0837$-$4135, J1224$-$6407 and J1745$-$3040) and a small number of pulsars seem to originate from outside the central $\pm 15 \degr$ (PSRs J1557$-$4258 and J2048$-$1616). In the latter case, \citet{2009Chatterjee} suggested that PSR J2048$-$1616 was born in the open cluster NGC 6604. While a few pulsar -- SNR associations are suggestive, which could indicate potential birth sites, care needs to be taken and associations might only be possible to claim in a statistical sense. In any case, our analysis confirms the plausibility of the proper motion measurements, and that most pulsars are born close to the Galactic plane and are moving away from it.

\subsection{Pulse widths}
\label{sec:PulseWidths}

\begin{figure}
	\centering
	\includegraphics[width=\columnwidth]{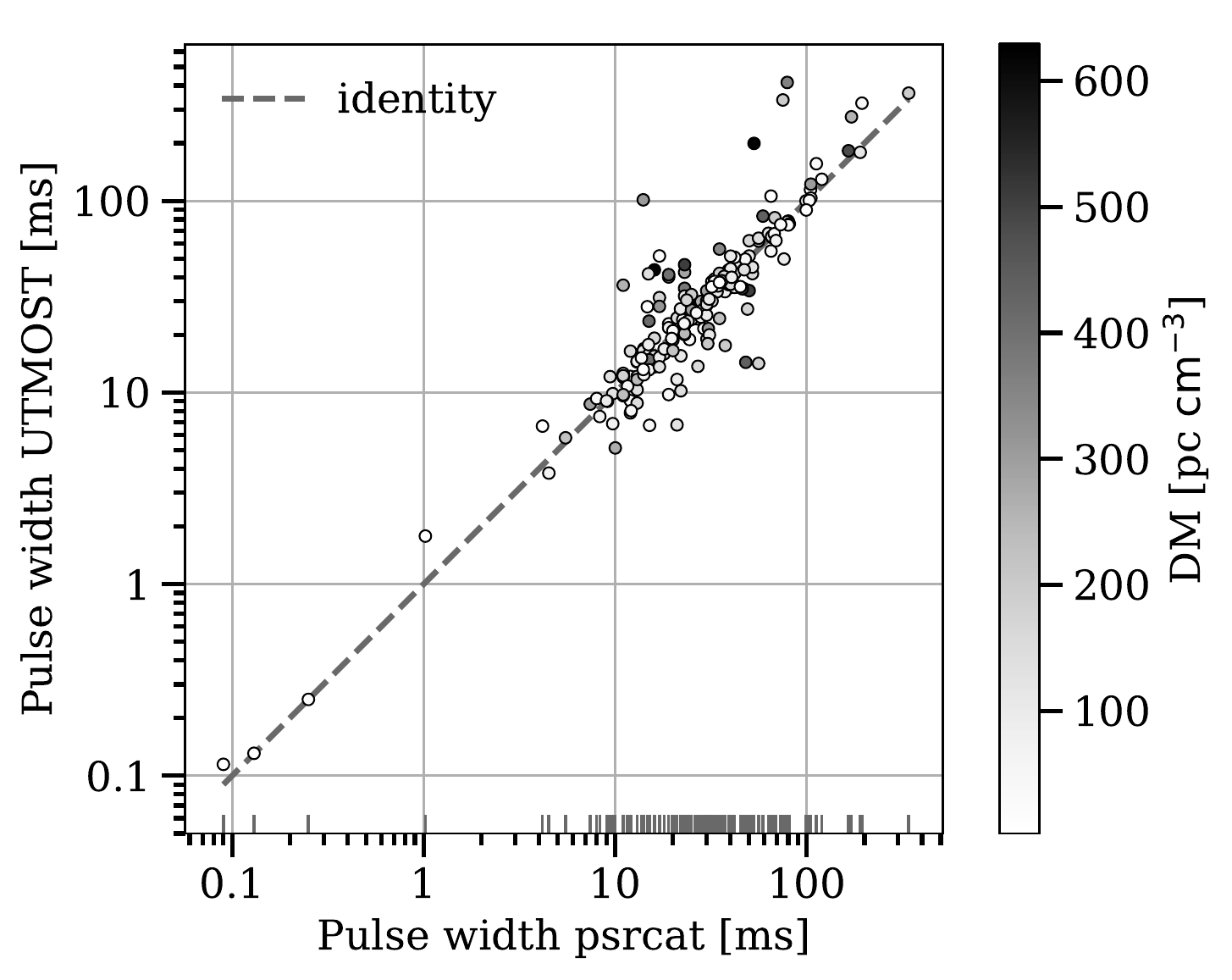}
	\caption{Comparison of measured pulse widths at 10 per cent maximum with those from the pulsar catalogue. The pulse widths measured at 843~MHz and primarily at 1.4~GHz are largely consistent, with the difference increasing with DM, as expected for profile scatter-broadening in the ISM.}
	\label{fig:PulseWidthComparison}
\end{figure}

\begin{figure}
	\centering
	\includegraphics[width=\columnwidth]{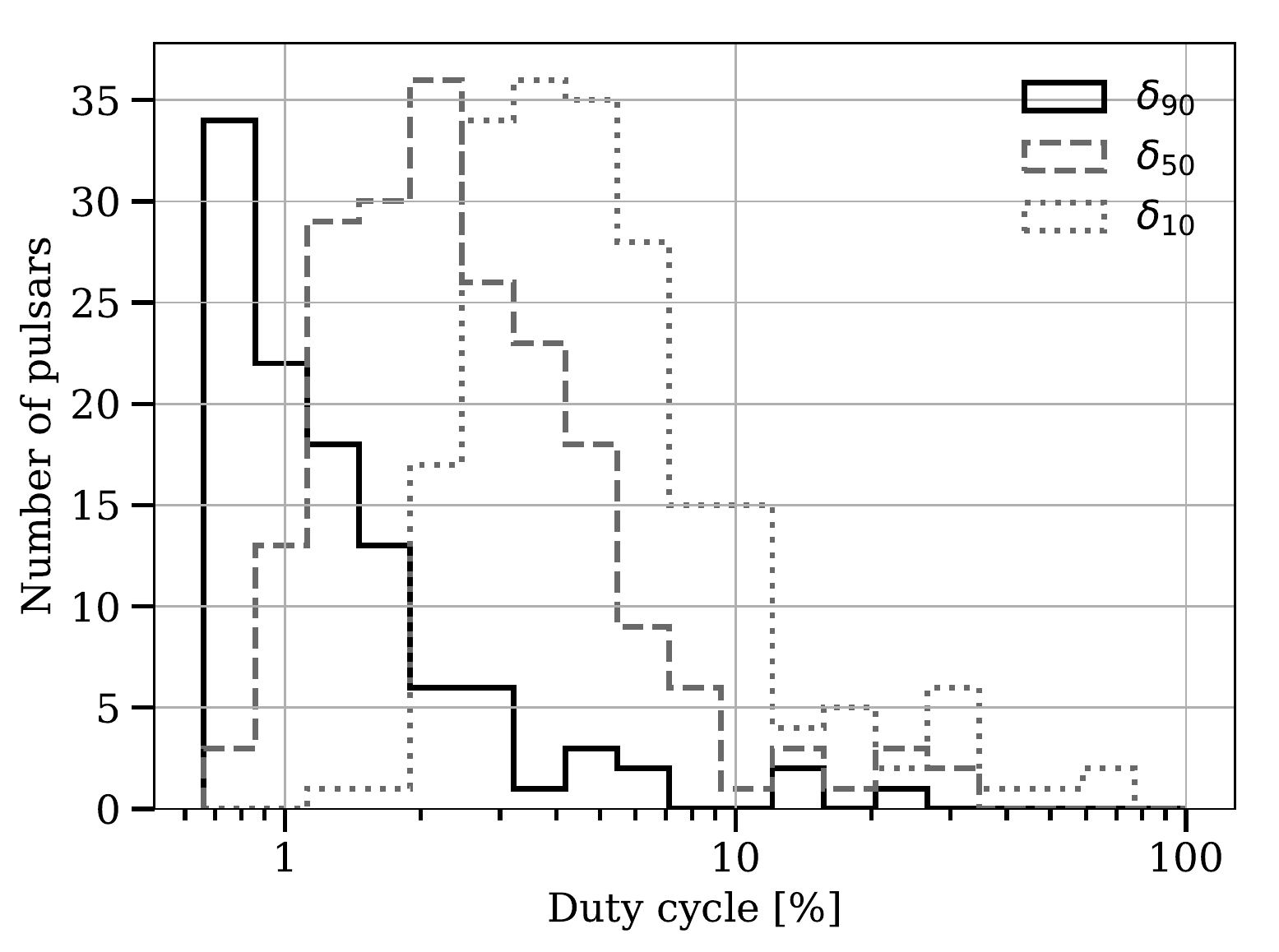}
	\caption{Histograms of measured duty cycles at 90, 50 and 10 per cent maximum, which show significant positive skew and are consistent with log-normal distributions.}
	\label{fig:DutyCycleDistributions}
\end{figure}

\begin{figure}
	\centering
	\includegraphics[width=\columnwidth]{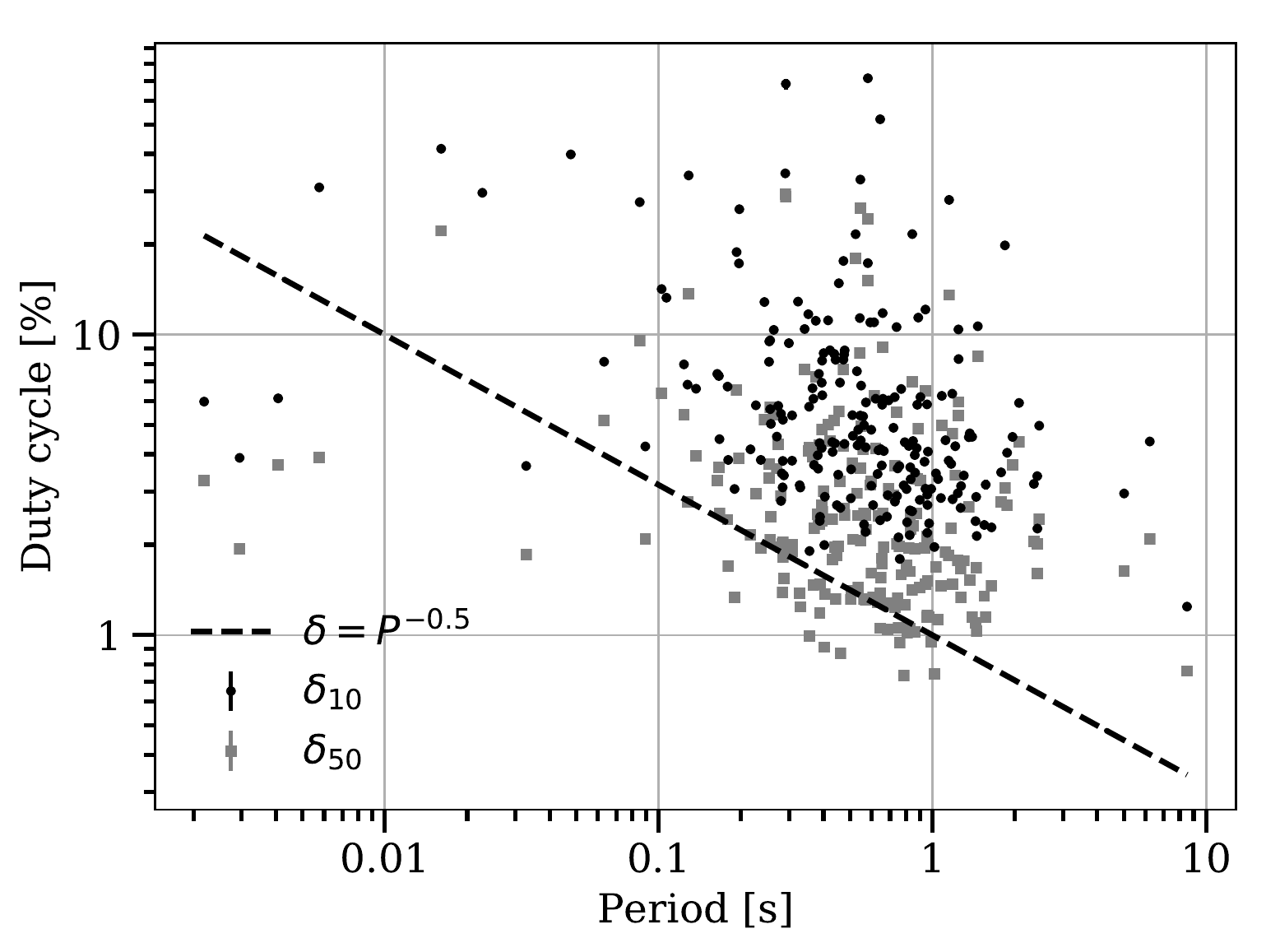}
	\caption{Plot of duty cycles at 50 and 10 per cent maximum versus pulsar period. We confirm the lower limit of $\delta_{10} = P^{-0.5}$ for the scaling of duty cycle with period for all pulsars with periods longer than about 40~ms.}
	\label{fig:DutyCycleScaling}
\end{figure}

A detailed analysis of pulse profiles using UTMOST data is currently restricted by the fact that only single-polarization data (mainly RCP) is available from the MOST. However, in Section~\ref{sec:PolarimetricResponseOfTheFeeds} we have presented a technique with which full-polarization pulse profiles (e.g.\ obtained at other telescopes or from the European Pulsar Network database of pulse profiles\footnote{\url{http://www.epta.eu.org/epndb}}; \citealt{1998Lorimer}) can be projected onto the MOST feeds. We have demonstrated this for the complex case of the MSP J0437$-$4715 and have simultaneously determined the best-fitting polarimetric parameters of the feeds. With this caveat in mind, we estimate the pulse widths at 90, 50 and 10 per cent maximum ($\text{W}_{90}, \text{W}_{50}, \text{W}_{10}$) for all pulsars analysed in this work either from high-S/N profiles or from the smoothed standard profiles. Our data set contains 7 pulsars with interpulses, for which we estimate the pulse width as the sum of all profile components. The pulse widths are listed in Appendix~\ref{sec:FluxDensitiesAndPulseWidths}.

We compare the 10 per cent pulse widths with the ones from the ATNF pulsar catalogue which are mainly from observations at 1.4~GHz. We find that the estimated pulse widths are largely consistent with the ones from the catalogue and that the difference increases with DM as expected because of profile scatter-broadening in the interstellar medium (ISM; e.g.\ \citealt{2004Bhat}; see Fig.~\ref{fig:PulseWidthComparison}). For the pulsars for which we measure smaller pulse widths in comparison with the catalogue, we suspect that this is a result of sampling a single polarization only or that the literature pulse widths were estimated from low-frequency data. Another possibility is that these pulsars show anomalous profile evolution with frequency.

The duty cycle distributions ($\delta_{90}$, $\delta_{50}$, $\delta_{10}$ with $\delta_i = W_i/P$) show significant positive skew (see Fig.~\ref{fig:DutyCycleDistributions}). We investigated the cumulative distribution functions and quantile--quantile (Q--Q) plots for the duty cycles at 10 and 50 per cent maximum, assuming various theoretical distributions, and find that a log-normal best describes the measurements. However, there are small deviations and the agreement is poorer, especially for the duty cycles at 50 per cent. If we exclude the MSPs with periods shorter than 30~ms, the agreement becomes slightly better, as might be expected. While the emission properties of normal pulsars and MSPs are similar, their beams and opening angles are known to differ, with MSPs generally having much larger beam opening angles and therefore $\delta_{10}$ measurements \citep{1998Kramer}. The best-fitting parameters are 1.0, 0.7 and 1.7 for the shape, location and scale parameters for $\delta_{50}$ and 1.0, 1.7, and 3.1 for $\delta_{10}$. The median duty cycles are 2.3 and 4.4 per cent.

Finally, we analyse the scaling of duty cycle with pulsar period (see Fig.~\ref{fig:DutyCycleScaling}). We confirm the well known lower limit of $\delta \propto P^{-0.5}$ (e.g.\ \citealt{1998Gould, 2016Pilia}) for the duty cycles at 10 per cent. All $\delta_{10}$ measurements are above this line up to a period of about 40~ms. Faster pulsars show more and more deviations from this lower limit. Physically this relates to a scaling of minimum beam longitude that cuts through the line of sight with the period, which is related to the minimum beam opening angle for orthogonal rotators with beams pointing towards the line of sight.

\subsection{Pulsars with interesting behaviour}
\label{sec:PulsarsWithInterestingBehaviour}

\begin{figure}
	\centering
	\includegraphics[width=\columnwidth]{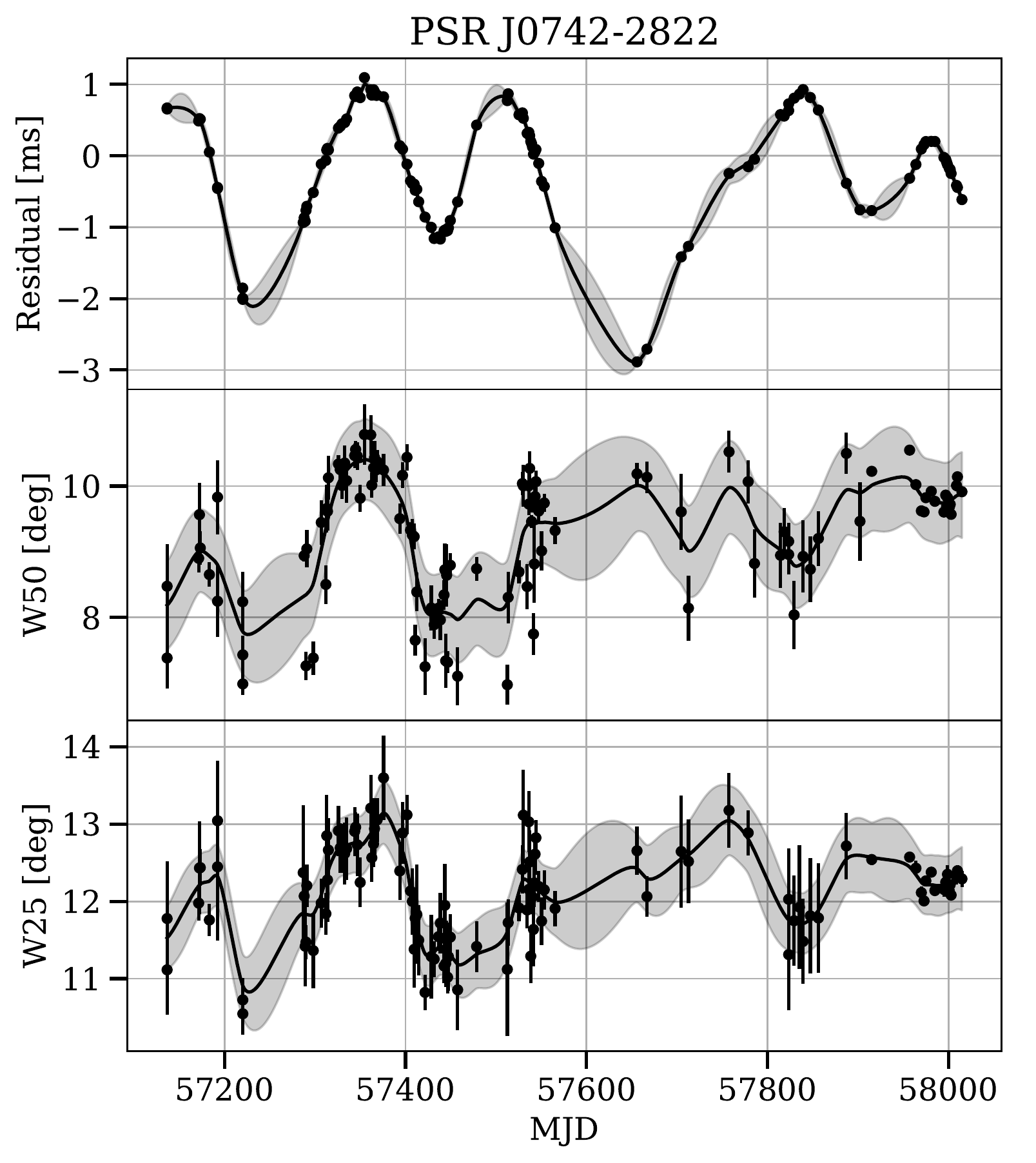}
	\caption{Correlation of timing residuals with 50 and 25 per cent pulse width in the case of the mode-changing pulsar J0742$-$2822. We also show Gaussian process regressions to the data (solid black lines) and their $1 \sigma$ uncertainty bands. The timing residuals and pulse widths are largely in phase until about MJD 57570, after which the behaviour changes, with the residuals and pulse widths being anti-correlated from MJD 57800 onwards.}
	\label{fig:ModeChanging}
\end{figure}

\begin{figure}
	\centering
	\includegraphics[width=\columnwidth]{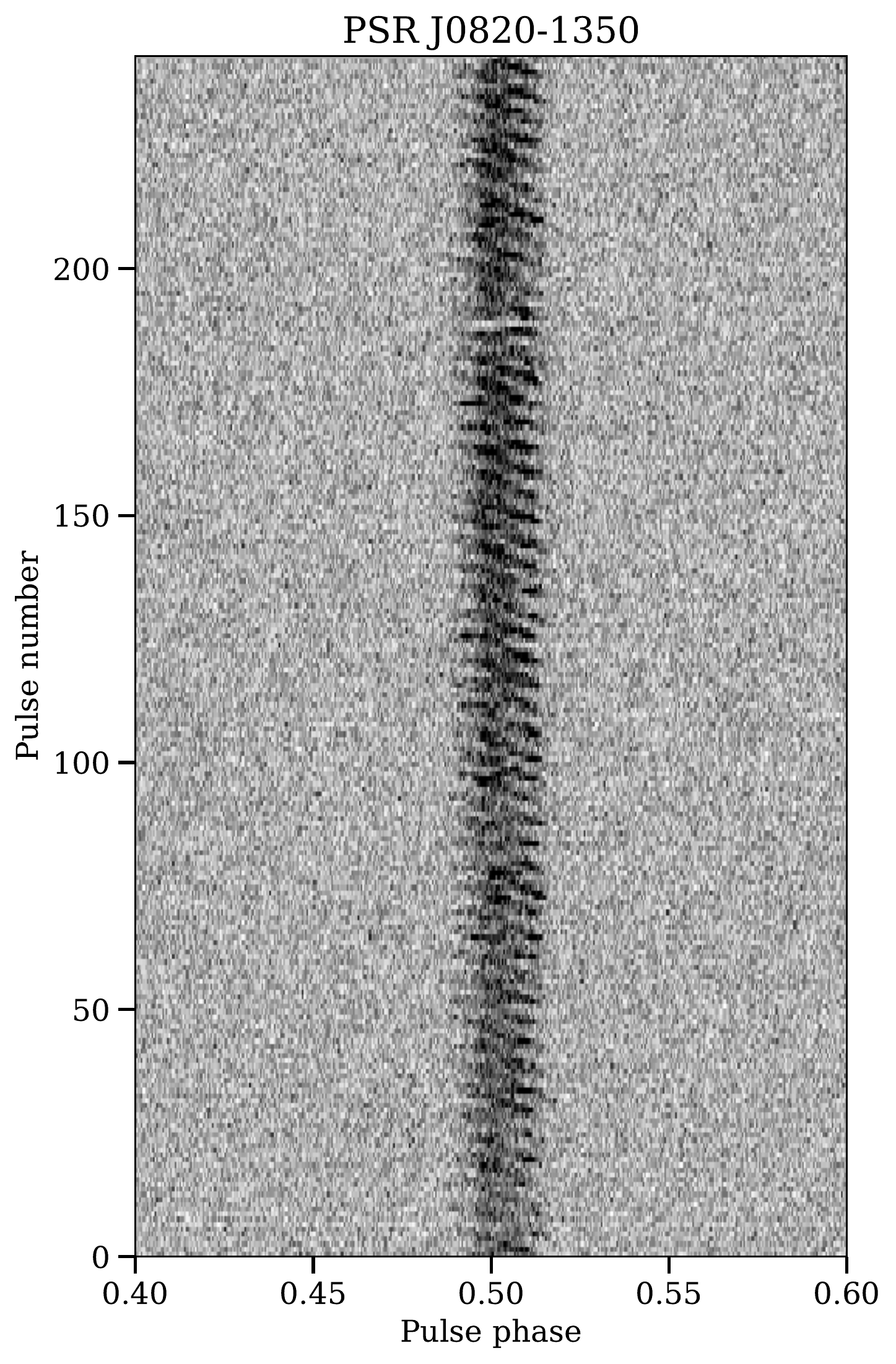}
	\caption{Single pulse stack of PSR J0820$-$1350, generated from consecutive UTMOST filterbank data. The greyscale represents pulse intensity in arbitrary units. The drifting sub-pulses and their separation into drift bands can clearly be seen. Horizontally the bands are separated by $P_2$ and vertically by $P_3$. The dynamic range has been reduced to show the drift bands more clearly.}
	\label{fig:DriftingSubpulses}
\end{figure}

In this section, we discuss a selection of pulsars that show interesting behaviour.

\begin{description}
	\item{PSR J0742$-$2822}: This pulsar is known to change between two profile states. Its spin-down rate is correlated with pulse shape variations, and the rate of changes seems to be influenced by glitch events \citep{2010Lyne, 2013Keith}. It is located in a bow shock nebula. We find a clear correlation between timing residuals and its 50 and 25 per cent pulse widths in the UTMOST data (Fig.~\ref{fig:ModeChanging}). A spin-down model including the position and terms up to the third frequency derivative have been subtracted. The observations are those that were obtained within $\pm 3$ hours from the meridian in order to avoid any instrumental effects. We also show a Gaussian process regression, for which we use a kernel consisting of a Mat\'{e}rn (3/2) covariance function, a constant and a white noise component \citep{2014Ivezic}. We may see a reversal of correlation behaviour: between the start of our observations and MJD 57570, where we have the densest sampling, the timing residuals and the pulse widths are largely in phase; the peaks and troughs line up temporally, which means that spin-down rate and pulse width are anti-correlated. From MJD 57800 onwards the behaviour seems to be reversed; peaks in timing residuals line up with troughs in pulse width. In between, the quasi-periodicity seems to be interrupted, for which the reason is unclear. The data do not show any clear evidence of a micro-glitch, but the sampling is rather sparse. A Lomb-Scargle periodogram of the residuals shows a highly significant ($\gg 5 \sigma$) peak at a period of 207 days with a full width at half-maximum (FWHM) of 40 days, with the significance determined by bootstrap resampling of the data. A second, only slightly less significant peak exists at a period of 155 days with at FWHM of about 24 days. This is slightly higher than the measurement by \citet{2010Lyne} of about $135 \pm 10$ days but within the combined uncertainty.

	\item{PSR J0820$-$1350}: This pulsar shows drifting subpulses together with pulse nulling at the 1 per cent level. Nulls seem to influence the subpulse drift, with a step in pulse longitude and a reduced drift rate after a null and an exponential recovery back to the original drift rate \citep{1983Lyne, 2006Weltevrede}. We show a stack of consecutive pulses in Fig.~\ref{fig:DriftingSubpulses}, generated from UTMOST filterbank data. The horizontal ($P_2$) and vertical ($P_3$) drift bands can easily be identified. We form a two-dimensional Fourier transform (FT) of the single pulse data in the pulse number and pulse longitude domain to determine the drift parameters more accurately. In particular, we determine the parameters from the maximum of the power spectral density, analogue to the technique described by \citet{2002Edwards}. We measure $P_2 = -6.4(2) \: \degr$ and $P_3 = 4.67(7) \: P$, where the uncertainty is determined from the bin width of the FT. The negative sign indicates negative drift. This is consistent with the measurement of $P_2 = -6.5_{-0.7}^{+0.2} \: \degr$ and $P_3 = 4.7(2) \: P$ by \citet{2006Weltevrede} at 21~cm. The pulsar provides a good example to demonstrate the single pulse capability of the UTMOST system for pulsar observations.
\end{description}

\section{Conclusions and future work}
\label{sec:Conclusions}

We presented the design of the UTMOST pulsar timing programme at the refurbished Molonglo Synthesis Radio Telescope and the first results obtained from analysing a subset of 205 pulsars with timing baselines of 1.4--3 yr. Our conclusions are the following:
\begin{enumerate}
	\item The UTMOST timing system is verified and stable to a precision of $5 \: \mu \text{s}$. The UTMOST data show good agreement with flux density, position and proper motion measurements from the literature.
	
	\item Profile and timing residual shifts due to sampling of a single polarization are negligible within $\pm 2$ hours from the meridian, where the vast majority of observations occur, compared with the current timing precision attainable. At angles beyond that, the shift can be modelled and can potentially be accounted for.
	
	\item Dynamic scheduling, as described in Appendix~\ref{sec:DynamicTelescopeScheduling}, can increase the observing efficiency of an instrument significantly (by a factor of 2--3) in comparison with schedule file-based static scheduling.
	
	\item The locally measured spin-down rates of pulsars deviate significantly from the inferred long-term spin-down, measured over multiple decades. However, the absolute differences are small (median of 0.13 per cent), indicating that the pulsar rotation is generally very stable. The local spin-down rates are most likely dominated by timing noise, mode-changes or recoveries from glitch events.
	
	\item By fitting linear functions to timing positions at multiple epochs spanning 48 yr, it is possible to derive proper motions that are comparable in precision to those from long-term timing observations. This technique allowed us to estimate proper motions for 60 pulsars, of which 24 are newly determined and most are significant improvements. Where interferometric measurements from the literature are available, they are consistent.
	
	\item The derived transverse velocities are best described by a model of two Gaussian velocity components centred at 139 and $463 \: \text{km} \: \text{s}^{-1}$ with standard deviations of $76$ and $124 \: \text{km} \: \text{s}^{-1}$ and abundance of 1.43/1. The model is preferred at the 84 per cent level over single-component models. However, the transverse velocities of 115 recycled and 156 non-recycled pulsars from the pulsar catalogue are best described by single exponential models with means of 103(10) and $235(19) \: \text{km} \: \text{s}^{-1}$.
	
	\item The vast majority of pulsars leave the Galactic plane, as expected, with 3 pulsars moving mainly inside the Galactic disk and 2 pulsars that seem to originate outside the central $\pm 15 \degr$ latitude.
	
	\item The pulse duty cycle distributions at 50 and 10 per cent maximum are best fit by log-normal distributions with median values of 2.3 and 4.4 per cent.
	
	\item The known mode-changing pulsar J0742$-$2822 shows transitions, in which the correlation behaviour between timing residual and pulse width changes from being correlated (in phase) to being anti-correlated.
	
	\item The single-pulse capability of the UTMOST system is such that the subpulse drift properties of PSR J0820$-$1350 can be determined reliably.
\end{enumerate}
This paper shows what is already possible with the current data set. Other studies will become feasible in the future, in particular when longer timing baselines are available. Examples are:
\begin{enumerate}
	\item Timing measurements of proper motions and potentially parallaxes
	
	\item Study of red pulsar timing noise. No attempt has been made so far to characterise the noise parameters, except in a few special cases
	
	\item Long-term analysis of spin-down rate changes including glitches
	
	\item Combination of our data with measurements at other frequencies, for example from the Parkes radio telescope, the Murchison Widefield Array (MWA), Australian Square Kilometre Array Pathfinder (ASKAP), MeerKAT, or others. The combined data will allow us to measure and correct for DM variations, which can affect the timing residuals at the highest precision
	
	\item Potential contribution to Pulsar Timing Array efforts to detect signals from gravitational wave emission in the pulsar timing frequency band, for example, to monitor DM changes.
\end{enumerate}

Together with upcoming timing programmes at MeerKAT in the southern and at the Canadian Hydrogen Intensity Mapping Experiment (CHIME) in the northern hemisphere, the UTMOST timing programme strengthens and complements the efforts at established facilities, providing all-sky coverage.

\section*{Acknowledgements}

Parts of this research were conducted by the Australian Research Council Centre of Excellence for All-sky Astrophysics (CAASTRO), through project number CE110001020, and the Australian Research Council grant Laureate Fellowship FL150100148. This work was performed on the gSTAR national facility at Swinburne University of Technology. gSTAR is funded by Swinburne and the Australian Government's Education Investment Fund.





\bibliographystyle{mnras}
\bibliography{utmost_timing_paper}




\appendix


\section{Dynamic telescope scheduling}
\label{sec:DynamicTelescopeScheduling}

Traditionally telescopes have largely been operated using static schedule files, or even by manual operators. However, we are approaching an era in which large arrays of (sufficiently) inexpensive individual telescopes start dominating, not only in radio astronomy (e.g.\ Giant Metrewave Radio Telescope (GMRT), Murchison Widefield Array (MWA), Long Wavelength Array (LWA), Low Frequency Array (LOFAR), Atacama Large Millimeter/submillimeter Array (ALMA), SKA low and mid), but also at higher energies (e.g.\ High Energy Stereoscopic System (H.E.S.S.), Very Energetic Radiation Imaging Telescope Array System (VERITAS), Cherenkov Telescope Array (CTA)). In addition, a few single-dish telescopes are currently being equipped with phased-array feeds (Parkes, Effelsberg, Lovell telescope). Features such as the formation of sub-arrays with vastly different sensitivities, or fields of view, the synthesis of multiple tied-array beams, their placement to achieve varying science cases, and the efficient application of sub-arrays, propel the need for advanced scheduling algorithms. \citet{2003Gomez} give a mathematical description of the scheduling problem and \citet{2011Buchner} considers dynamic scheduling of observation blocks with regard to the SKA.

As a consequence, observatories have started to employ dynamic schedulers, such as the Green Bank Telescope (GBT, \citealt{2009ONeil}). Because of their high observing frequencies of multiple tens of GHz, observations are significantly affected by adverse weather and this is particularly reflected in their scoring algorithm \citep{2009Balser}. In addition, they consider the efficient scheduling of observing blocks, or projects, not individual sources. The observations at Molonglo at 843~MHz are largely unaffected by weather and a strong emphasis is placed on minimising the slew times between individual sources and the wear and tear instead. This is to minimise the overall downtime of the telescope.

We have designed, implemented and tested a dynamic scheduling algorithm at the Molonglo telescope that has increased the source efficiency by a factor of 2.3 to 3 in comparison with schedule file based quasi-static operation. It is implemented in the \textsc{dynamic scheduler} part of the telescope system (see Fig.~\ref{fig:MolongloSoftware}). It tackles the problem of point source scheduling for a fully steerable telescope, which is a distinct and more complex problem than a survey scheduler that operates on a predefined pointing grid. It encompasses short-term scheduling (dynamic operation with a granularity of a few hours at most) and long-term scheduling by enforcing a user-defined cadence and observation priority scheme. Its aims are the following: (1) operate the telescope fully autonomously, (2) fulfil the requirements set in the user-defined observing strategy, (3) maximise the number of timing data points obtained in a single observing night, (4) reduce the wear and tear of the telescope by minimising slews, (5) dynamically react to changes in apparent source flux density (e.g.\ scintillation, or nulling), (6) dynamically react to interesting changes in a source (e.g.\ pulsar glitches, a long-term intermittent pulsar is in the on state) and (7) communicate any errors to the user in real-time.

The core of the scheduler is the scoring function (as opposed to cost function) that dynamically assigns a (positive) score to each source depending on the current UTC $t$, or local sidereal time (LST) $l$, the pointing position of the telescope $\vec{x}_\text{t} = (m_\text{t}, n_\text{t})$, the relative sensitivity of the telescope at that position, the position $\vec{x}_\text{s} = (m_\text{s}, n_\text{s})$ and the flux density of the source, the requested cadence and priority in the observing strategy and the source density around it. We modelled the scoring function iteratively by hand, based on extensive observing experience at Molonglo and the Parkes radio telescope. Consequently, it encodes and mimics the considerations of experienced human observers. We used the sources in the observing strategy in combination with our telescope model to refine the scoring function in sky rotation simulations. In particular, we define the score of source $i$ using a set of weights as:
\begin{equation}
	s_i (\vec{x}_\text{s}, \vec{x}_{t}, t, l) = \frac{\hat{s}_i}{\max \hat{s}_i},
\end{equation}
where
\begin{multline}
	\hat{s}_i (\vec{x}_\text{s}, \vec{x}_{t}, t, l) = w_\text{up}(l) \: w_\text{obs} \: w_\text{slew} (\vec{x}_\text{s}, \vec{x}_{t}, l) \: w_\text{tobs} \: w_\text{prio} \\
	w_\text{sens} \: w_\text{nn} \: w_\text{cad} (t, t_\text{last}),
	\label{eq:ScoringFunction}
\end{multline}
in which the weights parametrize the following:
\begin{enumerate}
	\item $w_\text{up} (t)$ is either zero or unity, depending on whether the source can be observed for at least the slew time to it, its nominal observing time and a small overhead.
	
	\item $w_\text{obs}$ is zero or unity, depending on whether the source was observed in the current observing run.
	
	\item $w_\text{slew} (\vec{x}_\text{s}, \vec{x}_{t}, l) = t_\text{slew, i}^{-2}$ reflects the slew time from the current telescope location to each source, taking into account the Earth's rotation, i.e.\ the movement of sources in the MD, NS plane and the telescope's slew rates.
		
	\item $w_\text{tobs} = t_\text{obs, i}^{-1}$ parametrizes the observation time necessary for a source to reach a given S/N. It takes into account the flux densities of sources at the telescope's centre frequency, the telescope parameters, such as bandwidth and gain, the sky temperatures and the pulsar duty cycles. The observation times are computed using the radiometer equation (equation \ref{eq:RadiometerEquation}).
	
	\item $w_\text{prio}$ lies in the range [0, 100] and indicates the priority that is given to a source. This weight provides a simple way to manually increase the score of a source and a hook for event processing software, such as a real-time glitch detector.
	
	\item $w_\text{sens} (\vec{x}_\text{t})$ reflects the telescope 2D sensitivity function, i.e.\ how the sensitivity of the telescope decreases away from the zenith. Simplistically this can be modelled as $\text{MD}^{-1}$, but we derive the full sensitivity function of Molonglo in Section~\ref{sec:MeridianAndNSGainCurve}.
	
	\item $w_\text{nn} (t)$ represents the next-neighbour density around a source, i.e.\ the number of neighbouring sources that are accessible at a given LST, as defined in (i), in a circle of diameter $5$, $10$, or $20 \degr$ around a given source in the MD, NS plane.
	
	\item $w_\text{cad} (t, t_\text{last})$ represents a weight that increases with the time since the last observation of a source $\theta$, measured in days, and is greater than unity when $\theta$ exceeds the requested cadence of the source $\delta_\text{req}$ in the observing strategy. We define it as:
	\begin{equation}
	w_\text{cad} (\theta) = \begin{cases}
	0			& \text{if} \: \theta < \theta_0 \\
	f(\theta)^2	& \text{otherwise}\\
	\end{cases},
	\end{equation}
	and $f(\theta) = \left( \theta - \theta_0 \right)/\left( \theta_1 - \theta_0 \right)$. A suitable choice of parameters is $\theta_0 = 4 \text{h}$ and $\theta_1 = 1.25 \: \delta_\text{req}$.
\end{enumerate}

All weights are normalised individually by the maximum weight among all sources that are accessible at the given LST and globally as indicated in equation~\ref{eq:ScoringFunction}. Certain weights change dynamically with time, telescope pointing position, or whenever an observation is completed, i.e.\ changes in time since the last observation. These are: $w_\text{up}$, $w_\text{obs}$, $w_\text{slew}$, $w_\text{sens}$ and $w_\text{cad}$. The weights $w_\text{tobs}$ and $w_\text{nn}$ are precomputed before each observing run and $w_\text{prio}$ can be changed manually. The trails of all sources in the MD, NS plane get precomputed as well.

The second part of the dynamic observing system is implemented in \textsc{automatic mode}. Depending on the user-defined observing mode on a per source level, the system can automatically skip to the next source, once a certain S/N or ToA uncertainty is reached, or abandon it, if an intermittent pulsar is off, or the source in a scintillation minimum for example. As a result, the observation times usually deviate from the nominal ones.

\subsection{The optimization problem}
\label{sec:TheOptimizationProblem}

The aim is to find the optimum path through the target cloud given the scores of each source at the LST of the observation. That is, we need to find the path that maximises the total sum of scores per observing run, or the optimum order in which sources should be observed, under various boundary conditions. The initial condition consists of telescope pointing position, LST and last-observed status of each source. The boundary conditions are: the best path needs to have a score and time efficiency that is within a few per cent of the maximum. Additionally, it needs to be within a few per cent of the shortest path in the MD, NS plane.

The computational problem is one of the well known NP-hard problems, because the number of possible paths, or the search parameter space, scales as $O(N!)$, where $N$ is the number of target sources. There is currently no algorithm known that can find the best path, without evaluating all possible solutions. The static problem, where the scores do not change dynamically with time or other external factors, has been studied extensively and is known as the travelling salesman (TSP), or vehicle routing problem (VRP) (e.g.\ \citealt{1956Flood, 2007Applegate}). Various algorithms exist that can find a path that is close to the optimum path, with varying degrees of efficiencies, such as 2-opt, 3-opt, simulated annealing, genetic, ant-colony, or particle swarm based algorithms \citep{2003Gomez}. \citet{2012Colme} summarise the optimization algorithms that are in use at various observatories for block-based scheduling. Our algorithm is modular enough to accommodate any of these optimizers. Our problem is slightly more complicated, as individual observation times can differ significantly from the nominal ones. Therefore it is not possible to precompute and the best path needs to be found within a couple of minutes, while the telescope is operating.

To do this we employ a heuristic optimization method that we term truncated stochastic brute force algorithm. It determines the best path in a piecewise way, currently 5 nodes, or sources, deep into the future. After the 5 sources have been observed it runs again with the new initial conditions. This process is iterated over during an observing run. To reduce the search space further we select the next node from the 10 highest-scoring nodes only. We do that in a pseudo-random way in each path evaluation. We can evaluate all $10^5$ possible paths, or a large fraction of them, in a few minutes on a small number of CPU cores. The reduction in search space provides a significant speed-up and ensures that we obtain paths that are within a few per cent of the local optimum.

\subsection{Performance of the algorithm}
\label{sec:PerformanceOfTheAlgorithm}

\begin{figure}
	\centering
	\includegraphics[width=\columnwidth]{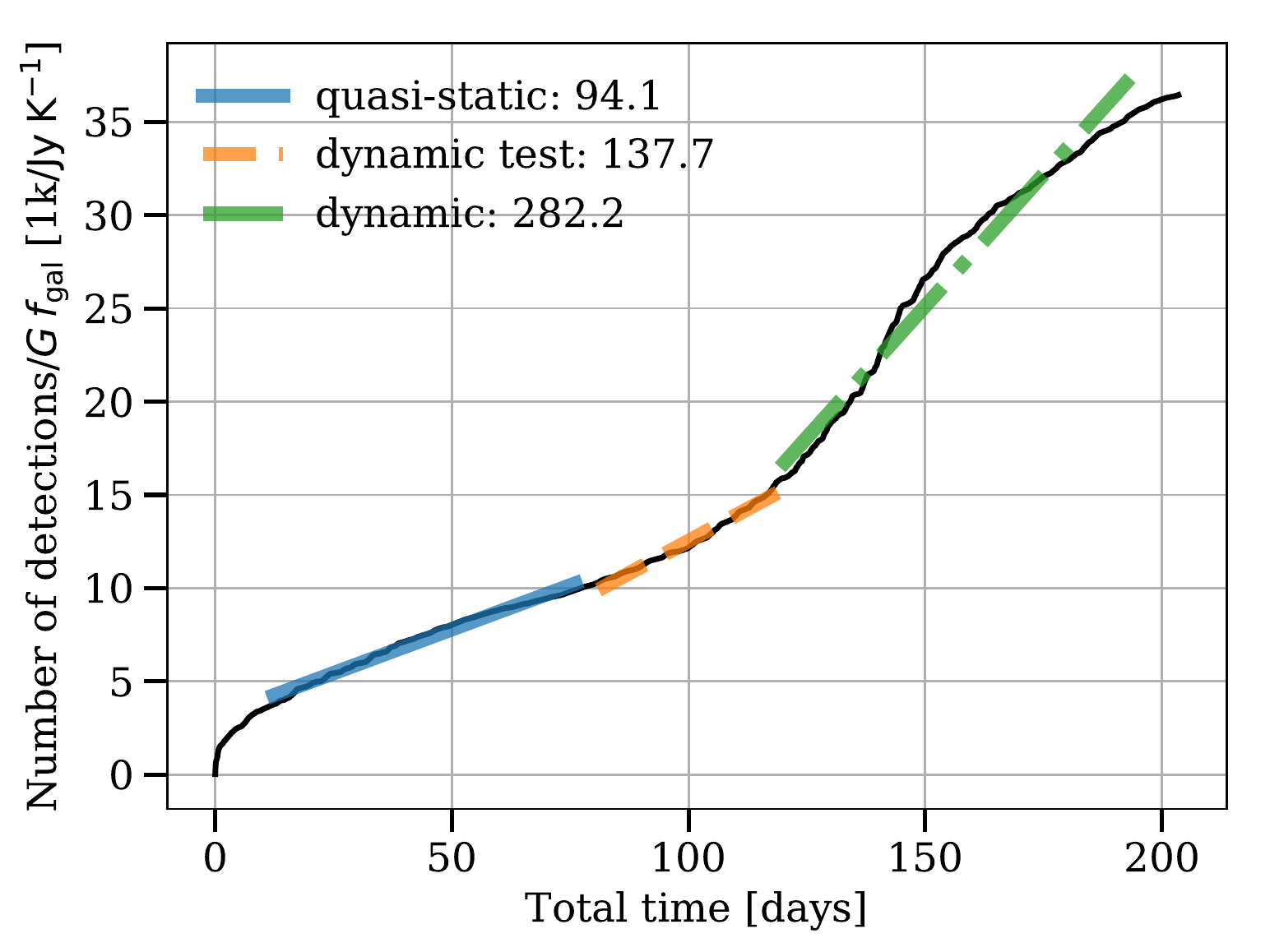}
	\includegraphics[width=\columnwidth]{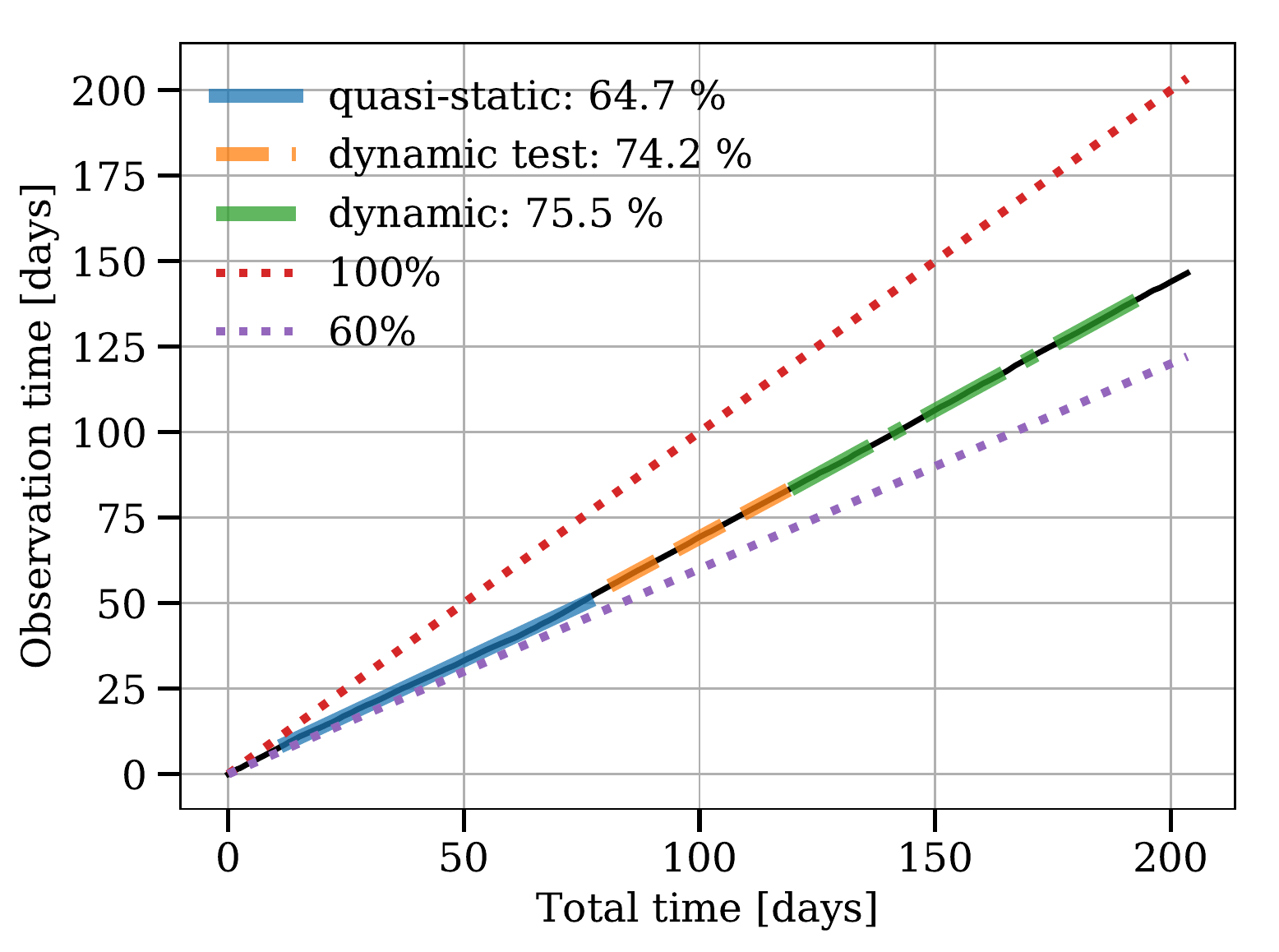}
	\caption{Plots of source (panel a) and time efficiency (panel b) over time. We perform linear fits in different regions of telescope operation as defined in the text. We find that the dynamic scheduler has increased the source efficiency by a factor of 2.3 to 3 and the time efficiency by about 11 per cent in comparison with schedule file based quasi-static operation. Both metrics are defined and more details are provided in Section~\ref{sec:PerformanceOfTheAlgorithm}.}
	\label{fig:SchedulerPerformance}
\end{figure}

Estimating the performance of a scheduling algorithm under realistic conditions is a difficult task. This is because a telescope like Molonglo undergoes significant changes in performance, due to the failure and repair of individual modules, extended periods of maintenance, and competition with other observing duties. Other influencing factors are tweaks to the scheduling algorithm, the observing strategy, changes of the RFI environment, or simply the fraction of time that the Galactic plane is visible outside maintenance hours during the test period. A synthetic benchmark using a telescope simulator eliminates most of these factors, but can only be as realistic as the simulator reflects the instrument.

For this reason, we evaluate the performance of the algorithm using real data obtained over a period of about 12 months, from 2016 May 17th to 2017 May 1st, of operation of the dynamic scheduler at the Molonglo telescope. The first two months were used to test the scheduler (dynamic test) after which it operated in production mode (dynamic). As baseline, we use the preceding 8 months, from 2015 September 1st to 2016 May 1st, in which we operated the telescope using static pre-generated schedule files (quasi-static). Although the source sequences were static, the software already had the ability to automatically skip to the next source once a certain S/N was reached, or could abandon it if it was impossible to reach it in that time. That means that it already had some dynamic features and we do not compare against a fully static (predefined source order and observation times) algorithm. We expect the efficiency improvement to be greater in such a comparison. The user-defined observing strategy, especially the list of pulsars to observe, has been largely constant since 2015 December, with a small adjustment in 2016 June, and therefore does not affect our conclusions in this section.

We define two performance metrics, the first one is the detection rate, i.e.\ the slope of the cumulative number of detections with an S/N greater than 7 over total observation time, which we denote as source efficiency. We evaluate two cases, one in which we include the observations of intermittent pulsars in this detection count, regardless of whether they were detected or not, and one in which they are excluded if they were not detected. This is because their emission states are completely independent of the scheduling algorithm. The second metric is the time efficiency, which is the observation time divided by the total time, which includes slew times. In both cases, we exclude maintenance time, or periods when the telescope was otherwise not observing pulsars. We normalise the source efficiency by the variable gain $G$ of the telescope as derived from observations of a set of flux density reference pulsars (see Section~\ref{sec:FluxDensityCalibration}) and additionally by the fraction of Galactic time $f_\text{gal}$ (defined as 7:00 to 20:00 hours LST) per day available outside of maintenance hours (8:00 to 17:00 AEST weekdays). The telescope sensitivity clearly affects the overall number of detections per unit time, but there is also a minimum observation time per source as defined in the observing strategy, which represents a hard upper limit to the source efficiency. This is mainly relevant for the brightest sources.

We show both performance metrics over time in Fig.~\ref{fig:SchedulerPerformance}. We find that the dynamic scheduler increased the source efficiency by a factor of 2.3 to 3 in comparison with schedule file based quasi-static operation, which is a significant improvement. The time efficiency however increased only by about 11 per cent to nearly 76 per cent. Possible reasons for this are that the time efficiency is the average over both Galactic and non-Galactic time and that the pulsar distribution is sparse outside of Galactic hours, which requires longer slews between sources. There is also a static cool-down time of the backend of between 30 and 60 seconds after each observation, which limits the maximum achievable time efficiency.

Apart from the increase in efficiency, we point out that the dynamic scheduler hugely reduced the human hours spent observing. In its current form, the scheduler only requires minimal manual intervention to start and stop an observing run and when unexpected telescope events are encountered, which are communicated via email to the observers.

\subsection{Future work}
\label{sec:FutureWork}

While the dynamic scheduler is feature-complete, further capabilities that we envision are:
\begin{enumerate}
	\item Zones of avoidance around Solar System objects such as the Sun, Jupiter, or various satellites.
	
	\item The ability to optimize on compound target objects, i.e.\ the ability to observe multiple pulsars in a single telescope pointing. This will become more and more important as the sensitivity of the telescope increases. We expect the source efficiency to roughly increase linearly with the number of pulsars observed per pointing.
	
	\item Trial and implementation of other optimization techniques, for example, an ant colony or genetic algorithm based optimizer.
\end{enumerate}

\section{Best-fitting ephemerides}
\label{sec:BestFittingPulsarEphemerides}

We present the best-fitting ephemerides for pulsars in binary systems in Table~\ref{tab:ParametersBinarySystems} and those for isolated pulsars in Table~\ref{tab:ParametersIsolatedPulsars}. The period, position and DM determination epoch is set to MJD 57600 for all pulsars. The full tables are available in the online version of this paper and can be downloaded in machine-readable form from the \textsc{VizieR} service at the Centre de Donn\'{e}es astronomiques de Strasbourg \citep{2000Ochsenbein}.

\begin{landscape}
\begin{table}
\caption{Rotational parameters of all pulsars in binary systems analysed in this work. The period, position and DM epoch is MJD 57600 for all pulsars. We present their astrometric positions in equatorial coordinates referenced to the ICRF, the spin frequencies and derivatives, the rms timing residuals in $\mu \text{s}$ and milliperiods, the reduced $\chi^2$, the degrees of freedom of the fit and the fitted time spans. We report uncertainties at the $1 \sigma$ level in units of the least significant digit. The flags indicate: w -- ToAs are corrected for underestimation.}
\label{tab:ParametersBinarySystems}
\begin{tabular}{lllllccccc}
\hline
PSRJ    & RAJ           & DECJ                          & $\nu$     & $\dot{\nu}$               & RMS   & $N_\text{ToA}$        & $\chi^2_\text{r}$, dof        & $\Delta t$    & Flags\\
        & (hh:mm:ss)    & ($\degr : \arcmin : \arcsec$) & (Hz)      & ($10^{-15} \: \text{s}^{-2}$)     & ($\mu \text{s}$), ($\text{m}P$)    &  &   & (yr)  &\\
\hline
J0437$-$4715 & 04:37:15.99744(1) & $-$47:15:09.7170(1) & $173.687945349246$(7) & $-1.7280$(2) & 4.3, 0.8 & 458 & 1.2, 445 & 3.1 & w\\
J0737$-$3039A & 07:37:51.2472(4) & $-$30:39:40.67(2) & $44.0540680812$(1) & $-3.412$(8) & 32.9, 1.5 & 55 & 0.9, 42 & 2.3 & --\\
J1141$-$6545 & 11:41:06.970(2) & $-$65:45:19.16(1) & $2.538715900675$(3) & $-27.7659$(2) & 161.3, 0.4 & 205 & 1.3, 194 & 2.5 & --\\
J1302$-$6350 & 13:02:47.62(2) & $-$63:50:08.7(2) & $20.93629909$(1) & $-1002$(1) & 378.2, 7.9 & 34 & 1.5, 24 & 2.1 & --\\
J1909$-$3744 & 19:09:47.4282(1) & $-$37:44:14.774(6) & $339.31568685584$(8) & $-1.611$(7) & 3.1, 1.1 & 48 & 1.1, 38 & 1.9 & w\\
J2145$-$0750 & 21:45:50.4559(2) & $-$07:50:18.590(7) & $62.29588781141$(2) & $-0.1145$(5) & 8.7, 0.5 & 136 & 1.3, 126 & 2.8 & --\\
J2222$-$0137 & 22:22:05.9856(6) & $-$01:37:15.81(2) & $30.47121332913$(2) & $-0.054$(1) & 17.3, 0.5 & 87 & 1.2, 75 & 2.1 & --\\
J2241$-$5236 & 22:41:42.03161(2) & $-$52:36:36.2492(2) & $457.31014943814$(1) & $-1.444$(1) & 3.5, 1.6 & 239 & 1.1, 227 & 2.4 & w\\
\hline
\end{tabular}
\end{table}
\end{landscape}

\begin{landscape}
\begin{table}
\caption{Rotational parameters of all isolated pulsars analysed in this work. The period, position and DM epoch is MJD 57600 for all pulsars. We present their astrometric positions in equatorial coordinates referenced to the ICRF, the spin frequencies and derivatives, the rms timing residuals in $\mu \text{s}$ and milliperiods, the reduced $\chi^2$, the degrees of freedom of the fit and the fitted time spans. We report uncertainties at the $1 \sigma$ level in units of the least significant digit. The flags indicate: w -- ToAs are corrected for underestimation.}
\label{tab:ParametersIsolatedPulsars}
\begin{tabular}{llllllccccc}
\hline
PSRJ    & RAJ           & DECJ                          & $\nu$     & $\dot{\nu}$               & $\ddot{\nu}$  & RMS   & $N_\text{ToA}$        & $\chi^2_\text{r}$, dof        & $\Delta t$    & Flags\\
        & (hh:mm:ss)    & ($\degr : \arcmin : \arcsec$) & (Hz)      & ($10^{-15} \: \text{s}^{-2}$)     & ($10^{-24} \: \text{s}^{-3}$)   & ($\mu \text{s}$), ($\text{m} P$)    &       &   & (yr)  &\\
\hline
J0034$-$0721 & 00:34:08.91(5) & $-$07:21:56(2) & $1.0605000639$(1) & $-0.448$(8) & -- & 2908.3, 3.1 & 62 & 6.9, 57 & 2.1 & --\\
J0134$-$2937 & 01:34:18.692(1) & $-$29:37:17.08(2) & $7.30131498121$(1) & $-4.178$(1) & -- & 114.3, 0.8 & 57 & 1.0, 52 & 1.7 & --\\
J0151$-$0635 & 01:51:22.74(1) & $-$06:35:03.9(3) & $0.68275004308$(1) & $-0.2067$(8) & -- & 615.0, 0.4 & 39 & 1.5, 34 & 2.1 & --\\
J0152$-$1637 & 01:52:10.852(6) & $-$16:37:53.9(1) & $1.200851165189$(7) & $-1.873$(2) & -- & 259.6, 0.3 & 39 & 1.0, 34 & 1.4 & --\\
J0206$-$4028 & 02:06:01.293(5) & $-$40:28:03.62(7) & $1.585913888562$(9) & $-3.0095$(7) & -- & 208.5, 0.3 & 26 & 1.4, 21 & 1.8 & --\\
J0255$-$5304 & 02:55:56.2925(9) & $-$53:04:21.275(9) & $2.233596324088$(1) & $-0.1561$(1) & -- & 77.1, 0.2 & 99 & 2.8, 94 & 2.2 & --\\
J0401$-$7608 & 04:01:51.75(3) & $-$76:08:13.0(1) & $1.83400703758$(3) & $-5.188$(5) & -- & 811.8, 1.5 & 36 & 1.6, 35 & 1.6 & --\\
J0418$-$4154 & 04:18:03.787(8) & $-$41:54:14.53(8) & $1.32079645096$(3) & $-2.293$(3) & -- & 350.9, 0.5 & 19 & 0.7, 14 & 1.6 & --\\
J0450$-$1248 & 04:50:08.77(2) & $-$12:48:07.0(5) & $2.2830308884$(1) & $-0.54$(1) & $-0.00090$(5) & 1763.3, 4.0 & 16 & 1.6, 15 & 1.7 & --\\
J0452$-$1759 & 04:52:34.115(1) & $-$17:59:23.30(4) & $1.821681556540$(2) & $-19.0946$(3) & -- & 185.5, 0.3 & 75 & 2.9, 70 & 2.2 & --\\
J0525+1115 & 05:25:56.500(3) & +11:15:19.1(2) & $2.82137066778$(1) & $-0.585$(2) & -- & 230.4, 0.6 & 24 & 0.8, 19 & 1.7 & --\\
J0533+0402 & 05:33:25.83(2) & +04:01:59.6(8) & $1.03840239455$(4) & $-0.171$(5) & -- & 1107.0, 1.1 & 15 & 0.9, 14 & 1.5 & --\\
J0536$-$7543 & 05:36:30.82(1) & $-$75:43:54.8(1) & $0.802660896055$(5) & $-0.3722$(8) & -- & 279.8, 0.2 & 49 & 1.6, 44 & 2.2 & --\\
J0601$-$0527 & 06:01:58.973(2) & $-$05:27:51.02(6) & $2.525443244407$(6) & $-8.3064$(9) & -- & 273.7, 0.7 & 75 & 0.8, 70 & 2.2 & --\\
J0627+0706 & 06:27:44.181(4) & +07:06:32.8(2) & $2.10134832230$(3) & $-131.618$(3) & -- & 283.4, 0.6 & 36 & 1.6, 35 & 1.7 & --\\
J0630$-$2834 & 06:30:49.371(8) & $-$28:34:42.2(2) & $0.80358373464$(2) & $-4.635$(1) & $-0.50$(1) & 591.8, 0.5 & 37 & 3.2, 31 & 2.2 & --\\
J0646+0905 & 06:46:30.96(5) & +09:05:50(2) & $1.1063007233$(1) & $-0.915$(9) & -- & 1669.3, 1.8 & 43 & 2.7, 38 & 1.7 & --\\
J0659+1414 & 06:59:48.196(3) & +14:14:19.4(3) & $2.59788422918$(2) & $-370.7985$(8) & -- & 818.8, 2.1 & 141 & 2.1, 136 & 3.0 & --\\
J0729$-$1836 & 07:29:32.290(4) & $-$18:36:42.21(7) & $1.96011842692$(1) & $-72.8593$(7) & -- & 1060.9, 2.1 & 160 & 13.5, 155 & 2.9 & --\\
J0738$-$4042 & 07:38:32.262(2) & $-$40:42:39.42(2) & $2.66723044134$(1) & $-9.7822$(4) & $-4.05$(7) & 153.1, 0.4 & 185 & $> 50$, 179 & 2.5 & --\\
J0742$-$2822 & 07:42:49.073(5) & $-$28:22:43.51(7) & $5.99612785971$(9) & $-603.377$(8) & $-22.4$(5) & 560.1, 3.4 & 121 & $> 50$, 114 & 2.4 & --\\
J0758$-$1528 & 07:58:29.055(5) & $-$15:28:08.46(6) & $1.46570344504$(1) & $-3.481$(1) & -- & 293.1, 0.4 & 63 & 0.9, 58 & 1.9 & --\\
J0809$-$4753 & 08:09:43.835(6) & $-$47:53:54.70(6) & $1.82747830437$(1) & $-10.274$(1) & -- & 329.1, 0.6 & 41 & 1.0, 36 & 1.8 & --\\
J0820$-$1350 & 08:20:26.402(3) & $-$13:50:56.40(8) & $0.807668884075$(2) & $-1.3731$(4) & -- & 181.0, 0.1 & 29 & 1.8, 24 & 2.4 & --\\
J0820$-$4114 & 08:20:15.45(2) & $-$41:14:35.3(3) & $1.83336353469$(6) & $-0.075$(5) & -- & 1153.2, 2.1 & 36 & 1.2, 31 & 1.9 & --\\
J0837$-$4135 & 08:37:21.1943(6) & $-$41:35:14.563(6) & $1.330449942221$(2) & $-6.26345$(6) & $0.031$(9) & 68.6, 0.1 & 94 & 7.8, 88 & 2.4 & --\\
J0840$-$5332 & 08:40:33.75(2) & $-$53:32:35.92(9) & $1.38770592258$(2) & $-3.149$(2) & -- & 590.6, 0.8 & 40 & 1.0, 35 & 1.9 & --\\
J0842$-$4851 & 08:42:05.33(3) & $-$48:51:21.1(2) & $1.55191490019$(2) & $-23.029$(3) & -- & 525.2, 0.8 & 13 & 1.5, 8 & 1.9 & --\\
J0846$-$3533 & 08:46:06.040(8) & $-$35:33:41.2(1) & $0.895978340344$(8) & $-1.288$(1) & -- & 587.6, 0.5 & 35 & 1.6, 30 & 1.9 & --\\
J0855$-$3331 & 08:55:38.40(2) & $-$33:31:39.0(2) & $0.78892977830$(1) & $-3.936$(2) & -- & 402.2, 0.3 & 23 & 0.7, 18 & 2.0 & --\\
J0907$-$5157 & 09:07:15.911(4) & $-$51:57:59.21(3) & $3.94387510244$(2) & $-28.556$(1) & -- & 321.2, 1.3 & 56 & 1.6, 51 & 2.0 & --\\
J0908$-$4913 & 09:08:35.429(4) & $-$49:13:05.06(4) & $9.3660112270$(1) & $-1324.723$(4) & $17.8$(6) & 510.6, 4.8 & 121 & $> 50$, 115 & 2.3 & --\\
J0922+0638 & 09:22:14.16(1) & +06:38:28.8(3) & $2.32217901670$(4) & $-73.937$(4) & $-3.3$(3) & 346.6, 0.8 & 41 & 3.5, 35 & 1.9 & --\\
J0924$-$5814 & 09:24:30.83(3) & $-$58:14:05.1(3) & $1.35225504153$(4) & $-9.002$(3) & -- & 1378.4, 1.9 & 27 & 1.0, 22 & 2.0 & --\\
J0934$-$5249 & 09:34:28.21(5) & $-$52:49:27.4(2) & $0.69214827100$(3) & $-2.231$(3) & -- & 1246.4, 0.9 & 23 & 2.1, 18 & 1.7 & --\\
J0942$-$5657 & 09:42:54.424(9) & $-$56:57:43.21(6) & $1.23737204382$(2) & $-60.639$(2) & -- & 227.4, 0.3 & 21 & 0.7, 16 & 1.5 & --\\
J0944$-$1354 & 09:44:28.97(1) & $-$13:54:41.9(2) & $1.75357327378$(3) & $-0.139$(4) & -- & 121.8, 0.2 & 23 & 1.0, 18 & 1.8 & --\\
J0953+0755 & 09:53:09.297(2) & +07:55:36.4(1) & $3.951547889072$(3) & $-3.5872$(2) & -- & 45.4, 0.2 & 25 & 7.7, 20 & 2.1 & --\\
J0955$-$5304 & 09:55:29.474(7) & $-$53:04:16.57(6) & $1.159928626936$(7) & $-4.7433$(7) & -- & 283.4, 0.3 & 29 & 1.1, 24 & 1.7 & --\\
J1001$-$5507 & 10:01:37.88(5) & $-$55:07:08.6(4) & $0.69607303565$(2) & $-25.031$(2) & -- & 4563.0, 3.2 & 77 & $> 50$, 72 & 2.4 & --\\
J1003$-$4747 & 10:03:21.528(7) & $-$47:47:01.2(1) & $3.25654170023$(3) & $-21.966$(2) & -- & 145.6, 0.5 & 11 & 0.8, 6 & 2.0 & --\\
J1012$-$5857 & 10:12:48.48(2) & $-$58:57:48.7(1) & $1.21962189421$(2) & $-26.469$(2) & -- & 727.9, 0.9 & 57 & 1.0, 52 & 2.3 & --\\
J1013$-$5934 & 10:13:31.835(8) & $-$59:34:26.63(5) & $2.25784124142$(1) & $-2.838$(1) & -- & 346.0, 0.8 & 39 & 0.8, 34 & 1.9 & --\\
J1017$-$5621 & 10:17:12.831(3) & $-$56:21:30.52(3) & $1.986244776367$(7) & $-12.387$(1) & -- & 145.5, 0.3 & 28 & 0.6, 23 & 1.6 & --\\
J1034$-$3224 & 10:34:19.49(2) & $-$32:24:27.2(3) & $0.86911889403$(3) & $-0.181$(3) & -- & 590.5, 0.5 & 10 & 0.2, 5 & 1.7 & --\\
\end{tabular}
\end{table}
\end{landscape}

\begin{landscape}
\begin{table}
\contcaption{}
\begin{tabular}{llllllccccc}
\hline
PSRJ    & RAJ           & DECJ                          & $\nu$     & $\dot{\nu}$               & $\ddot{\nu}$  & RMS   & $N_\text{ToA}$        & $\chi^2_\text{r}$, dof        & $\Delta t$    & Flags\\
        & (hh:mm:ss)    & ($\degr : \arcmin : \arcsec$) & (Hz)      & ($10^{-15} \: \text{s}^{-2}$)     & ($10^{-24} \: \text{s}^{-3}$)   & ($\mu \text{s}$), ($\text{m} P$)    &       &   & (yr)  &\\
\hline
J1042$-$5521 & 10:42:00.49(3) & $-$55:21:06.2(4) & $0.85406779038$(2) & $-4.901$(2) & -- & 640.5, 0.5 & 22 & 0.7, 17 & 1.9 & --\\
J1046$-$5813 & 10:46:18.81(2) & $-$58:13:52.0(1) & $2.70688676029$(8) & $-8.401$(6) & -- & 252.5, 0.7 & 21 & 0.8, 16 & 1.6 & --\\
J1048$-$5832 & 10:48:13.05(2) & $-$58:32:06.2(1) & $8.0824184675$(2) & $-6279.20$(2) & $147.0$(10) & 1403.9, 11.3 & 141 & $> 50$, 134 & 2.3 & --\\
J1056$-$6258 & 10:56:25.584(3) & $-$62:58:47.69(2) & $2.36714106186$(1) & $-20.0460$(3) & $0.88$(5) & 265.1, 0.6 & 127 & 3.6, 121 & 2.5 & --\\
J1057$-$5226 & 10:57:59.045(2) & $-$52:26:56.33(2) & $5.07318862008$(2) & $-150.272$(1) & $-0.70$(1) & 223.1, 1.1 & 92 & 7.6, 86 & 2.4 & --\\
J1059$-$5742 & 10:59:00.87(2) & $-$57:42:14.5(1) & $0.84387999092$(2) & $-3.062$(3) & -- & 792.6, 0.7 & 27 & 1.6, 22 & 1.5 & --\\
J1105$-$6107 & 11:05:26.14(1) & $-$61:07:48.94(7) & $15.8222513304$(2) & $-3966.60$(2) & -- & 369.5, 5.8 & 47 & 9.0, 42 & 1.6 & --\\
J1110$-$5637 & 11:10:00.40(1) & $-$56:37:32.6(1) & $1.79129810270$(2) & $-6.604$(3) & -- & 407.3, 0.7 & 32 & 1.2, 27 & 1.5 & --\\
J1116$-$4122 & 11:16:43.083(9) & $-$41:22:44.8(2) & $1.06026074420$(2) & $-8.952$(1) & -- & 256.7, 0.3 & 19 & 1.1, 14 & 1.9 & --\\
J1121$-$5444 & 11:21:19.09(1) & $-$54:44:05.1(1) & $1.86641502544$(3) & $-9.698$(3) & $-4.7$(2) & 278.1, 0.5 & 34 & 1.4, 28 & 1.8 & --\\
J1136+1551 & 11:36:03.072(7) & +15:51:15.5(2) & $0.841809871747$(4) & $-2.6442$(2) & -- & 234.9, 0.2 & 25 & 6.4, 20 & 3.0 & --\\
J1136$-$5525 & 11:36:02.16(2) & $-$55:25:06.9(2) & $2.74188009974$(6) & $-61.847$(3) & $-4.5$(4) & 861.8, 2.4 & 54 & 9.8, 48 & 2.4 & --\\
J1146$-$6030 & 11:46:07.722(6) & $-$60:30:59.49(6) & $3.65798554138$(1) & $-23.925$(1) & -- & 271.2, 1.0 & 40 & 1.0, 35 & 1.9 & --\\
J1157$-$6224 & 11:57:15.207(5) & $-$62:24:50.93(5) & $2.49671858339$(2) & $-24.5156$(7) & $-1.0$(1) & 261.6, 0.7 & 50 & 1.2, 44 & 2.1 & --\\
J1202$-$5820 & 12:02:28.325(5) & $-$58:20:33.64(5) & $2.20846545759$(2) & $-10.3898$(6) & $-1.09$(9) & 235.5, 0.5 & 45 & 1.0, 39 & 2.3 & --\\
J1210$-$5559 & 12:10:05.9861(9) & $-$55:59:03.87(1) & $3.574391887153$(2) & $-9.2671$(2) & -- & 39.7, 0.1 & 30 & 0.5, 25 & 2.1 & --\\
J1224$-$6407 & 12:24:22.274(1) & $-$64:07:53.834(9) & $4.619346760844$(3) & $-105.6997$(3) & -- & 99.0, 0.5 & 124 & 1.9, 119 & 2.1 & --\\
J1239$-$6832 & 12:39:58.8(1) & $-$68:32:30.0(5) & $0.76809485744$(5) & $-7.018$(7) & -- & 1039.2, 0.8 & 15 & 1.2, 10 & 1.4 & --\\
J1243$-$6423 & 12:43:17.101(1) & $-$64:23:23.73(1) & $2.574101117947$(2) & $-29.7838$(2) & -- & 163.3, 0.4 & 193 & $> 50$, 188 & 2.4 & --\\
J1253$-$5820 & 12:53:28.413(3) & $-$58:20:40.59(4) & $3.91392670018$(3) & $-32.257$(2) & -- & 176.8, 0.7 & 33 & 0.6, 28 & 1.6 & --\\
J1305$-$6455 & 13:05:23.39(2) & $-$64:55:28.3(2) & $1.74931666089$(4) & $-12.368$(1) & $-1.0$(3) & 721.0, 1.3 & 51 & 3.3, 45 & 2.4 & --\\
J1306$-$6617 & 13:06:38.14(3) & $-$66:17:22.1(3) & $2.1140406548$(1) & $-26.74$(1) & -- & 1340.6, 2.8 & 22 & 1.3, 17 & 1.5 & --\\
J1312$-$5402 & 13:12:04.71(6) & $-$54:02:42.6(6) & $1.37333511666$(7) & $-0.27$(1) & -- & 1638.3, 2.2 & 17 & 1.0, 12 & 1.6 & --\\
J1312$-$5516 & 13:12:53.52(2) & $-$55:16:48.0(2) & $1.17751983176$(3) & $-7.910$(3) & -- & 832.2, 1.0 & 39 & 0.9, 34 & 1.5 & --\\
J1319$-$6056 & 13:19:20.24(1) & $-$60:56:46.7(1) & $3.51675963951$(4) & $-18.895$(4) & -- & 400.2, 1.4 & 25 & 1.3, 20 & 1.7 & --\\
J1320$-$5359 & 13:20:53.921(3) & $-$53:59:05.00(4) & $3.57477758544$(1) & $-118.152$(1) & -- & 296.4, 1.1 & 75 & 0.8, 70 & 1.9 & --\\
J1326$-$5859 & 13:26:58.217(5) & $-$58:59:29.29(5) & $2.09207816446$(2) & $-14.1887$(8) & $1.4$(1) & 328.8, 0.7 & 61 & 2.1, 55 & 2.1 & --\\
J1326$-$6408 & 13:26:32.44(1) & $-$64:08:43.8(1) & $1.26155286489$(1) & $-4.921$(2) & -- & 396.5, 0.5 & 25 & 1.5, 20 & 1.4 & --\\
J1326$-$6700 & 13:26:02.73(2) & $-$67:00:49.2(2) & $1.84156961608$(6) & $-18.063$(4) & $-4.2$(6) & 1105.9, 2.0 & 47 & 2.7, 41 & 2.0 & --\\
J1327$-$6222 & 13:27:17.176(9) & $-$62:22:45.46(6) & $1.88704458069$(3) & $-66.917$(1) & $9.2$(2) & 602.4, 1.1 & 99 & 25.8, 93 & 2.1 & --\\
J1327$-$6301 & 13:27:07.46(1) & $-$63:01:15.45(8) & $5.08957797555$(6) & $-39.618$(6) & -- & 369.9, 1.9 & 24 & 1.1, 19 & 1.6 & --\\
J1328$-$4357 & 13:28:06.41(1) & $-$43:57:44.6(2) & $1.87722052787$(3) & $-10.760$(3) & -- & 440.1, 0.8 & 50 & 1.8, 45 & 1.7 & --\\
J1355$-$5153 & 13:55:58.685(7) & $-$51:53:54.0(1) & $1.55206115643$(1) & $-6.774$(1) & -- & 206.1, 0.3 & 28 & 0.9, 23 & 1.7 & --\\
J1359$-$6038 & 13:59:58.248(1) & $-$60:38:07.55(1) & $7.84261649184$(2) & $-389.3542$(7) & $1.6$(2) & 114.3, 0.9 & 114 & 6.8, 108 & 2.1 & --\\
J1401$-$6357 & 14:01:52.59(1) & $-$63:57:45.4(1) & $1.18651362796$(2) & $-23.769$(1) & $-2.4$(2) & 709.0, 0.8 & 59 & 15.5, 53 & 2.0 & --\\
J1420$-$5416 & 14:20:29.09(3) & $-$54:16:22.5(5) & $1.06863614348$(3) & $-0.265$(3) & -- & 666.6, 0.7 & 17 & 0.7, 12 & 1.7 & --\\
J1428$-$5530 & 14:28:26.246(9) & $-$55:30:50.1(1) & $1.75348642693$(3) & $-6.417$(2) & -- & 649.0, 1.1 & 44 & 4.1, 39 & 1.9 & --\\
J1430$-$6623 & 14:30:40.735(2) & $-$66:23:05.55(1) & $1.273166302706$(1) & $-4.50222$(8) & -- & 87.1, 0.1 & 69 & 2.1, 64 & 2.2 & --\\
J1435$-$5954 & 14:35:00.21(5) & $-$59:54:48.8(4) & $2.11418109792$(8) & $-6.911$(6) & -- & 1394.8, 2.9 & 21 & 1.5, 16 & 1.9 & --\\
J1440$-$6344 & 14:40:31.26(2) & $-$63:44:48.4(2) & $2.17577321653$(3) & $-5.306$(2) & -- & 388.2, 0.8 & 12 & 0.7, 7 & 1.7 & --\\
J1452$-$6036 & 14:52:51.86(5) & $-$60:36:32(1) & $6.4519415820$(6) & $-60.40$(7) & -- & 275.6, 1.8 & 17 & 0.4, 12 & 1.9 & --\\
J1453$-$6413 & 14:53:32.6546(6) & $-$64:13:16.047(5) & $5.571424351992$(5) & $-85.1832$(2) & $0.22$(3) & 52.4, 0.3 & 117 & 4.6, 111 & 2.2 & --\\
J1456$-$6843 & 14:55:59.923(3) & $-$68:43:39.50(2) & $3.796840090121$(6) & $-1.4264$(3) & -- & 106.2, 0.4 & 58 & 5.8, 53 & 2.6 & --\\
J1507$-$4352 & 15:07:34.156(3) & $-$43:52:03.8(1) & $3.48725495695$(3) & $-19.252$(3) & -- & 121.0, 0.4 & 14 & 0.8, 9 & 1.5 & --\\
J1507$-$6640 & 15:07:48.632(5) & $-$66:40:57.80(3) & $2.81170331278$(1) & $-9.108$(1) & -- & 103.8, 0.3 & 18 & 0.5, 13 & 1.5 & --\\
J1512$-$5759 & 15:12:43.027(7) & $-$58:00:00.2(1) & $7.7700147940$(2) & $-413.544$(5) & $-12$(2) & 437.1, 3.4 & 72 & 1.0, 66 & 1.7 & --\\
J1522$-$5829 & 15:22:42.26(1) & $-$58:29:03.1(1) & $2.52937565292$(2) & $-13.149$(2) & -- & 654.6, 1.7 & 33 & 1.2, 28 & 1.8 & --\\
J1527$-$3931 & 15:27:58.78(1) & $-$39:31:36.6(8) & $0.41363243245$(2) & $-3.265$(2) & -- & 707.4, 0.3 & 15 & 0.5, 10 & 1.6 & --\\
\end{tabular}
\end{table}
\end{landscape}

\begin{landscape}
\begin{table}
\contcaption{}
\begin{tabular}{llllllccccc}
\hline
PSRJ    & RAJ           & DECJ                          & $\nu$     & $\dot{\nu}$               & $\ddot{\nu}$  & RMS   & $N_\text{ToA}$        & $\chi^2_\text{r}$, dof        & $\Delta t$    & Flags\\
        & (hh:mm:ss)    & ($\degr : \arcmin : \arcsec$) & (Hz)      & ($10^{-15} \: \text{s}^{-2}$)     & ($10^{-24} \: \text{s}^{-3}$)   & ($\mu \text{s}$), ($\text{m} P$)    &       &   & (yr)  &\\
\hline
J1527$-$5552 & 15:27:40.75(1) & $-$55:52:08.3(3) & $0.95354468208$(2) & $-10.250$(2) & -- & 559.8, 0.5 & 21 & 1.6, 16 & 1.7 & --\\
J1534$-$5334 & 15:34:08.278(6) & $-$53:34:19.66(8) & $0.730523027416$(3) & $-0.7625$(2) & -- & 383.3, 0.3 & 48 & 1.5, 43 & 2.1 & --\\
J1539$-$5626 & 15:39:14.109(8) & $-$56:26:26.3(1) & $4.10852985575$(4) & $-81.898$(4) & -- & 522.9, 2.1 & 64 & 1.1, 59 & 1.8 & --\\
J1544$-$5308 & 15:44:59.838(2) & $-$53:08:46.93(4) & $5.60055059845$(1) & $-1.8901$(8) & -- & 151.7, 0.8 & 44 & 1.1, 39 & 2.1 & --\\
J1557$-$4258 & 15:57:00.252(2) & $-$42:58:12.39(4) & $3.03778582424$(1) & $-3.0435$(8) & -- & 133.8, 0.4 & 40 & 1.2, 35 & 1.8 & --\\
J1559$-$4438 & 15:59:41.529(1) & $-$44:38:45.74(3) & $3.89018703586$(3) & $-15.451$(1) & -- & 99.7, 0.4 & 48 & 5.7, 43 & 1.5 & --\\
J1600$-$5044 & 16:00:53.015(2) & $-$50:44:21.30(4) & $5.191975911315$(9) & $-136.452$(1) & -- & 258.2, 1.3 & 92 & 2.6, 87 & 2.1 & --\\
J1602$-$5100 & 16:02:18.73(1) & $-$51:00:06.1(2) & $1.15702413207$(3) & $-92.926$(2) & $6.0$(3) & 925.6, 1.1 & 83 & 16.6, 77 & 1.9 & --\\
J1604$-$4909 & 16:04:22.976(2) & $-$49:09:57.98(3) & $3.05419496502$(1) & $-9.503$(2) & $-1.3$(2) & 128.5, 0.4 & 52 & 1.5, 46 & 1.9 & --\\
J1605$-$5257 & 16:05:16.25(1) & $-$52:57:34.5(2) & $1.51972586120$(1) & $-0.592$(1) & -- & 953.8, 1.4 & 46 & 1.7, 41 & 2.1 & --\\
J1613$-$4714 & 16:13:28.99(2) & $-$47:14:26.9(4) & $2.61522196130$(6) & $-4.339$(4) & -- & 366.8, 1.0 & 18 & 0.8, 13 & 1.8 & --\\
J1633$-$5015 & 16:33:00.083(4) & $-$50:15:08.40(8) & $2.839736054539$(8) & $-30.5481$(8) & -- & 328.6, 0.9 & 45 & 1.0, 40 & 2.1 & --\\
J1644$-$4559 & 16:44:49.273(2) & $-$45:59:09.71(5) & $2.197424522481$(9) & $-97.0042$(3) & $4.33$(6) & 532.5, 1.2 & 553 & $> 50$, 547 & 2.9 & --\\
J1645$-$0317 & 16:45:02.033(2) & $-$03:17:57.20(8) & $2.57937109610$(1) & $-11.8159$(8) & $0.7$(1) & 186.9, 0.5 & 39 & $> 50$, 33 & 2.1 & --\\
J1651$-$4246 & 16:51:48.77(1) & $-$42:46:09.4(3) & $1.18472094030$(2) & $-6.682$(2) & -- & 1520.2, 1.8 & 70 & 2.1, 65 & 2.0 & --\\
J1651$-$5222 & 16:51:42.95(1) & $-$52:22:58.6(2) & $1.57465888880$(3) & $-4.492$(2) & -- & 608.3, 1.0 & 30 & 1.3, 25 & 1.7 & --\\
J1700$-$3312 & 17:00:52.78(8) & $-$33:12:52(7) & $0.7362090974$(1) & $-2.556$(6) & -- & 2288.3, 1.7 & 16 & 2.2, 11 & 1.7 & --\\
J1701$-$3726 & 17:01:18.41(3) & $-$37:26:25(2) & $0.40739535951$(2) & $-1.843$(3) & -- & 2197.1, 0.9 & 17 & 1.5, 12 & 1.7 & --\\
J1703$-$1846 & 17:03:51.09(3) & $-$18:46:16(4) & $1.24325189271$(3) & $-2.671$(3) & -- & 670.9, 0.8 & 11 & 1.5, 6 & 1.6 & --\\
J1703$-$3241 & 17:03:22.520(6) & $-$32:41:48.3(4) & $0.82522853905$(1) & $-0.4465$(8) & -- & 530.8, 0.4 & 38 & 1.4, 33 & 1.9 & --\\
J1703$-$4851 & 17:03:54.53(2) & $-$48:52:01.0(5) & $0.716124628386$(9) & $-2.602$(1) & -- & 537.3, 0.4 & 16 & 1.5, 11 & 2.2 & --\\
J1705$-$1906 & 17:05:35.967(2) & $-$19:06:39.6(2) & $3.344586792943$(8) & $-46.2571$(8) & -- & 139.7, 0.5 & 51 & 1.2, 46 & 2.0 & --\\
J1705$-$3423 & 17:05:42.362(3) & $-$34:23:43.4(2) & $3.91501633784$(1) & $-16.499$(2) & -- & 449.1, 1.8 & 76 & 1.2, 71 & 2.2 & --\\
J1707$-$4053 & 17:07:21.75(5) & $-$40:53:57(2) & $1.72111797510$(8) & $-5.681$(5) & -- & 1396.1, 2.4 & 16 & 0.6, 11 & 2.3 & --\\
J1708$-$3426 & 17:08:57.81(3) & $-$34:26:44(2) & $1.44484514048$(7) & $-8.779$(5) & -- & 1251.3, 1.8 & 15 & 2.6, 10 & 1.7 & --\\
J1709$-$1640 & 17:09:26.447(3) & $-$16:40:56.2(4) & $1.53125350177$(2) & $-14.8099$(5) & $0.9$(2) & 91.1, 0.1 & 20 & 0.7, 14 & 1.9 & --\\
J1709$-$4429 & 17:09:42.746(5) & $-$44:29:07.2(1) & $9.7542900321$(2) & $-8851.790$(7) & $203.0$(10) & 413.2, 4.0 & 67 & 12.1, 61 & 2.0 & --\\
J1711$-$5350 & 17:11:53.19(4) & $-$53:50:17.2(6) & $1.11205916018$(3) & $-19.207$(3) & -- & 891.5, 1.0 & 21 & 1.3, 16 & 1.6 & --\\
J1715$-$4034 & 17:15:40.9(2) & $-$40:34:21(5) & $0.48258947531$(4) & $-0.705$(3) & -- & 3770.8, 1.8 & 13 & 1.6, 8 & 2.1 & --\\
J1717$-$4054 & 17:17:52.36(3) & $-$41:03:17.8(7) & $1.12648201282$(3) & $-4.699$(3) & $-1.7$(2) & 1261.6, 1.4 & 25 & 1.5, 19 & 3.0 & --\\
J1720$-$1633 & 17:20:25.24(2) & $-$16:33:34(3) & $0.63873066515$(2) & $-2.375$(2) & -- & 1406.4, 0.9 & 28 & 4.2, 23 & 1.6 & --\\
J1720$-$2933 & 17:20:34.11(2) & $-$29:33:16(3) & $1.61173637045$(8) & $-1.942$(5) & -- & 785.2, 1.3 & 15 & 1.0, 10 & 1.7 & --\\
J1722$-$3207 & 17:22:02.959(2) & $-$32:07:45.3(2) & $2.095742100981$(7) & $-2.8338$(6) & -- & 215.6, 0.5 & 39 & 1.0, 34 & 1.8 & --\\
J1722$-$3712 & 17:22:59.239(6) & $-$37:12:03.7(5) & $4.23402576560$(6) & $-194.935$(8) & -- & 536.3, 2.3 & 48 & 9.5, 43 & 1.5 & --\\
J1739$-$2903 & 17:39:34.295(3) & $-$29:03:02.9(7) & $3.09704932885$(3) & $-75.532$(5) & -- & 332.9, 1.0 & 46 & 1.4, 41 & 1.5 & --\\
J1741$-$3927 & 17:41:18.079(5) & $-$39:27:38.2(2) & $1.95231526531$(2) & $-6.5261$(9) & $3.0$(2) & 234.1, 0.5 & 40 & 1.4, 34 & 1.9 & --\\
J1743$-$3150 & 17:43:36.70(1) & $-$31:50:21(1) & $0.414138298081$(6) & $-20.7152$(8) & -- & 1124.0, 0.5 & 38 & 0.7, 33 & 1.8 & --\\
J1744$-$1134 & 17:44:29.4172(1) & $-$11:34:54.71(1) & $245.42611956900$(5) & $-0.552$(4) & -- & 3.5, 0.9 & 11 & 1.8, 4 & 2.0 & --\\
J1745$-$3040 & 17:45:56.325(1) & $-$30:40:22.6(1) & $2.721563416149$(3) & $-79.0424$(3) & -- & 115.7, 0.3 & 69 & 6.7, 64 & 2.2 & --\\
J1748$-$1300 & 17:48:17.431(10) & $-$13:00:53.4(7) & $2.53720725132$(3) & $-7.800$(3) & -- & 195.9, 0.5 & 10 & 0.5, 5 & 1.8 & --\\
J1751$-$4657 & 17:51:42.196(2) & $-$46:57:26.53(5) & $1.347066944066$(1) & $-2.3551$(1) & -- & 124.7, 0.2 & 40 & 1.3, 35 & 2.3 & --\\
J1752$-$2806 & 17:52:58.713(6) & $-$28:06:33.9(7) & $1.77757096048$(1) & $-25.6971$(9) & -- & 964.3, 1.7 & 99 & $> 50$, 94 & 2.8 & --\\
J1759$-$2205 & 17:59:24.135(5) & $-$22:05:34(2) & $2.16928428071$(2) & $-51.084$(1) & -- & 159.3, 0.3 & 16 & 1.4, 11 & 1.8 & --\\
J1801$-$2920 & 18:01:46.86(1) & $-$29:20:36(1) & $0.924290873991$(8) & $-2.8163$(9) & -- & 354.2, 0.3 & 14 & 0.4, 9 & 1.8 & --\\
J1807$-$0847 & 18:07:38.0266(6) & $-$08:47:43.25(3) & $6.107713282169$(4) & $-1.0674$(3) & -- & 38.2, 0.2 & 41 & 1.0, 36 & 2.3 & --\\
J1807$-$2715 & 18:07:08.50(1) & $-$27:15:02(3) & $1.20804374589$(2) & $-17.816$(2) & -- & 372.0, 0.4 & 18 & 1.3, 13 & 1.9 & --\\
J1808$-$0813 & 18:08:09.37(5) & $-$08:13:03(2) & $1.1414938454$(1) & $-1.609$(8) & -- & 931.7, 1.1 & 12 & 1.2, 7 & 1.6 & --\\
J1816$-$2650 & 18:16:35.395(8) & $-$26:49:54(1) & $1.68666719259$(1) & $-0.1904$(9) & -- & 692.3, 1.2 & 42 & 1.4, 37 & 3.0 & --\\
\end{tabular}
\end{table}
\end{landscape}

\begin{landscape}
\begin{table}
\contcaption{}
\begin{tabular}{llllllccccc}
\hline
PSRJ    & RAJ           & DECJ                          & $\nu$     & $\dot{\nu}$               & $\ddot{\nu}$  & RMS   & $N_\text{ToA}$        & $\chi^2_\text{r}$, dof        & $\Delta t$    & Flags\\
        & (hh:mm:ss)    & ($\degr : \arcmin : \arcsec$) & (Hz)      & ($10^{-15} \: \text{s}^{-2}$)     & ($10^{-24} \: \text{s}^{-3}$)   & ($\mu \text{s}$), ($\text{m} P$)    &       &   & (yr)  &\\
\hline
J1817$-$3618 & 18:17:05.857(9) & $-$36:18:04.6(3) & $2.58384981884$(1) & $-13.506$(1) & -- & 200.7, 0.5 & 11 & 0.8, 6 & 1.7 & --\\
J1818$-$1422 & 18:18:23.87(5) & $-$14:22:45(4) & $3.4306484571$(2) & $-23.96$(1) & -- & 910.9, 3.1 & 18 & 0.6, 13 & 1.8 & --\\
J1820$-$0427 & 18:20:52.592(2) & $-$04:27:37.12(7) & $1.67201171086$(1) & $-17.6831$(3) & $-0.65$(6) & 171.8, 0.3 & 36 & 2.7, 30 & 2.3 & --\\
J1822$-$2256 & 18:22:59.0(1) & $-$22:56:38.0(4) & $0.533541177320$(9) & $-0.3861$(9) & -- & 816.6, 0.4 & 17 & 0.8, 12 & 2.1 & --\\
J1823$-$0154 & 18:23:52.16(1) & $-$01:54:10(2) & $1.31617369962$(5) & $-1.948$(4) & -- & 213.1, 0.3 & 10 & 0.5, 5 & 1.9 & --\\
J1823$-$3106 & 18:23:46.816(6) & $-$31:06:49.2(4) & $3.52042950501$(4) & $-36.377$(4) & -- & 147.7, 0.5 & 19 & 1.0, 14 & 1.5 & --\\
J1824$-$1945 & 18:24:00.4725(10) & $-$19:45:51.6(2) & $5.28154642995$(1) & $-146.1508$(4) & $-7.89$(7) & 103.1, 0.5 & 71 & 8.4, 65 & 2.2 & --\\
J1825$-$0935 & 18:25:30.594(2) & $-$09:35:21.2(2) & $1.30035460953$(1) & $-88.5289$(3) & $2.86$(6) & 433.9, 0.6 & 108 & 4.9, 102 & 2.3 & --\\
J1829$-$1751 & 18:29:43.167(9) & $-$17:51:01(1) & $3.25587939575$(8) & $-58.846$(2) & $-2.9$(4) & 630.5, 2.1 & 43 & 22.2, 37 & 2.2 & --\\
J1830$-$1135 & 18:30:01.7(1) & $-$11:35:09.0(1) & $0.16073093747$(5) & $-1.236$(3) & -- & 5251.3, 0.8 & 14 & 0.9, 9 & 1.8 & --\\
J1832$-$0827 & 18:32:37.020(5) & $-$08:27:03.8(2) & $1.544788609466$(7) & $-152.4912$(8) & -- & 454.2, 0.7 & 50 & 1.5, 45 & 1.9 & --\\
J1833$-$0338 & 18:33:41.889(1) & $-$03:39:04.43(7) & $1.456170355872$(3) & $-88.1273$(3) & -- & 241.2, 0.4 & 61 & 1.8, 56 & 1.7 & --\\
J1833$-$0827 & 18:33:40.264(5) & $-$08:27:32.0(4) & $11.7247184587$(1) & $-1261.91$(1) & -- & 342.5, 4.0 & 18 & 1.4, 13 & 1.7 & --\\
J1834$-$0426 & 18:34:25.632(10) & $-$04:26:15.2(7) & $3.44698922337$(2) & $-0.870$(3) & -- & 419.4, 1.4 & 24 & 1.2, 19 & 2.0 & --\\
J1836$-$1008 & 18:36:53.926(5) & $-$10:08:09.3(4) & $1.77708391440$(4) & $-37.177$(1) & $2.3$(3) & 456.0, 0.8 & 41 & 1.2, 35 & 2.1 & --\\
J1840$-$0809 & 18:40:33.37(2) & $-$08:09:03.7(8) & $1.04638272527$(2) & $-2.574$(2) & -- & 894.7, 0.9 & 29 & 0.7, 24 & 1.9 & --\\
J1841+0912 & 18:41:55.938(8) & +09:12:08.3(2) & $2.62246808519$(1) & $-7.489$(2) & -- & 246.5, 0.6 & 12 & 0.7, 7 & 1.9 & --\\
J1842$-$0359 & 18:42:26.49(2) & $-$04:00:02.1(7) & $0.54349459492$(3) & $-0.151$(1) & -- & 3046.3, 1.7 & 57 & 2.7, 52 & 2.9 & --\\
J1843$-$0000 & 18:43:27.97(2) & $-$00:00:39.8(6) & $1.13593208334$(3) & $-10.042$(2) & -- & 1265.6, 1.4 & 22 & 1.4, 17 & 2.0 & --\\
J1847$-$0402 & 18:47:22.842(2) & $-$04:02:14.4(1) & $1.672775771619$(7) & $-144.6410$(5) & -- & 455.9, 0.8 & 84 & 0.9, 79 & 2.3 & --\\
J1848$-$0123 & 18:48:23.597(2) & $-$01:23:58.8(1) & $1.516448575896$(5) & $-11.9846$(4) & -- & 321.6, 0.5 & 78 & 1.5, 73 & 2.3 & --\\
J1849$-$0636 & 18:49:06.47(1) & $-$06:37:06.1(7) & $0.689011346429$(7) & $-21.9581$(8) & -- & 435.6, 0.3 & 29 & 1.0, 24 & 1.9 & --\\
J1852$-$0635 & 18:52:57.49(2) & $-$06:36:01.8(8) & $1.90782618116$(2) & $-53.252$(2) & -- & 755.3, 1.4 & 36 & 0.9, 31 & 2.0 & --\\
J1854$-$1421 & 18:54:44.30(5) & $-$14:21:19(3) & $0.87214552616$(3) & $-3.180$(3) & -- & 735.8, 0.6 & 10 & 0.6, 5 & 1.8 & --\\
J1857+0212 & 18:57:43.66(2) & +02:12:41.1(7) & $2.4047071645$(2) & $-232.74$(1) & -- & 1261.9, 3.0 & 40 & 1.0, 35 & 2.0 & --\\
J1900$-$2600 & 19:00:47.544(6) & $-$26:00:44.2(7) & $1.633428124583$(7) & $-0.5478$(3) & -- & 261.7, 0.4 & 31 & 1.3, 26 & 2.9 & --\\
J1901+0331 & 19:01:31.770(2) & +03:31:04.96(7) & $1.525657449736$(5) & $-17.3017$(4) & -- & 336.7, 0.5 & 69 & 3.1, 64 & 1.9 & --\\
J1901$-$0906 & 19:01:53.021(9) & $-$09:06:11.9(7) & $0.561189668478$(4) & $-0.5160$(3) & -- & 403.5, 0.2 & 23 & 0.7, 18 & 2.0 & --\\
J1903+0135 & 19:03:29.986(1) & +01:35:38.53(4) & $1.371164754792$(4) & $-7.5674$(4) & -- & 177.1, 0.2 & 65 & 1.6, 60 & 1.7 & --\\
J1903$-$0632 & 19:03:37.950(7) & $-$06:32:22.1(3) & $2.31540809124$(1) & $-18.120$(1) & -- & 280.5, 0.6 & 28 & 0.9, 23 & 2.0 & --\\
J1909+0007 & 19:09:35.264(5) & +00:07:57.0(1) & $0.983329997580$(5) & $-5.3445$(5) & -- & 348.6, 0.3 & 33 & 0.7, 28 & 1.9 & --\\
J1909+0254 & 19:09:38.321(9) & +02:54:49.9(3) & $1.010269404842$(7) & $-5.6138$(7) & -- & 328.1, 0.3 & 22 & 0.6, 17 & 2.0 & --\\
J1909+1102 & 19:09:48.693(1) & +11:02:03.20(7) & $3.52556957552$(4) & $-32.8483$(5) & $3.5$(4) & 111.7, 0.4 & 43 & 0.9, 37 & 1.7 & --\\
J1910$-$0309 & 19:10:29.693(5) & $-$03:09:53.8(3) & $1.98174395505$(2) & $-8.608$(1) & -- & 254.8, 0.5 & 22 & 0.5, 17 & 1.9 & --\\
J1913$-$0440 & 19:13:54.182(2) & $-$04:40:47.6(1) & $1.210742185605$(2) & $-5.9780$(2) & -- & 141.8, 0.2 & 55 & 3.3, 50 & 2.4 & --\\
J1915+1009 & 19:15:29.991(3) & +10:09:43.37(6) & $2.471871539032$(9) & $-93.2267$(8) & -- & 298.4, 0.7 & 47 & 0.9, 42 & 1.8 & --\\
J1916+0951 & 19:16:32.352(5) & +09:51:25.1(2) & $3.70019376632$(6) & $-34.499$(6) & -- & 219.9, 0.8 & 21 & 0.7, 16 & 1.5 & --\\
J1916+1312 & 19:16:58.60(1) & +13:12:52.8(7) & $3.5480500850$(4) & $-46.967$(8) & $44$(3) & 290.0, 1.0 & 18 & 3.6, 12 & 1.8 & --\\
J1919+0021 & 19:19:50.682(7) & +00:21:39.6(2) & $0.785999270002$(5) & $-4.7424$(6) & -- & 399.8, 0.3 & 39 & 0.6, 34 & 2.0 & --\\
J1926+0431 & 19:26:24.46(1) & +04:31:31.9(4) & $0.93102927987$(2) & $-2.134$(1) & -- & 774.0, 0.7 & 59 & 1.3, 54 & 2.0 & --\\
J1932+1059 & 19:32:14.0570(8) & +10:59:33.38(2) & $4.41464565189$(2) & $-22.5455$(3) & $-1.21$(8) & 65.1, 0.3 & 50 & 8.9, 44 & 2.4 & --\\
J1941$-$2602 & 19:41:00.414(5) & $-$26:02:05.7(3) & $2.48226091397$(2) & $-5.894$(1) & -- & 127.1, 0.3 & 22 & 0.7, 17 & 1.5 & --\\
J1943$-$1237 & 19:43:25.45(2) & $-$12:37:44(1) & $1.028351509805$(8) & $-1.753$(1) & -- & 378.6, 0.4 & 20 & 1.2, 15 & 2.0 & --\\
J1946$-$2913 & 19:46:51.75(1) & $-$29:13:47.4(8) & $1.04226478934$(1) & $-1.619$(1) & -- & 506.3, 0.5 & 20 & 0.4, 15 & 1.8 & --\\
J1949$-$2524 & 19:49:25.63(5) & $-$25:24:12(4) & $1.04425628277$(2) & $-3.568$(2) & -- & 493.4, 0.5 & 11 & 0.4, 6 & 1.9 & --\\
J2006$-$0807 & 20:06:16.33(2) & $-$08:07:01(1) & $1.72155151632$(6) & $-0.133$(4) & -- & 820.4, 1.4 & 63 & 0.8, 58 & 2.0 & --\\
J2033+0042 & 20:33:31.1(1) & +00:42:23(5) & $0.19946542820$(2) & $-0.386$(2) & -- & 4303.6, 0.9 & 24 & 1.3, 19 & 1.9 & --\\
J2046$-$0421 & 20:46:00.16(1) & $-$04:21:25.9(6) & $0.64643778920$(1) & $-0.6148$(7) & -- & 931.6, 0.6 & 62 & 1.2, 57 & 2.0 & --\\
\end{tabular}
\end{table}
\end{landscape}

\begin{landscape}
\begin{table}
\contcaption{}
\begin{tabular}{llllllccccc}
\hline
PSRJ    & RAJ           & DECJ                          & $\nu$     & $\dot{\nu}$               & $\ddot{\nu}$  & RMS   & $N_\text{ToA}$        & $\chi^2_\text{r}$, dof        & $\Delta t$    & Flags\\
        & (hh:mm:ss)    & ($\degr : \arcmin : \arcsec$) & (Hz)      & ($10^{-15} \: \text{s}^{-2}$)     & ($10^{-24} \: \text{s}^{-3}$)   & ($\mu \text{s}$), ($\text{m} P$)    &       &   & (yr)  &\\
\hline
J2048$-$1616 & 20:48:35.77(4) & $-$16:16:47(2) & $0.509792367551$(3) & $-2.8489$(2) & -- & 409.6, 0.2 & 56 & 3.0, 51 & 2.8 & --\\
J2053$-$7200 & 20:53:47.283(9) & $-$72:00:42.45(4) & $2.92966112970$(2) & $-1.696$(1) & -- & 108.7, 0.3 & 17 & 0.6, 12 & 1.7 & --\\
J2116+1414 & 21:16:13.739(6) & +14:14:21.0(2) & $2.27193569861$(2) & $-1.491$(2) & -- & 235.0, 0.5 & 26 & 0.6, 21 & 1.9 & --\\
J2144$-$3933 & 21:44:12.05(5) & $-$39:33:59.0(9) & $0.117511188481$(3) & $-0.0063$(3) & -- & 2470.5, 0.3 & 39 & 1.1, 34 & 1.9 & --\\
J2155$-$3118 & 21:55:13.61(2) & $-$31:18:52.8(7) & $0.97087088286$(2) & $-1.169$(2) & -- & 433.6, 0.4 & 23 & 0.4, 18 & 1.8 & --\\
J2324$-$6054 & 23:24:27.15(4) & $-$60:54:05.8(3) & $0.425987202195$(9) & $-0.4676$(8) & -- & 799.5, 0.3 & 29 & 0.9, 24 & 1.9 & --\\
J2330$-$2005 & 23:30:26.974(5) & $-$20:05:29.5(2) & $0.608411174930$(2) & $-1.7144$(2) & -- & 444.9, 0.3 & 61 & 1.6, 56 & 2.2 & --\\
J2346$-$0609 & 23:46:50.51(4) & $-$06:10:01(1) & $0.846407381996$(7) & $-0.9691$(6) & -- & 436.8, 0.4 & 86 & 1.1, 81 & 2.0 & --\\
\hline
\end{tabular}
\end{table}
\end{landscape}

\section{Flux densities and pulse widths}
\label{sec:FluxDensitiesAndPulseWidths}

We present the calibrated median flux densities at  843~MHz of all pulsars analysed in this work in Table~\ref{tab:FluxWidths}, including their robust modulation indices. In addition to the statistical error, we added a 5 per cent systematic uncertainty to reflect the error introduced by the calibration and gain estimation. We also list their pulse widths at 50 and 10 per cent maximum ($\text{W}_{50}, \text{W}_{10}$) estimated either from high-S/N profiles or from the smoothed standard profiles. Our data set contains 7 pulsars with interpulses, for which we estimated the pulse width as the sum of all profile components.

\begin{table}
\caption{Flux densities at 843~MHz, the robust modulation indices of the flux density time series and the pulse widths at 50 and 10 per cent maximum for all pulsars analysed in this work. Pulsars with interpulses are marked with the superscript ``i". Their $W_{10}$ measurements are the sum of all pulse components. We report uncertainties at the $1 \sigma$ level.}
\label{tab:FluxWidths}
\begin{tabular}{lcccc}
\hline
PSRJ    &   $S_{843}$   & $m_\text{r}$      & $\text{W}_{50}$       & $\text{W}_{10}$\\
        &   (mJy)       &                   & ($\degr$)             & ($\degr$)\\
\hline
J0034$-$0721 & $6 \pm 1$ & 0.91 & 23.4 & 43.7\\
J0134$-$2937 & $9 \pm 1$ & 0.37 & 14.2 & 23.8\\
J0151$-$0635 & $6 \pm 2$ & 1.21 & 30.5 & 38.4\\
J0152$-$1637 & $3 \pm 2$ & 2.03 & 8.1 & 11.9\\
J0206$-$4028 & $2.2 \pm 0.5$ & 0.87 & 4.6 & 12.4\\
J0255$-$5304 & $7 \pm 2$ & 1.66 & 6.6 & 9.7\\
J0401$-$7608 & $5 \pm 2$ & 1.38 & 13.0 & 19.4\\
J0418$-$4154 & $2.2 \pm 0.3$ & 0.68 & 7.1 & 13.2\\
J0437$-$4715 & $300 \pm 50$ & 1.18 & 14.0 & 111.4\\
J0450$-$1248 & $4.0 \pm 0.7$ & 0.56 & 18.7 & 31.1\\
J0452$-$1759 & $50 \pm 10$ & 1.06 & 17.9 & 24.4\\
J0525+1115 & $5.7 \pm 0.9$ & 0.37 & 15.2 & 20.7\\
J0533+0402 & $3.6 \pm 0.6$ & 0.40 & 7.2 & 14.7\\
J0536$-$7543 & $23 \pm 3$ & 0.66 & 21.5 & 29.9\\
J0601$-$0527 & $6.2 \pm 0.7$ & 0.38 & 9.3 & 22.6\\
J0627+0706$^\text{i}$ & $5.8 \pm 0.7$ & 0.42 & -- & 30.9\\
J0630$-$2834 & $90 \pm 30$ & 1.12 & 19.3 & 37.5\\
J0646+0905 & $5.9 \pm 0.9$ & 0.54 & 11.8 & 22.3\\
J0659+1414 & $8 \pm 2$ & 0.72 & 15.5 & 26.7\\
J0729$-$1836 & $5.6 \pm 0.6$ & 0.50 & 13.4 & 19.4\\
J0737$-$3039A$^\text{i}$ & $8 \pm 1$ & 0.41 & -- & 106.9\\
J0738$-$4042 & $210 \pm 10$ & 0.25 & 26.1 & 40.1\\
J0742$-$2822 & $86 \pm 7$ & 0.44 & 9.2 & 16.2\\
J0758$-$1528 & $4.1 \pm 0.6$ & 0.75 & 4.6 & 8.9\\
J0809$-$4753 & $11 \pm 1$ & 0.29 & 7.4 & 16.0\\
J0820$-$1350 & $24 \pm 3$ & 0.36 & 6.4 & 10.7\\
J0820$-$4114 & $32 \pm 3$ & 0.30 & 94.8 & 118.3\\
J0837$-$4135 & $77 \pm 6$ & 0.33 & 3.8 & 7.6\\
J0840$-$5332 & $6.0 \pm 0.4$ & 0.22 & 10.5 & 17.6\\
J0842$-$4851 & $18 \pm 7$ & 0.64 & 4.9 & 187.8\\
J0846$-$3533 & $9.2 \pm 0.9$ & 0.35 & 6.8 & 16.0\\
J0855$-$3331 & $2.9 \pm 0.5$ & 0.62 & 6.0 & 9.5\\
J0907$-$5157 & $22 \pm 2$ & 0.27 & 12.0 & 34.2\\
J0908$-$4913$^\text{i}$ & $35 \pm 3$ & 0.57 & -- & 47.8\\
J0922+0638 & $10 \pm 2$ & 0.86 & 8.7 & 15.8\\
J0924$-$5814 & $13 \pm 2$ & 0.58 & 19.9 & 38.2\\
J0934$-$5249 & $4.7 \pm 0.4$ & 0.43 & 6.0 & 10.4\\
J0942$-$5657 & $3.6 \pm 0.7$ & 0.81 & 3.7 & 8.6\\
J0944$-$1354 & $5 \pm 2$ & 1.34 & 4.7 & 7.9\\
J0953+0755 & $210 \pm 40$ & 0.64 & 13.3 & 29.3\\
J0955$-$5304 & $6.7 \pm 0.8$ & 0.58 & 3.7 & 14.3\\
J1001$-$5507 & $25 \pm 2$ & 0.24 & 4.0 & 8.6\\
J1003$-$4747 & $3 \pm 1$ & 0.82 & 7.2 & 13.7\\
J1012$-$5857 & $4.4 \pm 0.4$ & 0.39 & 7.0 & 15.4\\
J1013$-$5934 & $5.8 \pm 0.6$ & 0.41 & 4.7 & 29.7\\
J1017$-$5621 & $8.0 \pm 0.8$ & 0.36 & 4.7 & 10.3\\
J1034$-$3224 & $19 \pm 4$ & 0.73 & 48.8 & 101.2\\
J1042$-$5521 & $5.6 \pm 0.6$ & 0.39 & 8.2 & 13.4\\
J1046$-$5813 & $4.1 \pm 0.3$ & 0.30 & 8.2 & 13.3\\
J1048$-$5832 & $13.5 \pm 0.9$ & 0.40 & 19.5 & 28.7\\
J1056$-$6258 & $59 \pm 4$ & 0.30 & 15.9 & 32.0\\
J1057$-$5226$^\text{i}$ & $20 \pm 2$ & 0.69 & -- & 94.2\\
J1059$-$5742 & $5.1 \pm 0.8$ & 0.93 & 5.3 & 10.2\\
J1105$-$6107 & $2.4 \pm 0.2$ & 0.45 & 18.6 & 29.3\\
J1110$-$5637 & $5.4 \pm 0.4$ & 0.37 & 14.9 & 19.3\\
J1116$-$4122 & $7 \pm 4$ & 1.70 & 5.3 & 11.1\\
J1121$-$5444 & $5.1 \pm 0.3$ & 0.20 & 5.2 & 17.4\\
J1136+1551 & $90 \pm 70$ & 1.91 & 10.1 & 13.4\\
\end{tabular}
\end{table}

\begin{table}
\contcaption{}
\begin{tabular}{lcccc}
\hline
PSRJ    &   $S_{843}$   & $m_\text{r}$      & $\text{W}_{50}$       & $\text{W}_{10}$\\
        &   (mJy)       &                   & ($\degr$)             & ($\degr$)\\
\hline
J1136$-$5525 & $14 \pm 1$ & 0.36 & 14.1 & 23.9\\
J1141$-$6545 & $8.5 \pm 0.6$ & 0.52 & 8.7 & 15.1\\
J1146$-$6030 & $8.5 \pm 0.7$ & 0.39 & 15.5 & 20.9\\
J1157$-$6224 & $34 \pm 2$ & 0.25 & 10.8 & 31.3\\
J1202$-$5820 & $10.1 \pm 0.8$ & 0.32 & 7.1 & 12.3\\
J1210$-$5559 & $4.9 \pm 0.4$ & 0.43 & 7.1 & 10.1\\
J1224$-$6407 & $19 \pm 1$ & 0.47 & 7.8 & 15.0\\
J1239$-$6832 & $3.5 \pm 0.8$ & 0.73 & 6.4 & 12.2\\
J1243$-$6423 & $119 \pm 7$ & 0.27 & 5.3 & 8.9\\
J1253$-$5820 & $8.3 \pm 0.7$ & 0.37 & 7.5 & 20.3\\
J1302$-$6350$^\text{i}$ & $4.7 \pm 0.6$ & 0.78 & -- & 143.4\\
J1305$-$6455 & $9.0 \pm 0.7$ & 0.32 & 8.1 & 21.4\\
J1306$-$6617 & $13 \pm 2$ & 0.53 & 27.6 & 63.4\\
J1312$-$5402 & $3.7 \pm 0.7$ & 0.76 & 13.2 & 22.3\\
J1312$-$5516 & $8.5 \pm 0.8$ & 0.47 & 8.3 & 15.9\\
J1319$-$6056 & $4.4 \pm 0.4$ & 0.37 & 7.3 & 18.8\\
J1320$-$5359 & $7.1 \pm 0.5$ & 0.38 & 10.5 & 19.6\\
J1326$-$5859 & $36 \pm 2$ & 0.30 & 9.1 & 31.9\\
J1326$-$6408 & $8.2 \pm 0.7$ & 0.36 & 4.5 & 15.8\\
J1326$-$6700 & $25 \pm 2$ & 0.31 & 31.4 & 40.9\\
J1327$-$6222 & $72 \pm 4$ & 0.21 & 10.7 & 27.2\\
J1327$-$6301 & $15 \pm 1$ & 0.38 & 14.0 & 62.2\\
J1328$-$4357 & $7.6 \pm 0.9$ & 0.66 & 9.0 & 15.4\\
J1355$-$5153 & $4.3 \pm 0.4$ & 0.39 & 3.8 & 8.7\\
J1359$-$6038 & $40 \pm 2$ & 0.33 & 10.0 & 24.6\\
J1401$-$6357 & $20 \pm 1$ & 0.36 & 5.1 & 9.3\\
J1420$-$5416 & $4.6 \pm 0.4$ & 0.34 & 7.0 & 13.6\\
J1428$-$5530 & $15 \pm 1$ & 0.49 & 9.0 & 15.2\\
J1430$-$6623 & $33 \pm 3$ & 0.49 & 2.6 & 11.3\\
J1435$-$5954 & $4.7 \pm 0.5$ & 0.55 & 15.3 & 29.7\\
J1440$-$6344 & $7 \pm 1$ & 0.84 & 11.7 & 24.9\\
J1452$-$6036 & $4.3 \pm 0.7$ & 0.66 & 23.2 & 62.0\\
J1453$-$6413 & $53 \pm 3$ & 0.38 & 6.1 & 13.8\\
J1456$-$6843 & $90 \pm 30$ & 2.07 & 19.5 & 37.4\\
J1507$-$4352 & $8 \pm 2$ & 0.67 & 5.6 & 12.2\\
J1507$-$6640 & $3.5 \pm 0.3$ & 0.39 & 3.6 & 6.9\\
J1512$-$5759 & $25 \pm 1$ & 0.27 & 49.4 & 122.1\\
J1522$-$5829 & $12.9 \pm 0.8$ & 0.28 & 17.4 & 29.5\\
J1527$-$3931 & $2.8 \pm 0.4$ & 0.35 & 5.8 & 8.1\\
J1527$-$5552 & $5.2 \pm 0.5$ & 0.36 & 4.0 & 11.9\\
J1534$-$5334 & $21 \pm 2$ & 0.66 & 5.5 & 16.9\\
J1539$-$5626 & $8.4 \pm 0.6$ & 0.45 & 18.8 & 46.3\\
J1544$-$5308 & $11 \pm 1$ & 0.58 & 8.7 & 24.2\\
J1557$-$4258 & $10.3 \pm 0.6$ & 0.20 & 4.5 & 11.2\\
J1559$-$4438 & $82 \pm 6$ & 0.27 & 8.9 & 18.2\\
J1600$-$5044 & $50 \pm 3$ & 0.24 & 23.6 & 67.8\\
J1602$-$5100 & $20 \pm 1$ & 0.27 & 7.0 & 12.5\\
J1604$-$4909 & $16 \pm 1$ & 0.44 & 5.0 & 11.4\\
J1605$-$5257 & $28 \pm 3$ & 0.63 & 32.7 & 42.5\\
J1613$-$4714 & $3.8 \pm 0.4$ & 0.39 & 8.5 & 12.9\\
J1633$-$5015 & $28 \pm 2$ & 0.23 & 14.8 & 42.1\\
J1644$-$4559 & $1040 \pm 60$ & 0.32 & 20.0 & 53.4\\
J1645$-$0317 & $80 \pm 10$ & 0.79 & 4.3 & 8.6\\
J1651$-$4246 & $68 \pm 4$ & 0.15 & 25.1 & 77.9\\
J1651$-$5222 & $9.9 \pm 0.9$ & 0.39 & 9.0 & 14.9\\
J1700$-$3312 & $4.8 \pm 0.6$ & 0.44 & 9.6 & 16.4\\
J1701$-$3726 & $11 \pm 1$ & 0.48 & 8.7 & 17.9\\
J1703$-$1846 & $5 \pm 1$ & 0.56 & 6.1 & 11.0\\
J1703$-$3241 & $21 \pm 1$ & 0.24 & 12.3 & 15.3\\
J1703$-$4851 & $4.6 \pm 0.8$ & 0.59 & 4.1 & 16.5\\
J1705$-$1906$^\text{i}$ & $19 \pm 2$ & 0.65 & -- & 33.8\\
J1705$-$3423 & $11.0 \pm 0.8$ & 0.35 & 20.6 & 34.5\\
J1707$-$4053 & $52 \pm 5$ & 0.38 & 87.4 & 257.2\\
\end{tabular}
\end{table}

\begin{table}
\contcaption{}
\begin{tabular}{lcccc}
\hline
PSRJ    &   $S_{843}$   & $m_\text{r}$      & $\text{W}_{50}$       & $\text{W}_{10}$\\
        &   (mJy)       &                   & ($\degr$)             & ($\degr$)\\
\hline
J1708$-$3426 & $8.1 \pm 0.7$ & 0.30 & 11.0 & 21.8\\
J1709$-$1640 & $17 \pm 2$ & 0.49 & 6.5 & 13.2\\
J1709$-$4429 & $23 \pm 1$ & 0.27 & 23.0 & 51.1\\
J1711$-$5350 & $3.4 \pm 0.4$ & 0.40 & 5.2 & 10.1\\
J1715$-$4034 & $4.7 \pm 0.7$ & 0.60 & 15.9 & 21.3\\
J1717$-$4054 & $8 \pm 1$ & 1.12 & 17.5 & 41.1\\
J1720$-$1633 & $3.4 \pm 0.4$ & 0.55 & 4.1 & 11.4\\
J1720$-$2933 & $7 \pm 1$ & 0.54 & 15.0 & 22.0\\
J1722$-$3207 & $15.4 \pm 0.9$ & 0.18 & 9.5 & 15.6\\
J1722$-$3712 & $7.5 \pm 0.6$ & 0.38 & 7.0 & 13.8\\
J1739$-$2903$^\text{i}$ & $11 \pm 1$ & 0.56 & -- & 46.4\\
J1741$-$3927 & $11.6 \pm 0.8$ & 0.25 & 7.5 & 16.6\\
J1743$-$3150 & $7.3 \pm 0.6$ & 0.37 & 7.2 & 12.2\\
J1744$-$1134 & $7 \pm 3$ & 1.14 & 13.3 & 22.1\\
J1745$-$3040 & $46 \pm 4$ & 0.52 & 5.3 & 22.0\\
J1748$-$1300 & $7 \pm 1$ & 0.46 & 9.8 & 24.9\\
J1751$-$4657 & $15 \pm 3$ & 0.85 & 7.2 & 10.5\\
J1752$-$2806 & $160 \pm 10$ & 0.53 & 4.7 & 8.4\\
J1759$-$2205 & $3.7 \pm 0.6$ & 0.48 & 3.1 & 9.5\\
J1801$-$2920 & $6.3 \pm 0.8$ & 0.48 & 18.0 & 22.5\\
J1807$-$0847 & $37 \pm 3$ & 0.30 & 11.8 & 26.7\\
J1807$-$2715 & $2.8 \pm 0.4$ & 0.63 & 5.9 & 9.4\\
J1808$-$0813 & $4.0 \pm 0.4$ & 0.29 & 9.2 & 15.1\\
J1816$-$2650 & $7.4 \pm 0.6$ & 0.24 & 11.4 & 39.6\\
J1817$-$3618 & $7.8 \pm 0.8$ & 0.20 & 8.4 & 15.7\\
J1818$-$1422 & $26 \pm 8$ & 1.04 & 103.8 & 246.4\\
J1820$-$0427 & $27 \pm 2$ & 0.24 & 5.8 & 11.3\\
J1822$-$2256 & $12 \pm 2$ & 0.56 & 9.7 & 14.6\\
J1823$-$0154 & $2.0 \pm 0.2$ & 0.25 & 3.4 & 6.5\\
J1823$-$3106 & $11.1 \pm 0.8$ & 0.20 & 6.5 & 13.7\\
J1824$-$1945 & $24 \pm 2$ & 0.35 & 4.8 & 11.0\\
J1825$-$0935 & $24 \pm 2$ & 0.70 & 5.7 & 23.8\\
J1829$-$1751 & $21 \pm 2$ & 0.30 & 6.9 & 19.4\\
J1830$-$1135 & $3.0 \pm 0.4$ & 0.47 & 7.5 & 15.9\\
J1832$-$0827 & $8.1 \pm 0.7$ & 0.45 & 5.6 & 15.0\\
J1833$-$0338 & $8.5 \pm 0.5$ & 0.25 & 3.8 & 10.5\\
J1833$-$0827 & $20 \pm 1$ & 0.22 & 34.4 & 99.5\\
J1834$-$0426 & $35 \pm 3$ & 0.35 & 105.9 & 124.0\\
J1836$-$1008 & $14 \pm 1$ & 0.26 & 9.1 & 18.0\\
J1840$-$0809 & $5.7 \pm 0.6$ & 0.42 & 7.7 & 21.1\\
J1841+0912 & $5.8 \pm 0.7$ & 0.30 & 9.1 & 14.3\\
J1842$-$0359 & $21 \pm 2$ & 0.36 & 11.1 & 71.4\\
J1843$-$0000 & $9.0 \pm 0.6$ & 0.19 & 11.9 & 21.1\\
J1847$-$0402 & $12.7 \pm 0.7$ & 0.20 & 11.7 & 17.4\\
J1848$-$0123 & $30 \pm 2$ & 0.34 & 9.1 & 22.0\\
J1849$-$0636 & $4.9 \pm 0.5$ & 0.42 & 3.7 & 7.7\\
J1852$-$0635 & $11 \pm 1$ & 0.39 & 64.6 & 77.8\\
J1854$-$1421 & $4 \pm 3$ & 1.98 & 6.6 & 13.7\\
J1857+0212 & $6.2 \pm 0.6$ & 0.42 & 18.0 & 40.2\\
J1900$-$2600 & $38 \pm 4$ & 0.30 & 22.5 & 39.6\\
J1901+0331 & $26 \pm 2$ & 0.20 & 6.2 & 21.1\\
J1901$-$0906 & $7 \pm 1$ & 0.59 & 10.0 & 12.5\\
J1903+0135 & $19 \pm 1$ & 0.19 & 4.4 & 10.0\\
J1903$-$0632 & $5.5 \pm 0.7$ & 0.48 & 6.4 & 14.7\\
J1909+0007 & $3.2 \pm 0.4$ & 0.62 & 2.7 & 7.1\\
J1909+0254 & $4.2 \pm 0.3$ & 0.21 & 3.4 & 11.0\\
J1909+1102 & $10 \pm 1$ & 0.41 & 5.0 & 11.2\\
J1909$-$3744 & $3.2 \pm 0.4$ & 0.67 & 7.0 & 14.0\\
J1910$-$0309 & $5.1 \pm 0.5$ & 0.33 & 5.0 & 12.8\\
J1913$-$0440 & $21 \pm 2$ & 0.50 & 3.8 & 7.7\\
J1915+1009 & $3.7 \pm 0.3$ & 0.26 & 4.9 & 10.4\\
J1916+0951 & $5.9 \pm 0.6$ & 0.33 & 12.9 & 16.5\\
J1916+1312 & $5.3 \pm 0.7$ & 0.37 & 7.2 & 12.4\\
\end{tabular}
\end{table}

\begin{table}
\contcaption{}
\begin{tabular}{lcccc}
\hline
PSRJ    &   $S_{843}$   & $m_\text{r}$      & $\text{W}_{50}$       & $\text{W}_{10}$\\
        &   (mJy)       &                   & ($\degr$)             & ($\degr$)\\
\hline
J1919+0021 & $4.1 \pm 0.3$ & 0.35 & 4.8 & 11.3\\
J1926+0431 & $3.5 \pm 0.3$ & 0.50 & 5.2 & 10.3\\
J1932+1059 & $60 \pm 20$ & 1.52 & 10.7 & 21.0\\
J1941$-$2602 & $6.2 \pm 0.8$ & 0.50 & 3.3 & 7.2\\
J1943$-$1237 & $2.8 \pm 0.2$ & 0.13 & 4.1 & 8.5\\
J1946$-$2913 & $3.4 \pm 0.4$ & 0.42 & 5.5 & 9.8\\
J1949$-$2524 & $2.2 \pm 0.6$ & 0.60 & 4.2 & 7.9\\
J2006$-$0807 & $6.9 \pm 0.8$ & 0.64 & 54.4 & 62.3\\
J2033+0042 & $2.5 \pm 0.5$ & 1.02 & 5.9 & 10.7\\
J2046$-$0421 & $4.4 \pm 0.5$ & 0.59 & 4.9 & 8.4\\
J2048$-$1616 & $23 \pm 6$ & 1.33 & 13.3 & 16.4\\
J2053$-$7200 & $8 \pm 2$ & 1.04 & 27.6 & 37.6\\
J2116+1414 & $3.5 \pm 0.4$ & 0.38 & 7.1 & 15.6\\
J2144$-$3933 & $2.2 \pm 0.4$ & 0.99 & 2.7 & 4.5\\
J2145$-$0750 & $15 \pm 3$ & 1.24 & 79.9 & 149.8\\
J2155$-$3118 & $2.1 \pm 0.3$ & 0.66 & 6.1 & 12.4\\
J2222$-$0137 & $2.7 \pm 0.4$ & 0.74 & 6.7 & 13.2\\
J2241$-$5236 & $3.6 \pm 0.6$ & 1.72 & 11.8 & 21.6\\
J2324$-$6054 & $2.3 \pm 0.4$ & 0.96 & 7.4 & 11.5\\
J2330$-$2005 & $4 \pm 1$ & 1.51 & 5.2 & 8.2\\
J2346$-$0609 & $5.2 \pm 0.7$ & 0.51 & 16.9 & 22.9\\
\hline
\end{tabular}
\end{table}


\bsp	
\label{lastpage}
\end{document}